\documentclass[prb,twocolumn,superscriptaddress,showpacs]{revtex4}

\usepackage{graphicx}
\usepackage{amsmath}

\newcommand{\lvec}{\boldsymbol{l}}
\newcommand{\nvec}{\boldsymbol{n}}
\newcommand{\mvec}{\boldsymbol{m}}
\newcommand{\kvec}{\boldsymbol{k}}
\newcommand{\rvec}{\boldsymbol{r}}
\newcommand{\avec}{\boldsymbol{a}}

\begin{document}

\title{Bipolarons from long range interactions: Singlet and triplet
pairs in the screened Hubbard-Fr\"ohlich model on the chain}

\author{J.P.Hague}
\affiliation{Department of Physics and Astronomy, The Open University, Walton Hall, Milton Keynes, MK7 6AA, UK}
\author{P.E.Kornilovitch}
\affiliation{Hewlett-Packard Company, 1000 NE Circle Blvd, Corvallis, Oregon 97330, USA}

\begin{abstract}
We present details of a continuous-time quantum Monte-Carlo algorithm
for the screened Hubbard-Fr\"ohlich bipolaron. We simulate the
bipolaron in one dimension with arbitrary interaction range in the
presence of Coulomb repulsion, computing the effective mass,
binding energy, total number of phonons associated with the
bipolaron, mass isotope exponent and bipolaron radius in a
comprehensive survey of the parameter space. We discuss the role of
the range of the electron-phonon interaction, demonstrating the
evolution from Holstein to Fr\"ohlich bipolarons and we compare the
properties of bipolarons with singlet and triplet pairing. Finally, we
present simulations of the bipolaron dispersion. The band width of the
Fr\"ohlich bipolaron is found to be broad, and the decrease in bandwidth as the two polarons bind into a bipolaron is
found to be far less rapid than in the case of the Holstein
interaction. The properties of bipolarons formed from long range electron-phonon interactions, such as light strongly bound bipolarons and intersite pairing when Coulomb repulsion is large, are found to be
robust against screening, with qualitative differences between Holstein and screened Fr\"ohlich bipolarons found even for interactions
screened within a single lattice site.

\pacs{71.38.Mx, 71.38.Fp, 71.10.Fd}

\end{abstract}




\date{23rd April 2009}

\maketitle

\section{Introduction}

Electron-phonon interactions have significant effects in quasi
one-dimensional solids and can lead to the formation of polarons. As
shown by angle resolved photo-emission spectroscopy, the
quasi-particles of the one-dimensional cuprate SrCuO$_2$
\cite{kim2006a} and the conductor K$_{0.3}$MoO$_{3}$
\cite{perfetti2002a} are significantly modified by interactions with
phonons. Strong electron-phonon interactions are also expected to be
relevant in low dimensional organic semiconductors
\cite{physicsoforganicsemiconductors}. The large phonon mediated
attractions in low-dimensional materials may also lead to pairing of
polarons into bipolarons. In one-dimensional (1D) systems, bipolarons
can have either singlet or triplet symmetry, and the presence of
triplet bipolarons has been observed in disordered conducting polymers
using electron spin-resonance measurements \cite{chauvet1994a}. It is
the purpose of this article to present a comprehensive set of
simulations of the properties of 1D bipolarons that are bound with
long-range electron-phonon coupling.

The polaron problem was originally formulated by Landau and Pekar to
describe the effects of lattice polarization on the properties of
electrons \cite{landau1933a}. As an electron moves through a material,
it polarizes the underlying medium of ions and electrons leading to a
dipole field and thus an electron-phonon interaction. Typically the
interaction causes a cloud of phonons to follow a single electron, and
the combination of the particles is called a polaron. The mass of
polarons may be significantly larger than that of the individual
electrons \cite{kornilovitch1998a}. It is now well established that
polarons are formed in many materials, and are probably one of the
most common excitations in the solid state.

More controversial is the existence of bound pairs of polarons known
as bipolarons. If the electron-phonon interaction is sufficiently
attractive, then it is possible for two polarons to pair
\cite{alexandrovandmott}. It is necessary that phonon-mediated
attraction must overcome the inherent Coulomb repulsion between
electrons in order for the pair to form. Studies of the continuum
limit of the bipolaron have shown that bipolaron may be stable, but
only in a very small region of the parameter space of retarded
attraction and Coulomb repulsion \cite{verbist1991a}.

With the goal of understanding the properties of bipolarons, we have
developed a quantum Monte Carlo technique for numerically
exact simulation of two electrons interacting via both Coulomb
repulsion and phonon mediated attraction \cite{kornilovitch1998a,
hague2007a}. Our approach is formulated in continuous time, and
therefore has the advantage that there are no finite size effects from
the Trotter decomposition. We are also able to consider general long
range forms for the electron-phonon problem. In 1D, the ground state
properties of bipolarons forming through local and nearest neighbor
interactions can be found using an advanced variational technique
which predicted that long range electron-phonon interactions may lead
to light pairs in 1D \cite{bonca2000a, bonca2001b}. We have
subsequently determined that bipolarons interacting via very long
range attractive potentials on triangular lattices may be
exceptionally light \cite{hague2007a,hague2007b}.

\begin{figure*}
\begin{center}
\includegraphics[height=50mm]{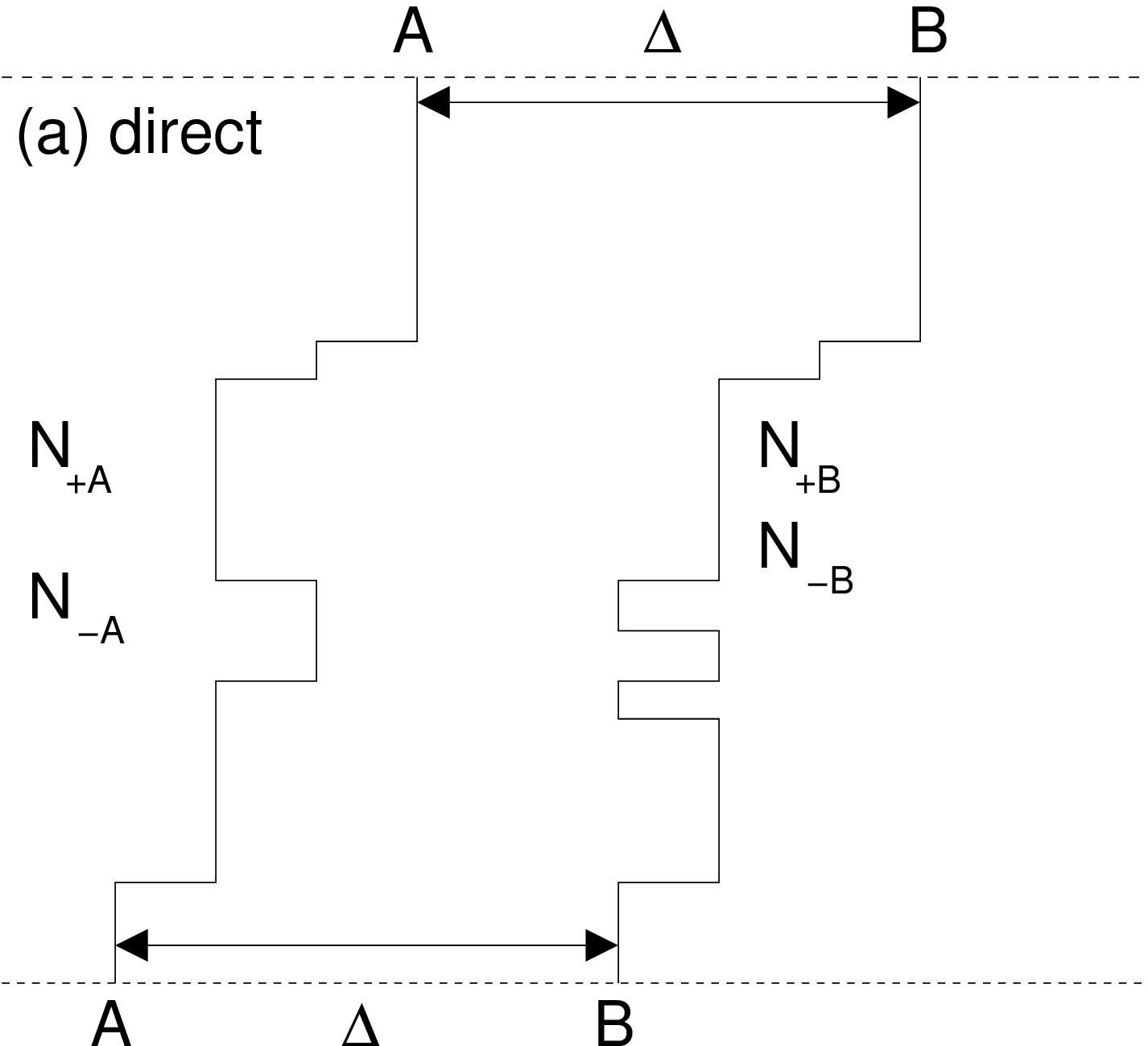}
\includegraphics[height=50mm]{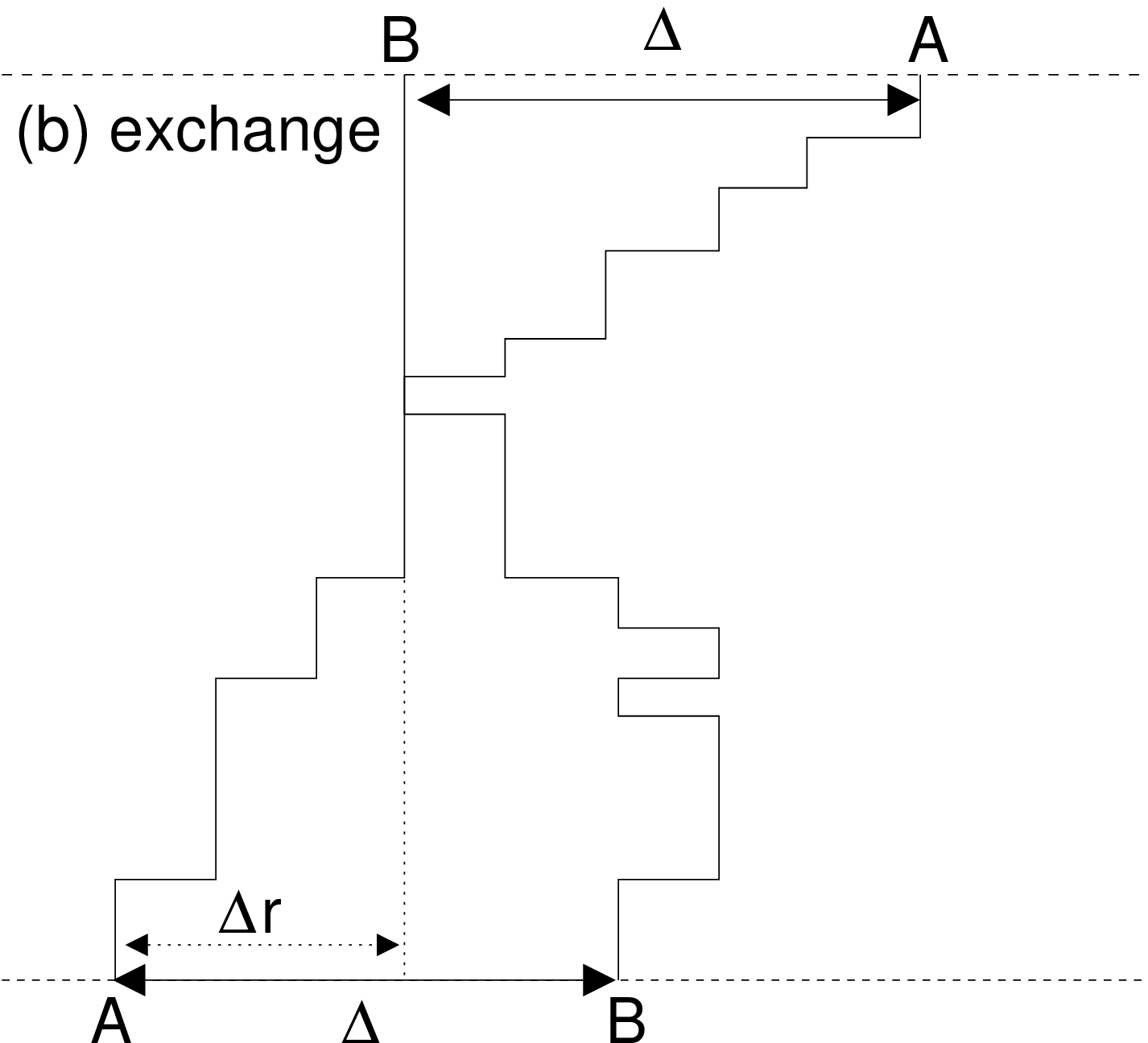}
\includegraphics[height=50mm]{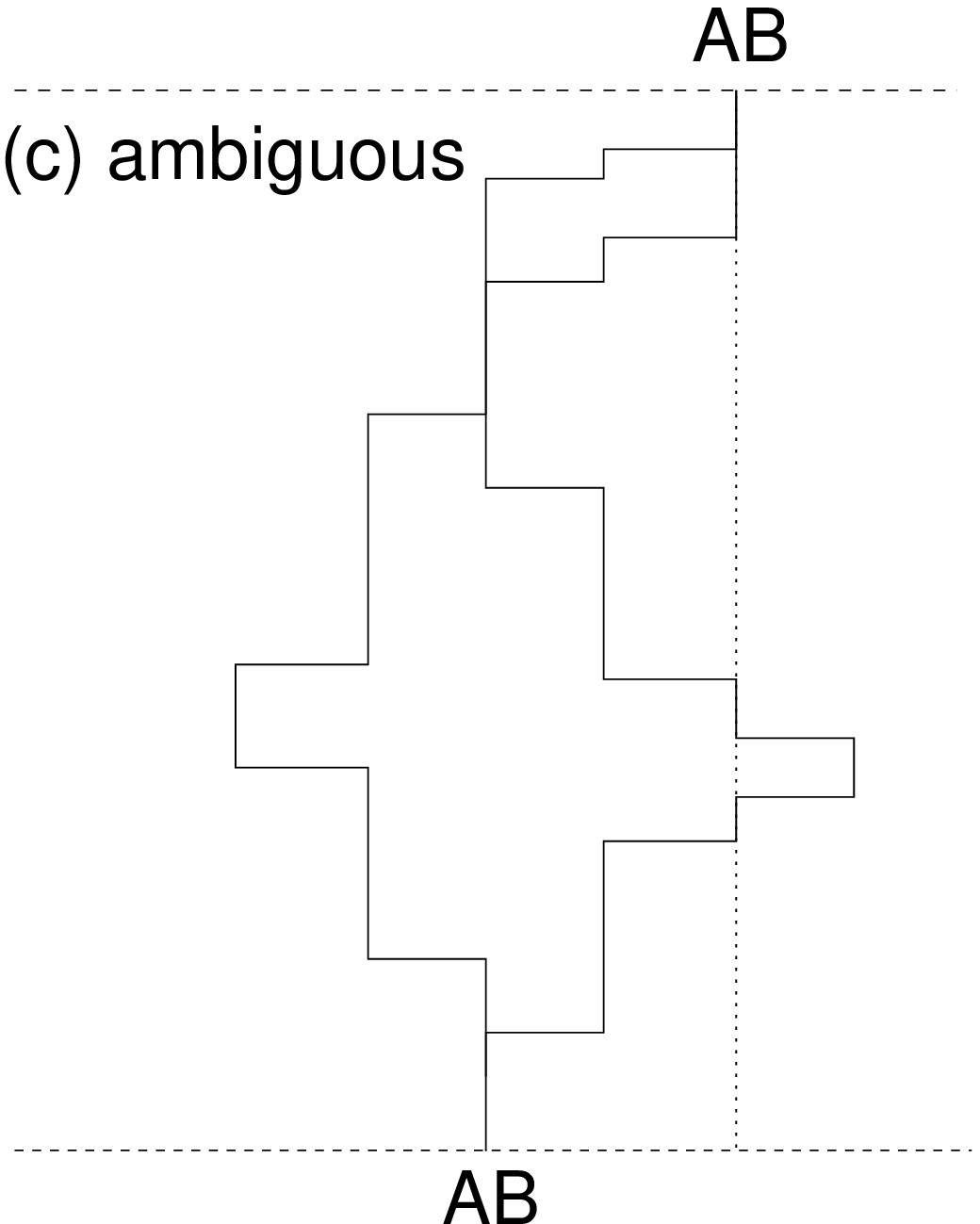}
\end{center}
\caption{Example paths in 1D, showing the notation and concepts of the
bipolaron simulation. (a) Paths in the direct configuration. The ends
of the paths are separated by a distance $\Delta$. There are $N_{+A}$
kinks and $N_{-A}$ antikinks on the 1st path and equivalent numbers
for the 2nd (B) path. (b) In the exchanged configuration, the paths
cross. A feature of the 1D problem is that there is always at least
one shared segment between kinks in the exchanged configuration. Thus
if the Coulomb repulsion is infinite, there can be no exchanged
configuration. (c) There is a third configuration where the endpoints of
the paths sit on the same site, which we call the ambiguous
configuration. When the singlet properties are calculated, all path
configurations are associated with a factor 1. For triplet properties,
direct configurations have a sign 1, exchanged a sign -1 and ambiguous
have sign 0.}
\label{fig:examplepath}
\end{figure*}

Our algorithm can compute numerically exact bipolaron
dispersions. Previously, dispersions of the Hubbard-Holstein model
have been computed on small lattices, using exact-diagonalization
techniques at relatively strong coupling \cite{wellein1996a}. The
intermediate coupling dispersions of a modified Hubbard-Holstein model
have been computed using an approximate variational technique
\cite{eagles2007a}. In principle, approximate dispersions could also
be extracted from the spectral functions in
Ref. \onlinecite{hohenadler2005a}. Bipolaron dispersions computed by
El Shawish for the Hubbard-Holstein model using the same method as in Ref. \cite{shawish2003a} were also presented in
Ref. \onlinecite{hohenadler2005a}.

Our aim here is to simulate the singlet and triplet pairs that form in
systems described by a generic Hubbard-Fr\"ohlich Hamiltonian with
local Coulomb repulsion and long-range electron-phonon interaction
\cite{alexandrov2002a}. A crucial feature of this model Hamiltonian is
that it treats both strong electronic repulsion and attraction from
electron-phonon interaction on equal footing. The Hamiltonian has the
following form,
\begin{eqnarray}
H & = & - t \sum_{\langle \nvec \nvec' \rangle\sigma}
c^{\dagger}_{\nvec'\sigma} c_{\nvec\sigma} + \sum_{
\nvec\nvec'\sigma}\frac{V(\nvec,\nvec')}{2} c^{\dagger}_{\nvec\sigma}
c_{\nvec\sigma}c^{\dagger}_{\nvec'\bar{\sigma}} c_{\nvec'\bar{\sigma}}\nonumber\\&& +
\sum_{\mvec} \frac{\hat{P}^{2}_{\mvec}}{2M} + 
\sum_{\mvec} \frac{\xi^{2}_{\mvec} M\omega^2}{2} -
\sum_{\nvec\mvec\sigma} f_{\mvec}(\nvec)
c^{\dagger}_{\nvec\sigma} c_{\nvec\sigma} \xi_{\mvec}\nonumber
\: .
\label{eqn:hamiltonian}
\end{eqnarray}
Each ion has a displacement $\xi_{\mvec}$. Sites labels are $\nvec$ or
$\mvec$ for electrons and ions respectively. $c$ operators annihilate
electrons. The phonons are Einstein oscillators with frequency
$\omega$ and mass $M$.  $\langle\nvec\nvec'\rangle$ denote pairs of
nearest neighbor sites, and $\hat{P}_{\mvec}$ are the momentum
operators of the ions. The instantaneous interaction $V(\nvec,\nvec')$
has an on-site component $U$ which we always treat as repulsive and an
off-site component $V$. We consider 3 forms for the electron ion force
function, $f$. The Holstein interaction is
$f_{\mvec}(\nvec)=\kappa\delta_{\nvec\mvec}$, leading to a
Hubbard-Holstein model. The near-neighbor model introduced by Bon\v{c}a
and Trugman has the form $f_{\mvec}(\nvec)=
\kappa[\delta_{\nvec,\mvec+1/2} + \delta_{\nvec,\mvec-1/2}]/2$. The
Fr\"ohlich interaction force is long-range,
$f_{\mvec}(\nvec)=\kappa\left[(\mvec-\nvec)^2+1\right]^{-3/2}\exp(-|\mvec-\nvec|/R_{sc})$,
where $\kappa$ is a constant and $R_{sc}$ is the screening radius. It is useful to introduce the dimensionless constant $\lambda=\kappa^2/2M\omega^2 W$ (polaron shift normalized by the half band width, $W=2t$) which characterises the strength of the electron-phonon coupling.

The Lang-Firsov transformation can be used to understand the extreme
anti-adiabatic limit of electron-phonon models, mapping onto a
Hubbard-like system with instantaneous interaction between electrons
\cite{langfirsov}. The Hubbard-Holstein model maps directly to a
Hubbard model with a new interaction $\tilde{U} = U – 2W\lambda$. Since
pairs are bound in 1D for any attractive interaction, the critical
lambda for singlet formation is $\lambda_C = U/2W$.

In the anti-adiabatic limit, the near neighbor model becomes a $UV$
model with $\bar{U}=U-2W\lambda$ and $\bar{V}=2W\lambda\phi(\avec)$. The
near-neighbor model may be solved analytically to determine the
binding condition $\bar{V}>2\bar{U}t/(\bar{U}+4t)$ for singlet
and $\bar{V}>2t$ for triplet pairing
\cite{kornilovitch2004a} (we use the symbols $\bar{U}$ and $\bar{V}$
when discussing the UV model to distinguish from the $U$ in equation
\ref{eqn:hamiltonian}). Singlet and triplet pairs are degenerate for
very large $U$. Since the wavefunction of triplet states has a node at
zero separation, triplet states do not depend on $U$. The high phonon
frequency limit of the Fr\"ohlich model has a long range instantaneous
attractive tail, which makes determination of the binding condition
quite complicated \cite{kornilovitch1995a}. Here, we will investigate
bound states for finite phonon frequencies.

A feature of the models with long-range electron-phonon interaction is that several different configurations of bipolarons can pair. Bipolarons may be of a singlet type, strongly bound into an onsite pair denoted S0, or an intersite pair separated by one lattice site denoted S1, which has a zero in the pair wavefunction for zero separation. Triplets are never formed in the Hubbard-Holstein model, but are permitted in the models with long range interaction.

This article is organized as follows: The continuous time quantum
Monte-Carlo algorithm for bipolarons is outlined in section
\ref{sec:algorithm}. In sections \ref{sec:singlet} and
\ref{sec:triplet}, we simulate singlet and triplet bipolaron
properties. Dispersions and densities of states are presented in
section \ref{sec:dispersion}. We summarize in section
\ref{sec:summary}. The appendix contains complete details of the
Monte-Carlo algorithm.

\section{Algorithm}
\label{sec:algorithm}

The continuous time quantum Monte-Carlo algorithm for bipolarons is an
extension of a similar path-integral method for simulating the polaron
\cite{kornilovitch1998a}. An integration over phonon degrees of
freedom leads to an effective action. In the two particle case, the
action is a functional of two polaron paths in imaginary time
\cite{hague2007b},
\begin{widetext}
\begin{eqnarray}
A[\rvec(\tau)] & = & \frac{z\lambda\bar{\omega}}{2\Phi_0(0,0)}
\int_0^{\bar\beta} \int_0^{\bar\beta} d \tau d \tau'
e^{-\bar{\omega} \bar\beta/2}\left( e^{\bar{\omega}(\bar\beta/2-|\tau-\tau'|)} +
                               e^{-\bar{\omega}(\bar\beta/2-|\tau-\tau'|)} \right)
\sum_{ij}\Phi_0[\rvec_i(\tau),\rvec_j(\tau')]  \\
 & + & \frac{z\lambda\bar{\omega}}{\Phi_0(0,0)}
 \int_0^{\bar\beta} \int_0^{\bar\beta} d \tau d \tau' e^{- \bar{\omega} \tau}
 e^{-\bar{\omega}( \bar\beta - \tau')}
 \sum_{ij}\left( \Phi_{\Delta\rvec}[\rvec_i(\tau),\rvec_j(\tau')] -
 \Phi_0[\rvec_i(\tau),\rvec_j(\tau')]\right) -\int_0^{\beta}V[\rvec_1(\tau),\rvec_2(\tau)]\,d\tau \: .\nonumber
\label{eqn:action}
\end{eqnarray}
\end{widetext}
The retarded interaction mediated by the phonons is characterized by
$\Phi_{\Delta\rvec}[\rvec(\tau),
\rvec(\tau')]=\sum_{\mvec}f_{\mvec}[\rvec(\tau)]f_{\mvec+\Delta\rvec}[\rvec(\tau')]$,
where the vector $\Delta\rvec=\rvec(\beta)-\rvec(0)$ is an offset
between the end configurations. $\Delta\rvec$ enables the computation
of momentum dependent properties \cite{kornilovitch1998a}, and is
schematically introduced in figure \ref{fig:examplepath}.  The indices
$i$ and $j=1,2$ denote which fermion path is being considered. The
dimensionless variables $\bar{\omega}=\hbar \omega/t$ and
$\bar{\beta}=t/k_BT$. An instantaneous Coulomb repulsion,
$V[\rvec_1,\rvec_2]$, is also included. The weight of a configuration 
is given by $\exp(A)$.

Our electron paths are continuous in time, with hops between sites
(kinks) introduced or removed from the path on each Monte-Carlo
step. In contrast to the one-particle algorithm, kinks and antikinks
must be updated in pairs to maintain boundary conditions in imaginary
time. Another significant difference between one- and two-particle
algorithms is particle exchange, leading to bipolarons with singlet
and triplet symmetries. Measurements of the ground-state singlet
bipolaron are not subject to sign problems. It is also interesting to
measure the properties of triplet bipolarons, which depend on the sign
that the wavefunction picks up under exchange. Estimators for the
triplet bipolaron will be discussed later.


We use a correlated weighting scheme for kink addition and removal,
and a simple weighting scheme for path selection and kink type.

\renewcommand{\labelenumi}{R\arabic{enumi}}
\begin{enumerate}
\item Choose a kink with direction (type) $\lvec$ from all possible
hops with equal probability $P_{\lvec}=1/N_k$ and determine the anti
kink as $-\lvec$. $N_k$ is the number of possible hops between sites
(normally the number of nearest neighbors). $P_{\lvec}$ always cancels
in the balance equations.
\label{rule1}
\item Assign path A with probability $1/2$ from both paths, and assign
path B as the other path.
\label{rule2}
\item Choose shift type for the first kink with equal probability
$P_S=1/2$. There are two shift types, the path above the kink
insertion point can be shifted through $\lvec$, or the path below the
insertion point can be shifted through $-\lvec$.
\label{rule3}
\item When selecting the first kink to remove (which has type
$\lvec$), do so with probability $1/N_{A\lvec}(I)$. Where $I$ is the
configuration before removal. That kink is at time $\tau$.
\label{rule4}
\item When selecting the second kink of type $\lvec$ to remove from
path $A$, do so with probability
$p(\tau',\tau)/\sum_{i=1}^{N_{A\lvec}(I)}p(\tau_i,\tau)$. Where $I$ is
the configuration before removal.
\label{rule4a}
\item Always insert the first kink at $\tau$ with probability density $p(\tau) = 1/\beta$.
\label{rule5}
\item Always insert the second kink at $\tau'$ with probability $p(\tau',\tau)$.
\label{rule5a}
\item Choose shift type for kink 2 from kink 1, depending if shifts are correlated (no change in interpath distance) or anticorrelated.
\label{rule6}
\end{enumerate}

The probability $p(\tau',\tau)$ can be chosen either as unity,
recovering previous insertion and removal rules \cite{hague2007a}, or
$p(\tau',\tau)$ can be larger for kinks that are separated by smaller
time differences, potentially leading to improved acceptance ratios at
strong coupling. We use the weighting
$p(\Delta\tau)=\theta(\alpha-\Delta\tau)/2\alpha+1/2\beta$, where
$\theta$ is the step function and $\alpha$ is a parameter that can be
made smaller to improve acceptance ratios ($\Delta\tau=\tau'-\tau$ is
mapped onto the interval $[0,\beta)$ and $0<\alpha<\beta$).


In the two particle case, it is necessary to operate on two kinks
 simultaneously to ensure that the end configurations of the paths
 satisfy the boundary conditions. We have previously discussed how the
 algorithm works without exchange (applicable to ladder systems where
 particles sit on different legs). Four extra Monte-Carlo update rules
 can be used in the exchanged configuration, bringing the potential
 number of binary updates to eight. The new rules are required to
 ensure ergodicity in the exchanged configuration. Here we summarize
 all the possible updates:
\renewcommand{\labelenumi}{\Roman{enumi}}
\begin{enumerate}
\item Two kinks of the same type $\lvec$ are added to (or removed
from) two different paths.
\item A kink-antikink pair is added to (or removed from) one of the two
paths.
\item A kink of type $\lvec$ is inserted into one path, and another
kink of the same type $\lvec$ is removed from the same path. This type
of update shifts kinks in imaginary time. It is not essential.
\item A kink of type $\lvec$ is added to one path, and an antikink $-\lvec$ is
removed from the other path (this is also not essential).
\item Addition (or removal) of kink $\lvec$ and antikink $-\lvec$ on different paths.
\item Kink type $\lvec$ is inserted on one path and kink type $\lvec$ is removed from the other.
\item Addition of kink $\lvec$ and removal of antikink $-\lvec$ on the
same path.
\item Addition (or removal) of a pair of kinks of type $\lvec$ on a
single path.
\end{enumerate}

In the direct configuration, updates (I-IV) may be used, and the shift
types of the two kinks may be identical (correlated) or opposite
(anticorrelated). It is convenient to think of the seperate correlated
and anticorrelated updates as 8 different updates. In the exchanged
configuration, updates (I-IV) are used unchanged, except that the
shifts on both paths must be correlated (i.e. both top, or both
bottom) to maintain the correct boundary conditions in time, thus I-IV
only count as 4 updates. The remaining 4 updates (V-VIII) are used
only with anticorrelated shifts, thus in total, there are 8 possible
updates in the exchanged configuration. Updates (V-VIII) are essential
to access the whole configuration space as they change the interpath
distance $\boldsymbol{\Delta}$ of the exchanged
configuration. Finally in the ambiguous configuration, all the
possible updates may be used. To simplify the scheme and its testing,
we use a minimal update set of I, II, V and VIII. Details of the
probabilities for binary updates are given in the Appendix.


The algorithm can be made more efficient by introducing Monte-Carlo
updates to change the paths from direct into exchanged
configurations. In the path-integral formalism, an exchange of
particles corresponds to swapping the ends of the paths at
$\tau=\beta$. An example of an exchanged configuration is shown in
Fig. \ref{fig:examplepath}. To make the exchange in 1D, the
$\tau=\beta$ end of path B must be shifted by $-\Delta$ lattice sites,
and that of path A through $+\Delta$ sites. This can be achieved by
inserting $\Delta-n$ kinks and removing $n$ antikinks from path A, and
inserting $\Delta-m$ antikinks and removing $m$ kinks from path
B. Details of how to take the continuous time limit of this type of
update are provided in the appendix. This kind of exchange update is
relatively slow, and we attempt this with a small probability of the
order of 10\% on any Monte-Carlo step.

There is another way of carrying out exchange updates. If the paths
occupy the same lattice site at the same point in imaginary time, then
the paths have a segment in common. The paths can be broken at that
segment, with the bottom of path A attached to the top of path B and
the bottom of path B to the top of path A. This type of exchange can
speed up the computation of triplet properties since the update is
very fast and it can be attempted regularly without loss of speed. In
our algorithm, we attempt a common segment exchange with probability
$p_{CS}$ every time an exchange is attempted. Once the attempt is
started, we test for the existence of a common segment. If there are
no shared segments, then the update is rejected and no further update
is attempted on that Monte-Carlo step. Otherwise, the common segment
exchange is accepted with probability $P(C\rightarrow D) = {\rm
min}\left\{1,\exp(A(D) - A(C))\right\}$.

There is another update that can be useful if the action has a similar value for path configurations with small and large separation (as is
typically the case when the bipolaron is only just bound). Then
it can be appropriate to attempt to shift one of the paths through a longer distance when the paths are in a direct configuration. This update enables more rapid sampling of the path configurations and is
especially useful for making accurate measurements of the inter-particle
separation.

After a warmup period, we make a series of measurements. The
measurements are made every few updates to avoid correlation between
Monte-Carlo steps. We also use a blocking procedure in conjunction
with bootstrap resampling for error estimation. In situations where
the measurements take a long time (comparable with the computation of
the action) - such as the estimators for the total energy and isotope
exponent, this sparse sampling acts to speed up the algorithm, since
correlated measurements do not improve estimation of the averages.

Estimators relevant to the current manuscript are the total singlet energy,
\begin{equation}
E = -\lim_{\beta\rightarrow\infty} \left[\left\langle \frac{\partial A}{\partial\beta} \right\rangle
- \frac{1}{\beta}\left\langle \sum_i N_i \right\rangle\right] \: ,
\label{eq:nine}
\end{equation}
where $N_i$ is the number of kinks of type $i$, and angular brackets
denote ensemble averaging.  The number of phonons associated with the
bipolaron,
\begin{equation}
N_{\mathrm{ph}} = - \lim_{\beta\rightarrow\infty}\frac{1}{\bar{\beta}\langle s \rangle}\left\langle s\left. 
\frac{\partial A}{\partial \bar{\omega}}\right|_{\lambda\bar{\omega}}\right\rangle \: ,
\label{eq:ten}
\end{equation}
where the derivative is taken keeping $\lambda \bar\omega$ constant. $s=1$ when considering singlet states, and $s=1,0,-1$ for measurements of triplet states, as described in Fig. \ref{fig:examplepath}.
The bipolaron band energy spectrum can be computed from
\begin{equation}
\epsilon(\kvec) - \epsilon(0)= - \lim_{\beta\rightarrow\infty} \frac{1}{\beta} \ln
\left( \frac{\langle s \cos ({\kvec} \cdot \Delta{\rvec}) \rangle}{\langle s \rangle} \right) \: ,
\label{eqn:spectrum}
\end{equation}
where ${\bf k}$ is the quasi momentum.  By expanding this expression in small
${\bf k}$, the $i$-th component of the inverse effective mass is found to be
\begin{equation}
\frac{1}{m^*_i} = \lim_{\beta\rightarrow\infty}\frac{1}{\beta \hbar^2} \frac{\langle s ( \Delta {\rvec}_i)^2 \rangle}{\langle s \rangle} \: .
\label{eq:twelve}
\end{equation}
We compute the bipolaron radius as the root mean square distance between paths,
\begin{equation}
R_{bp} =\frac{1}{\langle s \rangle}\left\langle s \sqrt{\frac{1}{\beta}\int_{0}^{\beta}(\rvec(\tau)_{1}-\rvec(\tau)_2)^2
d\tau}\right\rangle
\end{equation}
The mass isotope coefficient, $\alpha_{m^*_i} = d \ln m^*_i / d \ln M$,
is calculated as follows
\begin{eqnarray}
\alpha_{m^* _i} & = & \lim_{\beta\rightarrow\infty}\frac{\bar{\omega}}{2} 
\frac{1}{\langle s (\Delta {\rvec}_i)^2 \rangle}
\left[\left\langle s (\Delta {\rvec}_i)^2 
\left. \frac{\partial A}{\partial \bar{\omega}} \right\vert_{\lambda}
\right\rangle\right. \nonumber\\
& & \left. - \frac{\langle s (\Delta {\rvec}_i)^2 \rangle}{\langle s \rangle}
\left\langle s \left. \frac{\partial A}{\partial \bar{\omega}} \right\vert_{\lambda}
\right\rangle \right] \: .
\label{eq:thirteen}
\end{eqnarray}
The singlet-triplet splitting can also be determined,
\begin{equation}
\Delta_{st}  = -\frac{1}{\beta}\ln \left(\langle s\rangle\right)
\end{equation}
Since $\langle s\rangle$ is always less than 1, the singlet-triplet
splitting is always positive and the energy of the triplet state is
higher in energy, as expected since the ground state wavefunction of
two particles should have no nodes.

An alternative way of computing spectra and density of states is also
available. We take the measurement $\langle \delta_{\rvec,\Delta\rvec}
\rangle$ for values of $\rvec$ consistent with a few lattice
spacings. The magnitude of the measurement drops off very rapidly
with $\rvec$. Then the whole spectrum can be computed for any $\kvec$
point, without deciding which $\kvec$ to investigate before the
algorithm starts. The expression for the dispersion calculated in this
way is,
\begin{equation}
\epsilon_{\kvec}-\epsilon_{0} = -\ln\left (\sum_{\rvec} \cos(\kvec.\rvec) \frac{\langle
s \delta_{\rvec,\Delta\rvec} \rangle}{\langle s \rangle}\right)/\beta
\end{equation}
This form is particularly useful for determining the density of
bipolaron states. 

In the following, we have restricted the electron paths to remain
withing 200 lattice sites of each other by enforcing a pairwise
infinite instantaneous potential well dependent on the inter-particle
separation. This stops paths from moving apart indefinitely. The 200
site well is sufficiently large to ensure that finite size effects are
negligible compared with other considerations such as the statistical
error in the Monte Carlo averaging.

\begin{figure*}
\includegraphics[height=85mm,angle=270]{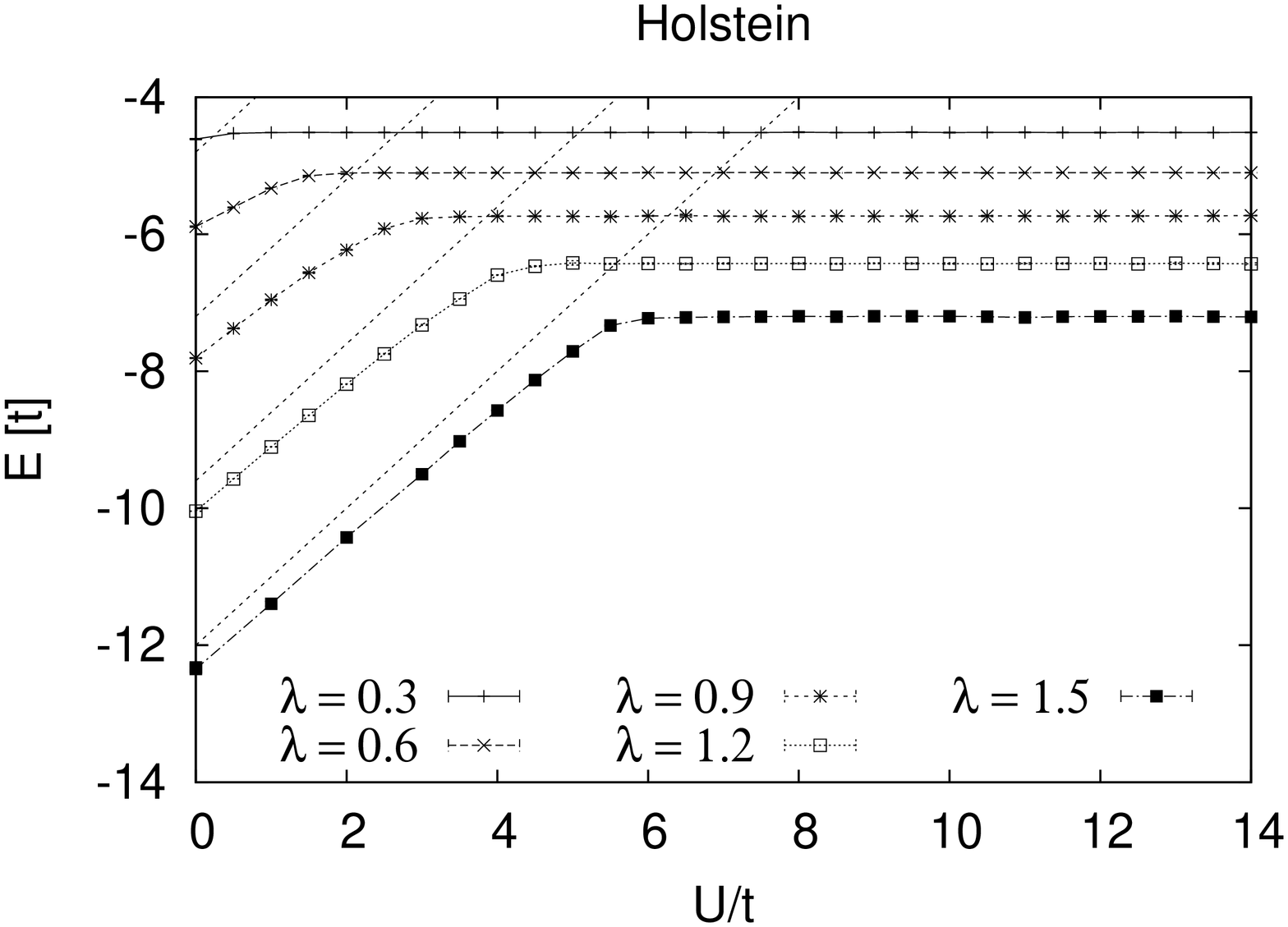}
\includegraphics[height=85mm,angle=270]{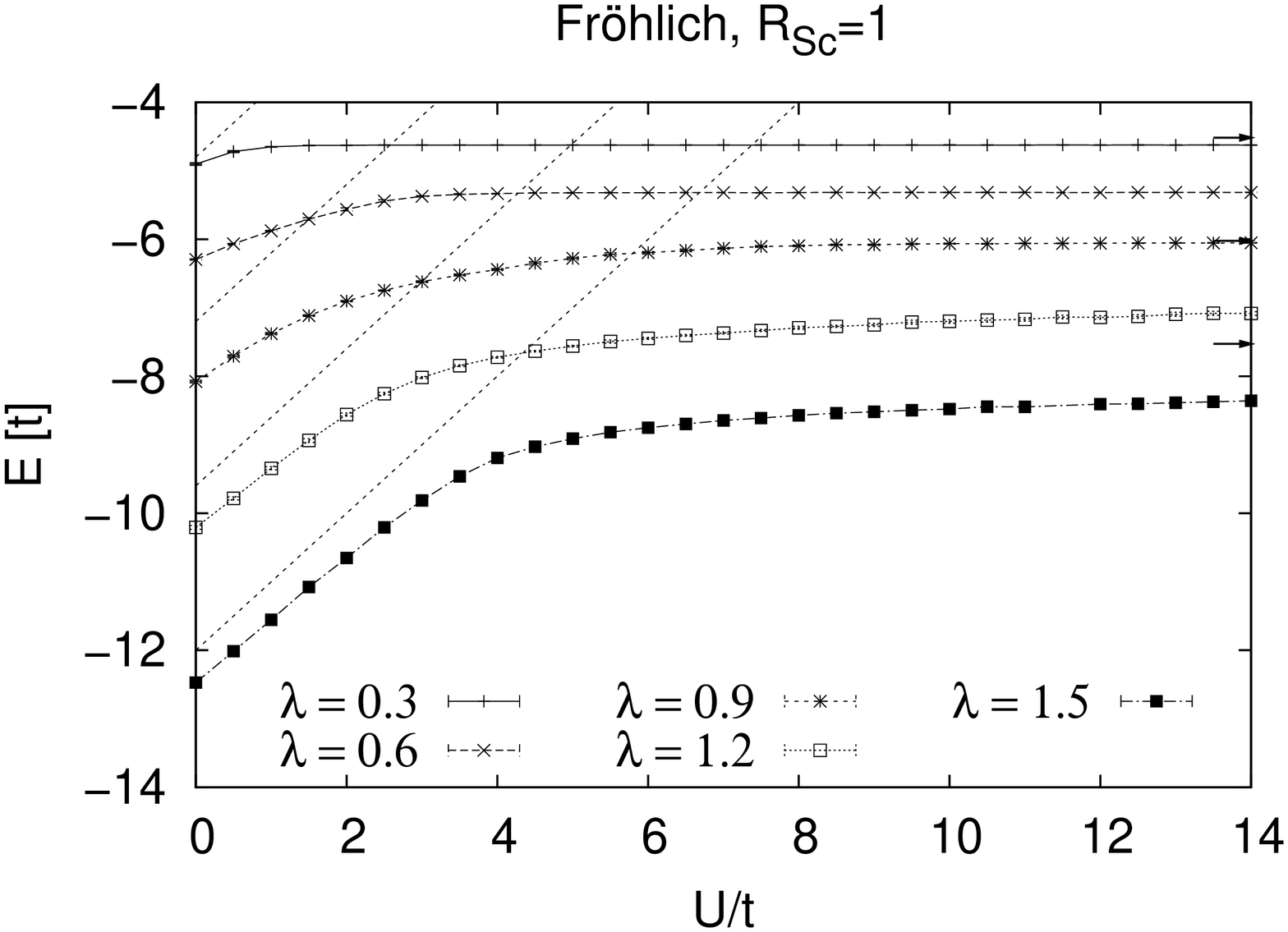}
\includegraphics[height=85mm,angle=270]{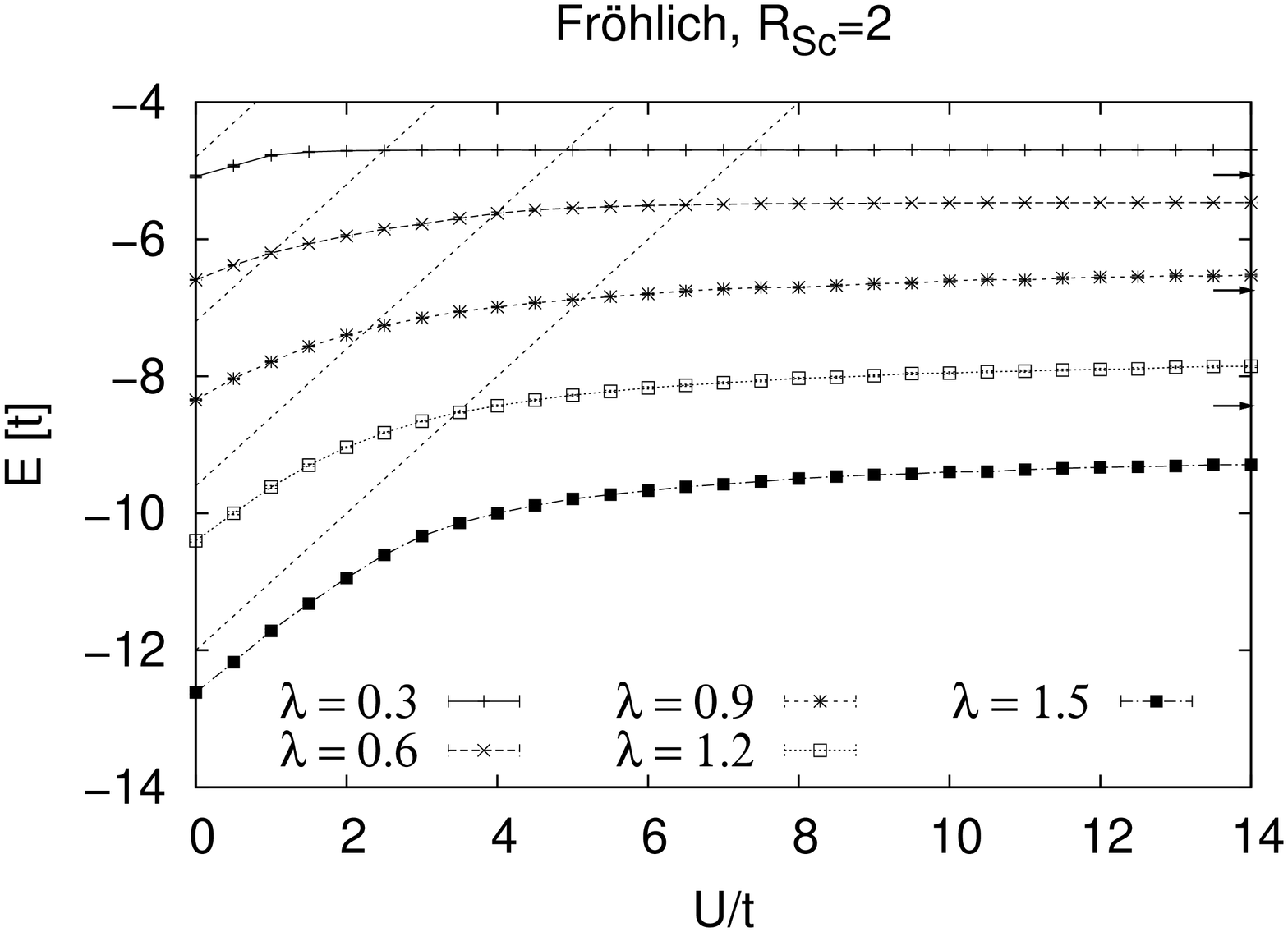}
\includegraphics[height=85mm,angle=270]{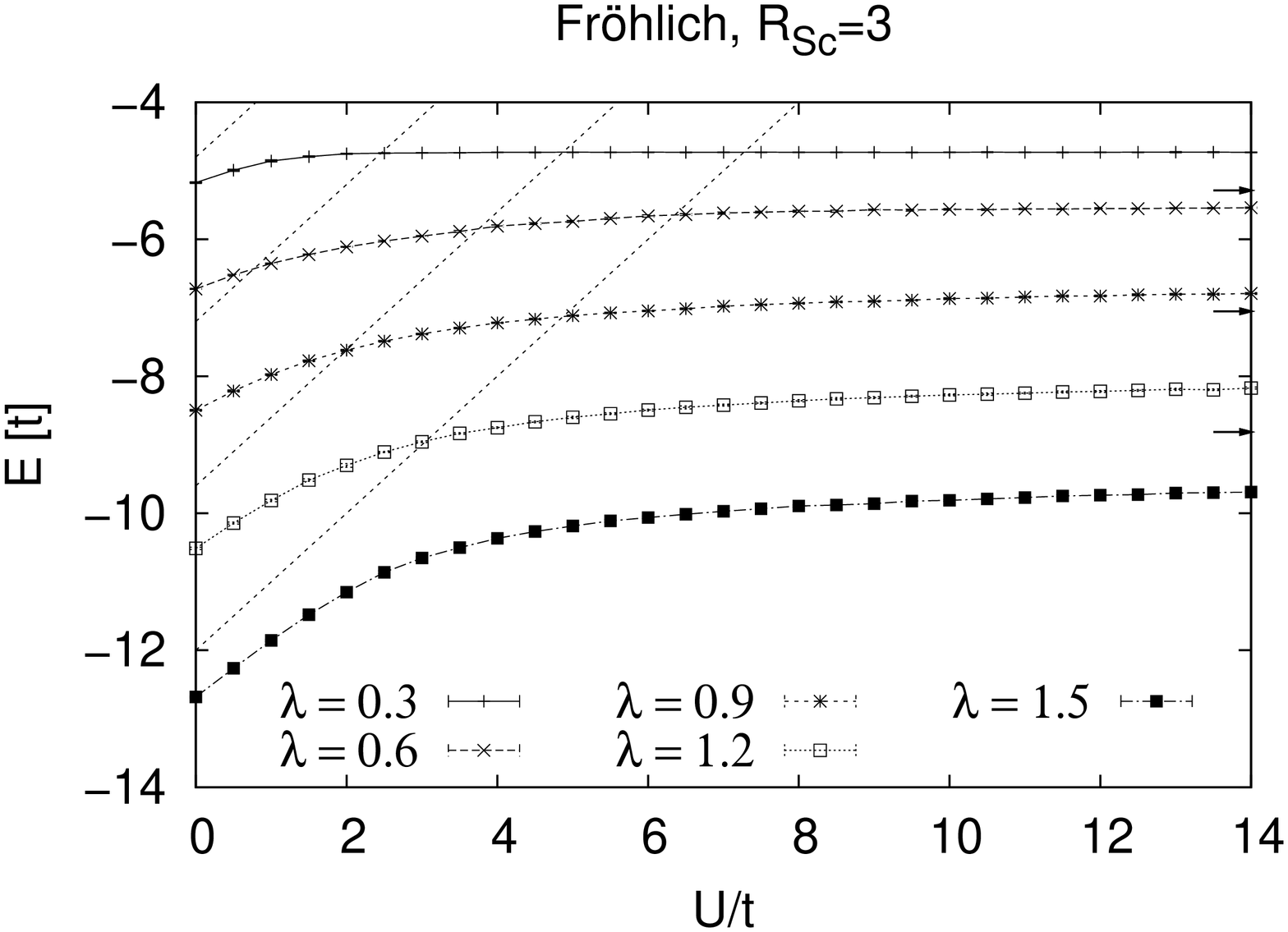}
\includegraphics[height=85mm,angle=270]{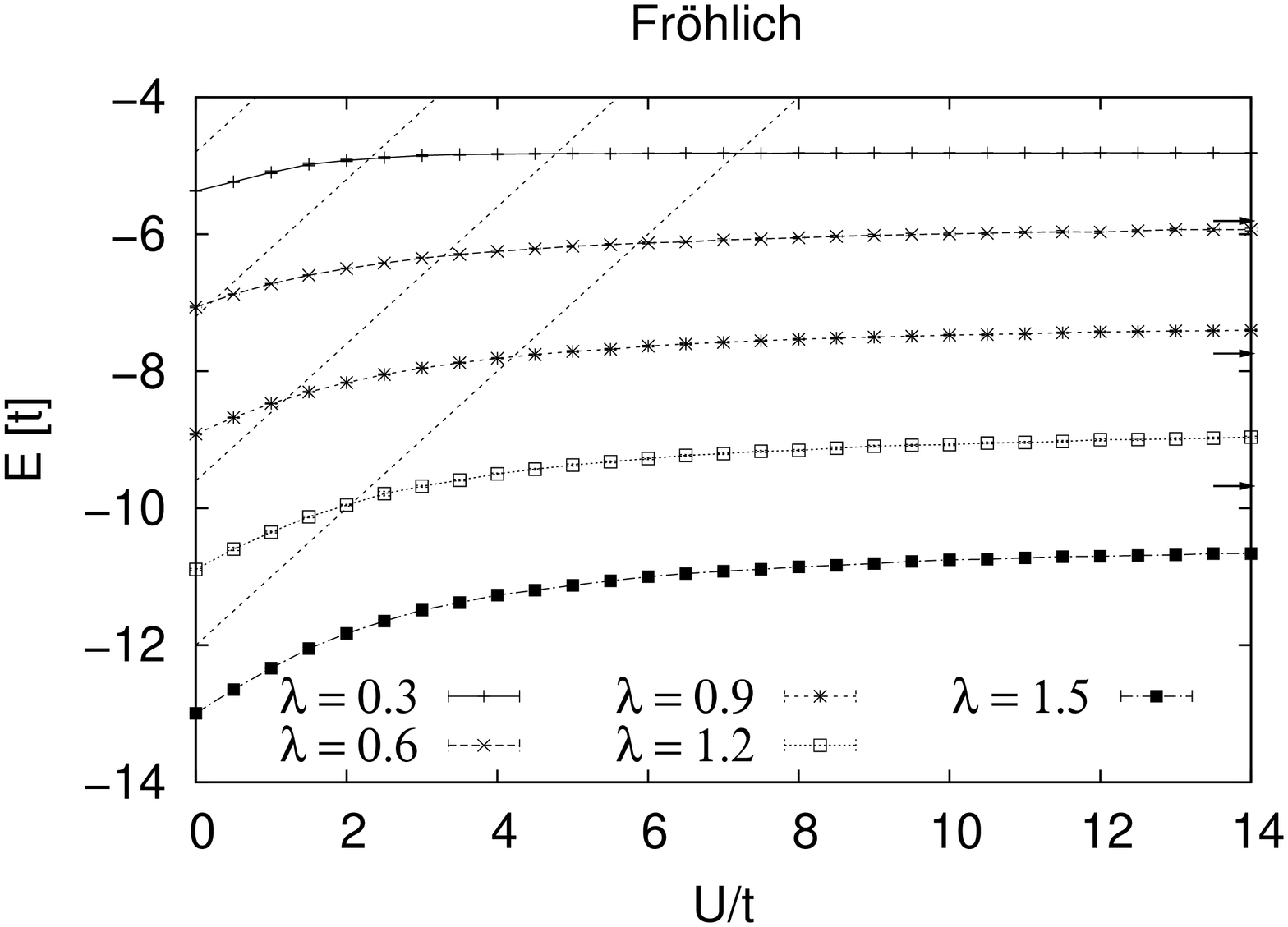}
\includegraphics[height=85mm,angle=270]{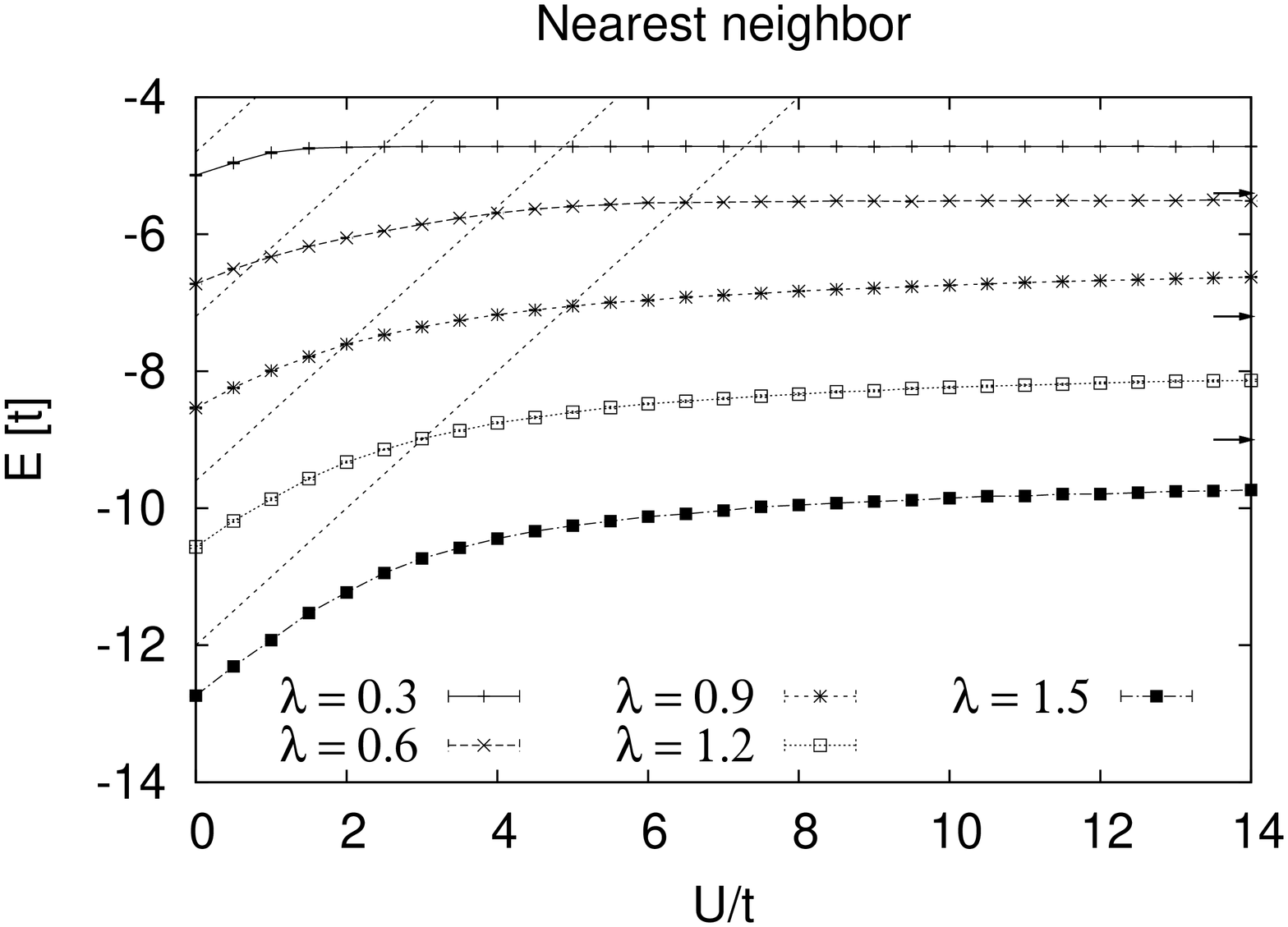}
\caption{Total energy associated with the singlet bipolaron for the 3
different models. The limits corresponding to strongly bound S0 bipolarons are shown as dotted lines, and arrows on the right axis correspond to strongly bound S1 bipolarons. The surprising result here is that the long range
tails in the Hubbard-Fr\"ohlich model with $R_{Sc}=3$ lead to similar
results to the near-neighbor model proposed by Bon\v{c}a and
Trugman. In the results for the Hubbard-Holstein model, there is a
sudden change in the gradient of the plot around the critical
coupling. This is not apparent for either of the models with long
range coupling. $\bar{\beta}=14$ in all figures in this section.}
\label{fig:totalenergy}
\end{figure*}

\begin{figure*}
\includegraphics[height=85mm,angle=270]{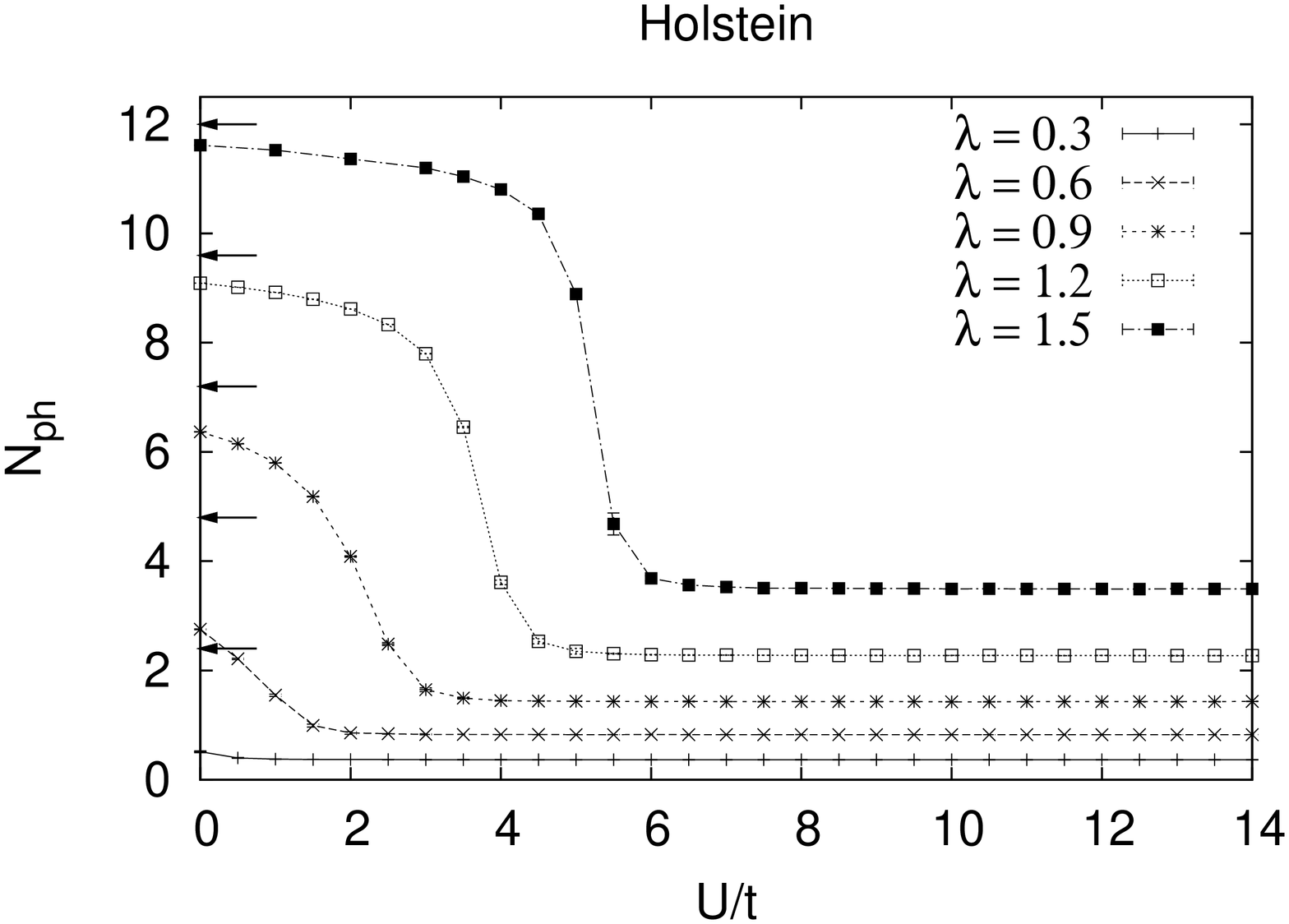}
\includegraphics[height=85mm,angle=270]{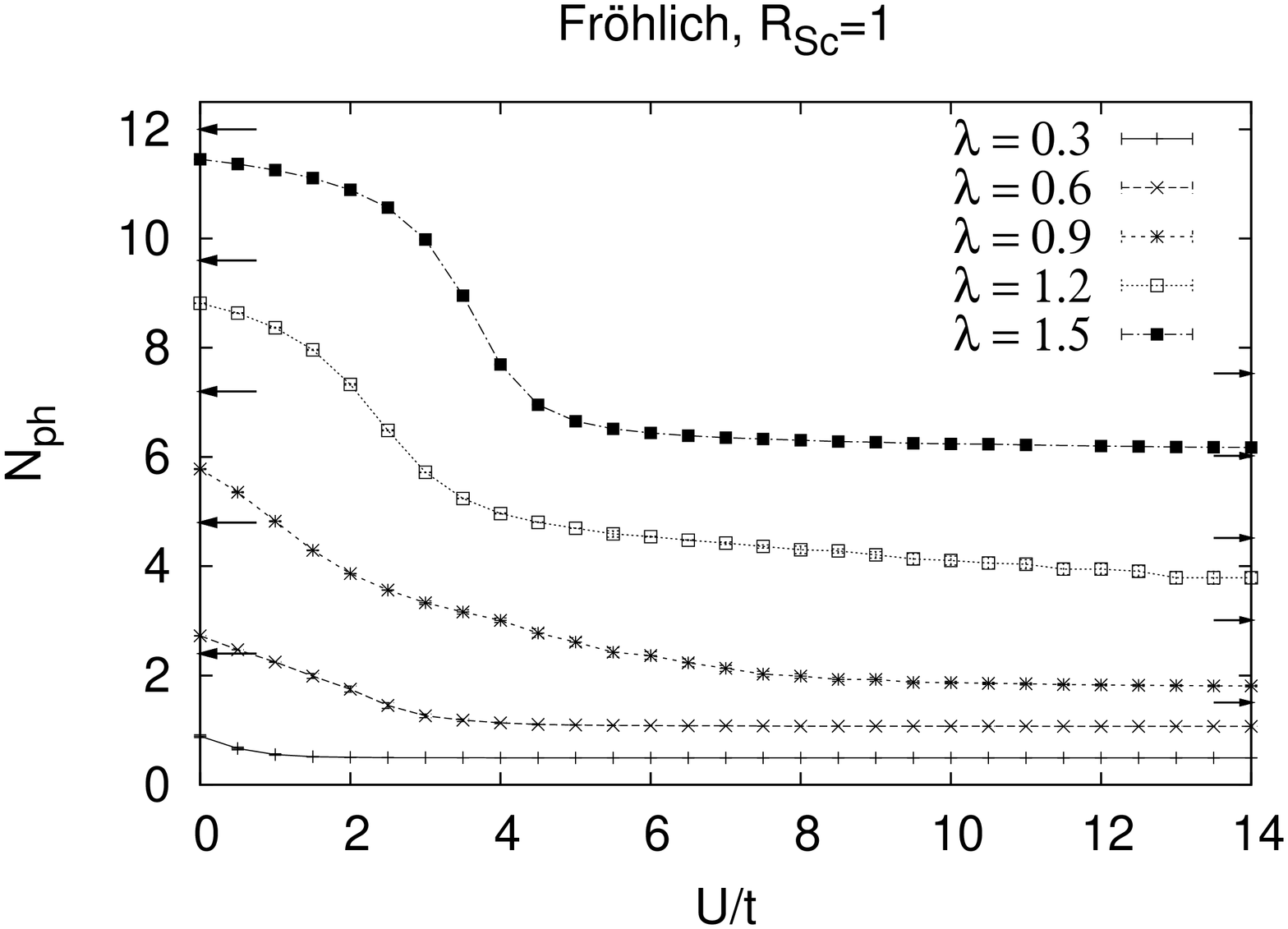}
\includegraphics[height=85mm,angle=270]{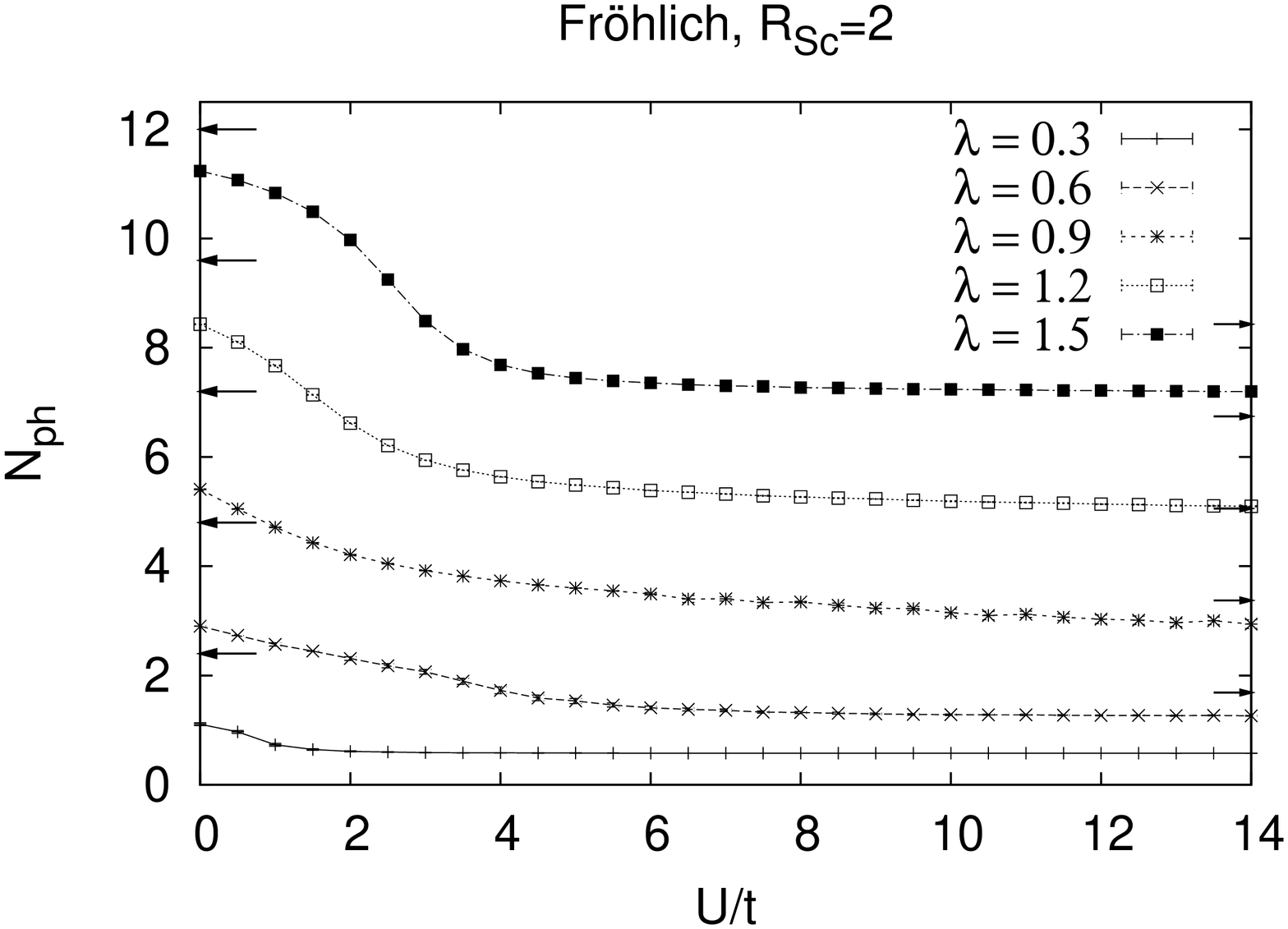}
\includegraphics[height=85mm,angle=270]{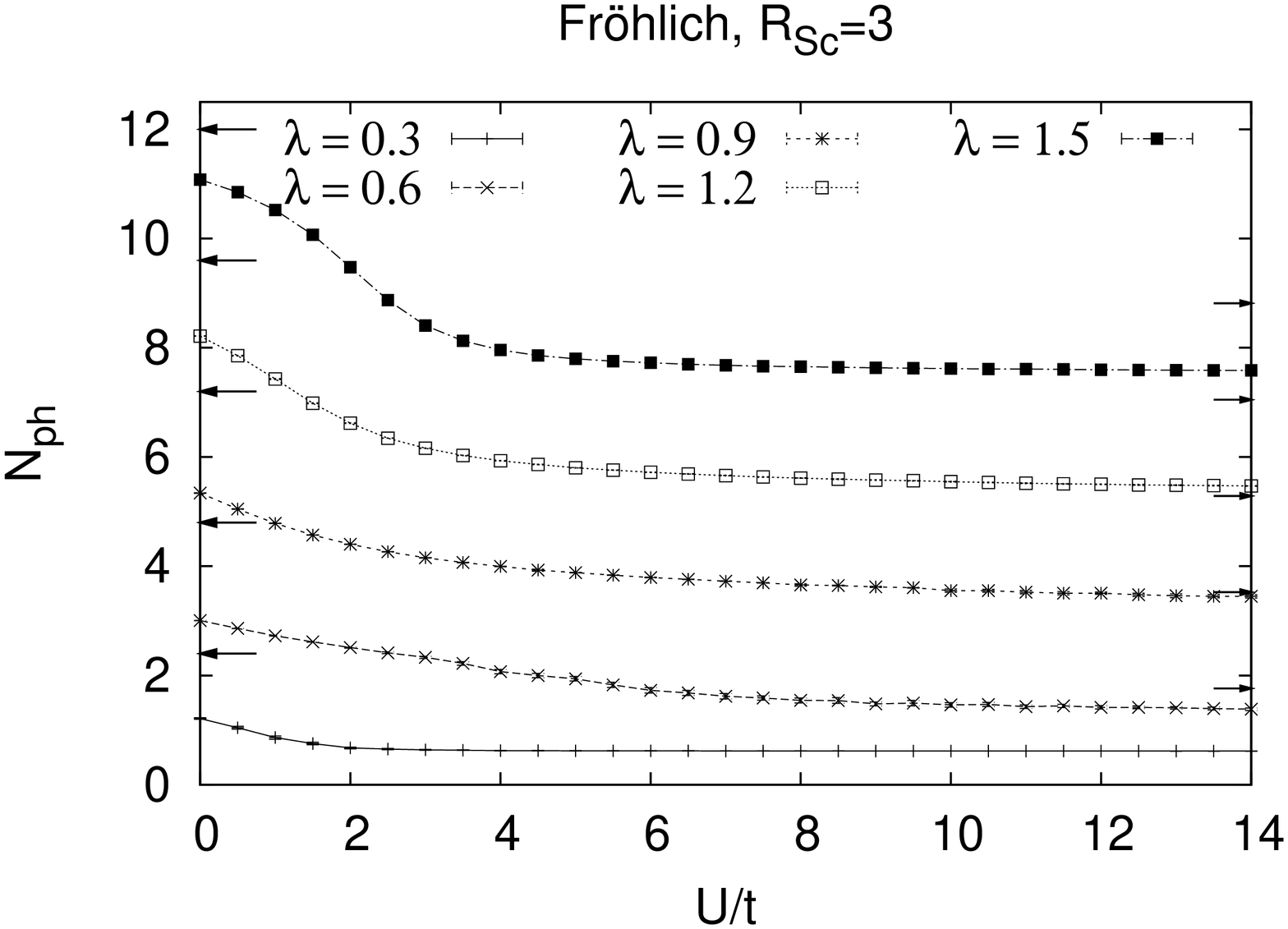}
\includegraphics[height=85mm,angle=270]{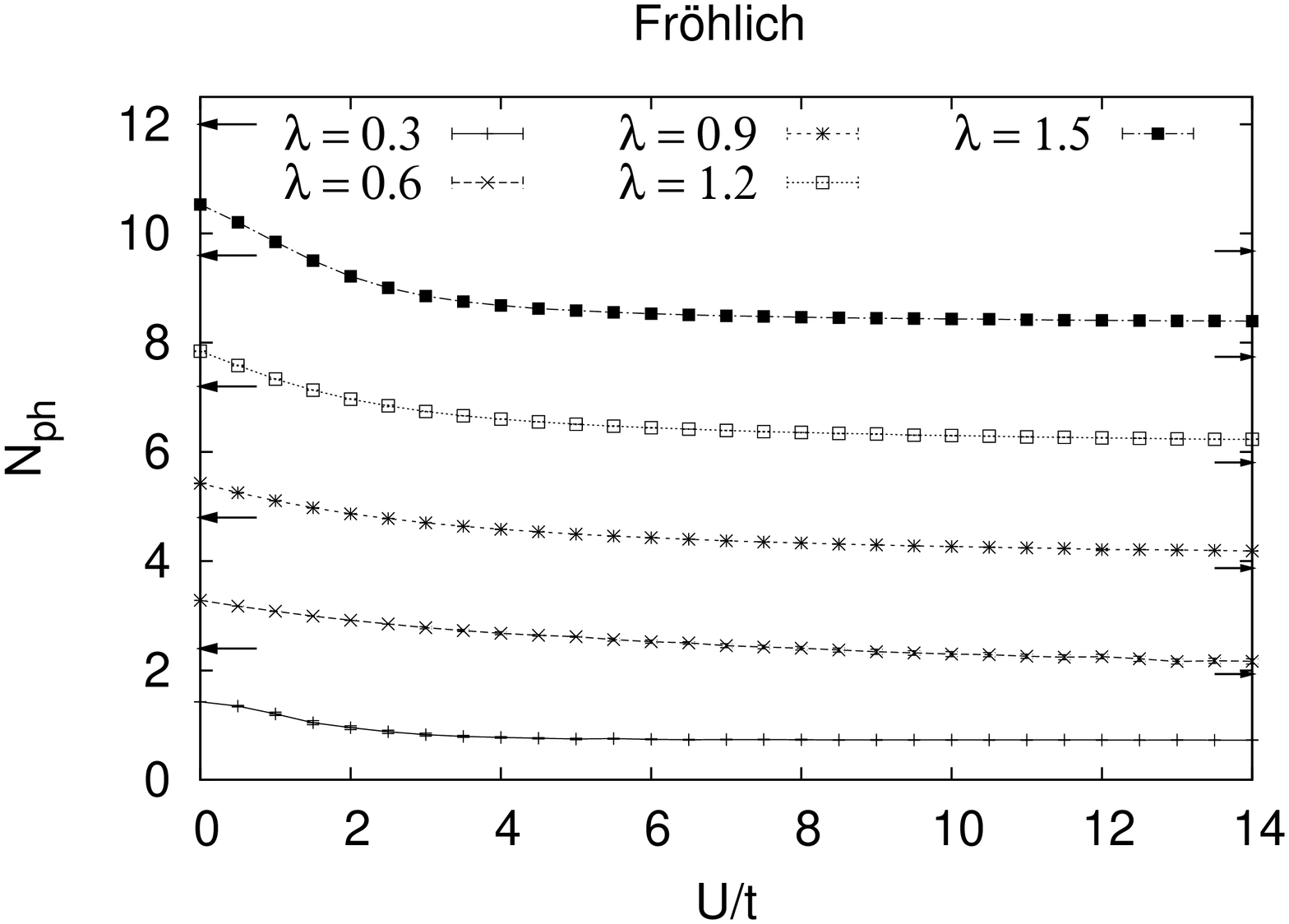}
\includegraphics[height=85mm,angle=270]{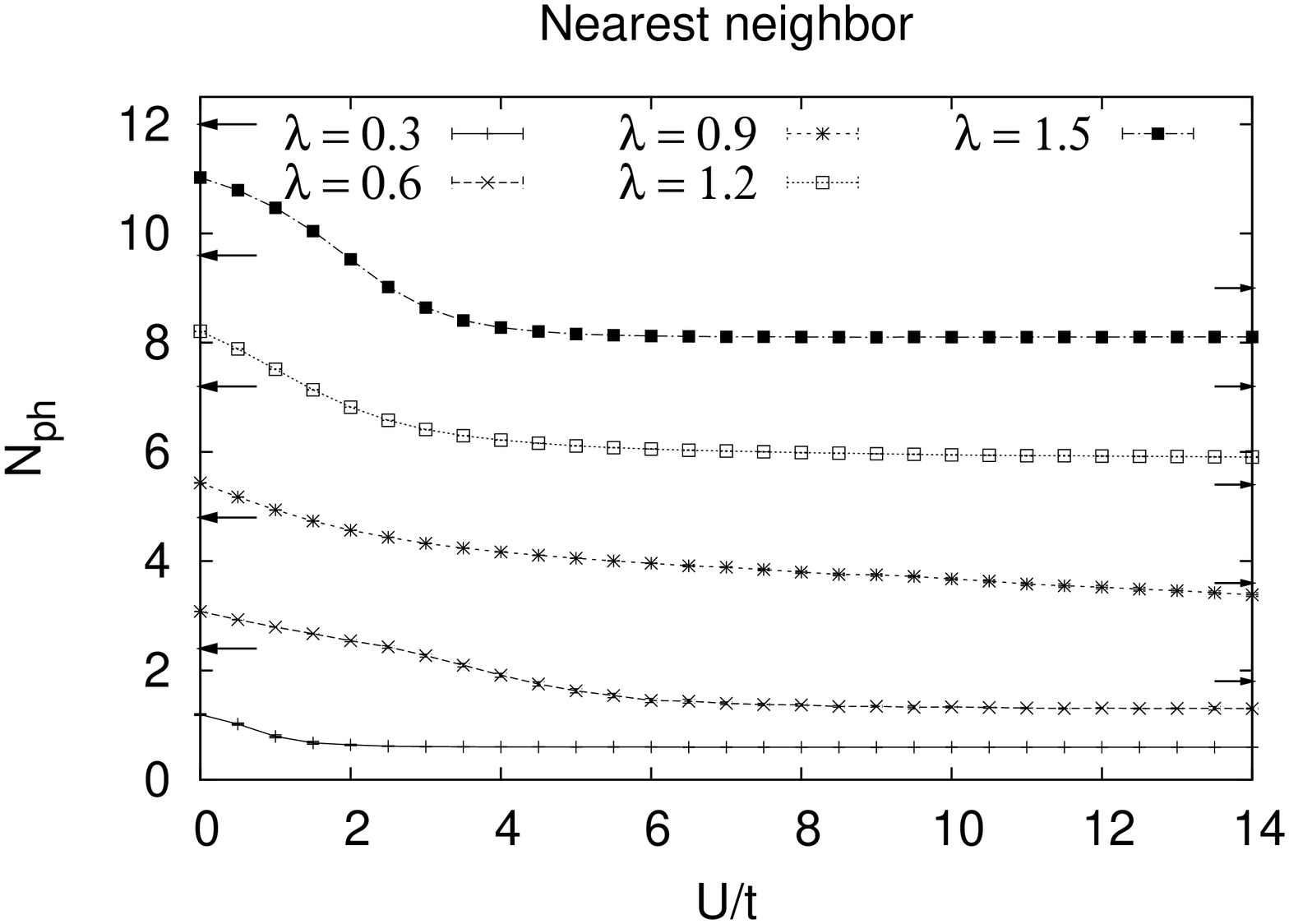}
\caption{Total number of phonons associated with the singlet
bipolaron. The limits corresponding to strongly bound S0 bipolarons are shown as arrows on the left axis, and arrows on the right axis correspond to strongly bound S1 bipolarons. The similarity between the two models with long range
interaction and the distinction with the Holstein bipolaron can
immediately be seen. The nearest neighbor and lattice-Fr\"ohlich models
show only quantitative differences. In particular, the crossover from
the weakly to strongly bound states is gentle in comparison with the
Holstein interaction, where the bipolaron is suddenly bound at a
critical coupling, causing a sudden increase in the number of phonons
associated with the bipolaron.}
\label{fig:numberofphonons}
\end{figure*}

\begin{figure*}
\includegraphics[height=85mm,angle=270]{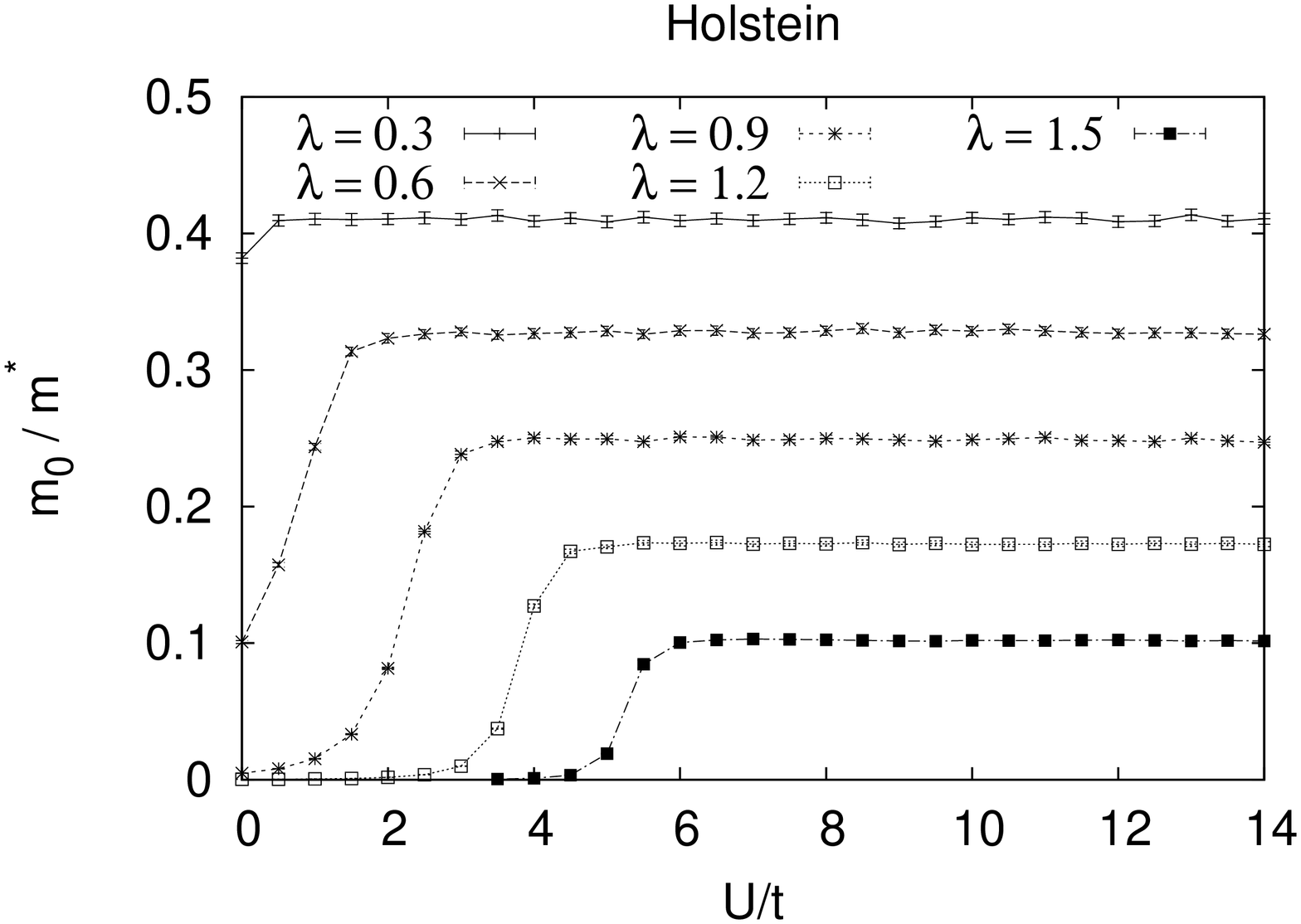}
\includegraphics[height=85mm,angle=270]{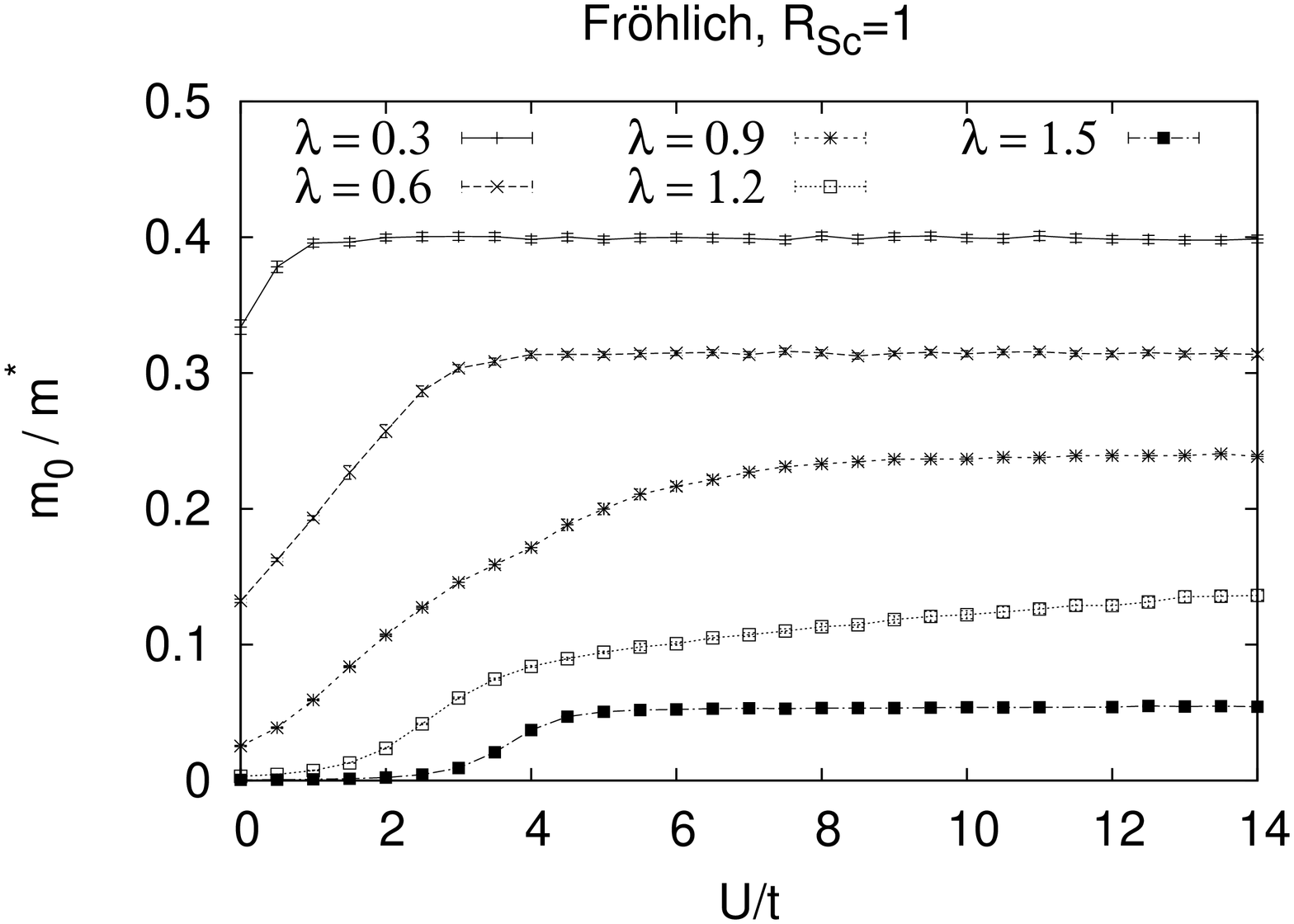}
\includegraphics[height=85mm,angle=270]{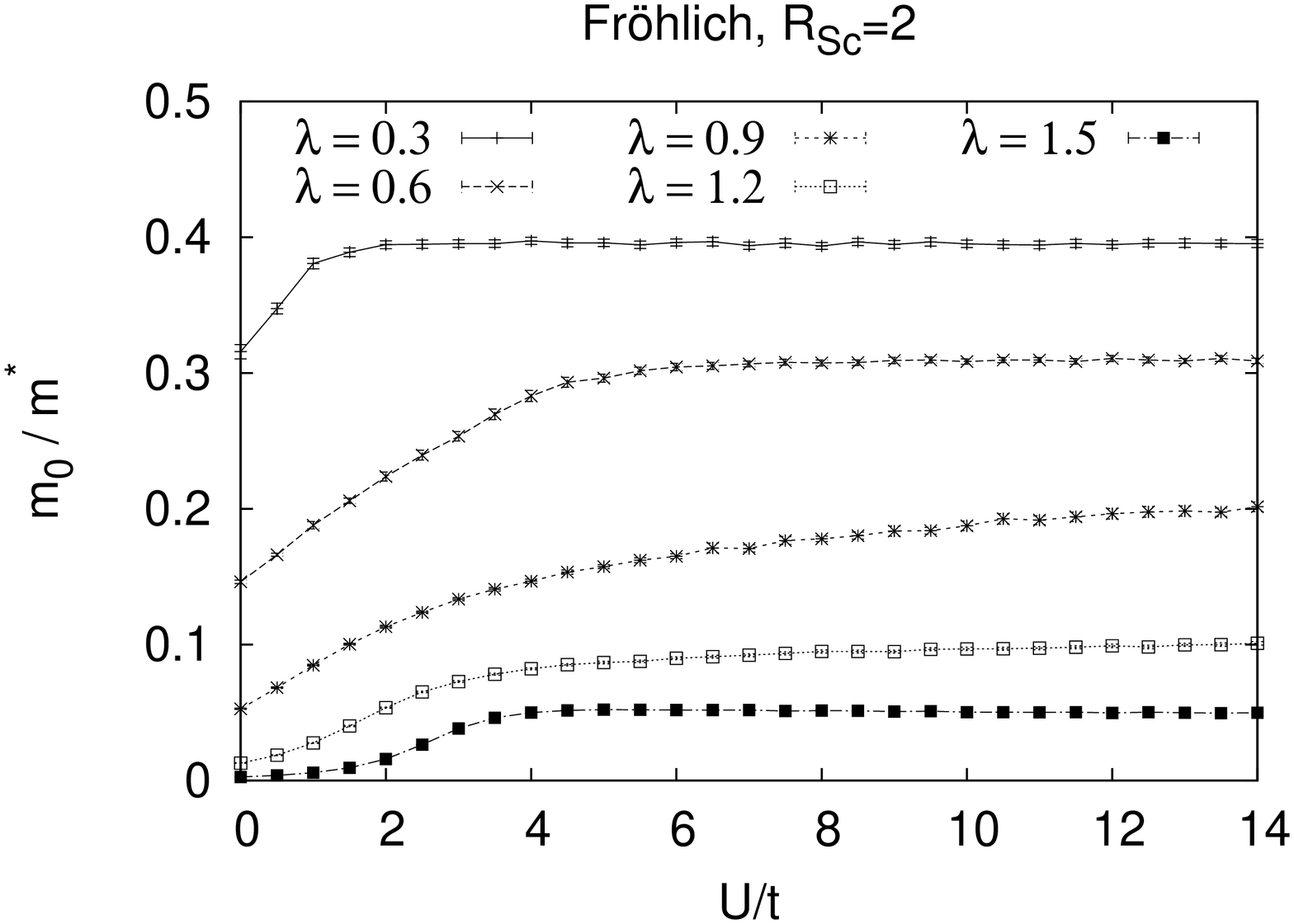}
\includegraphics[height=85mm,angle=270]{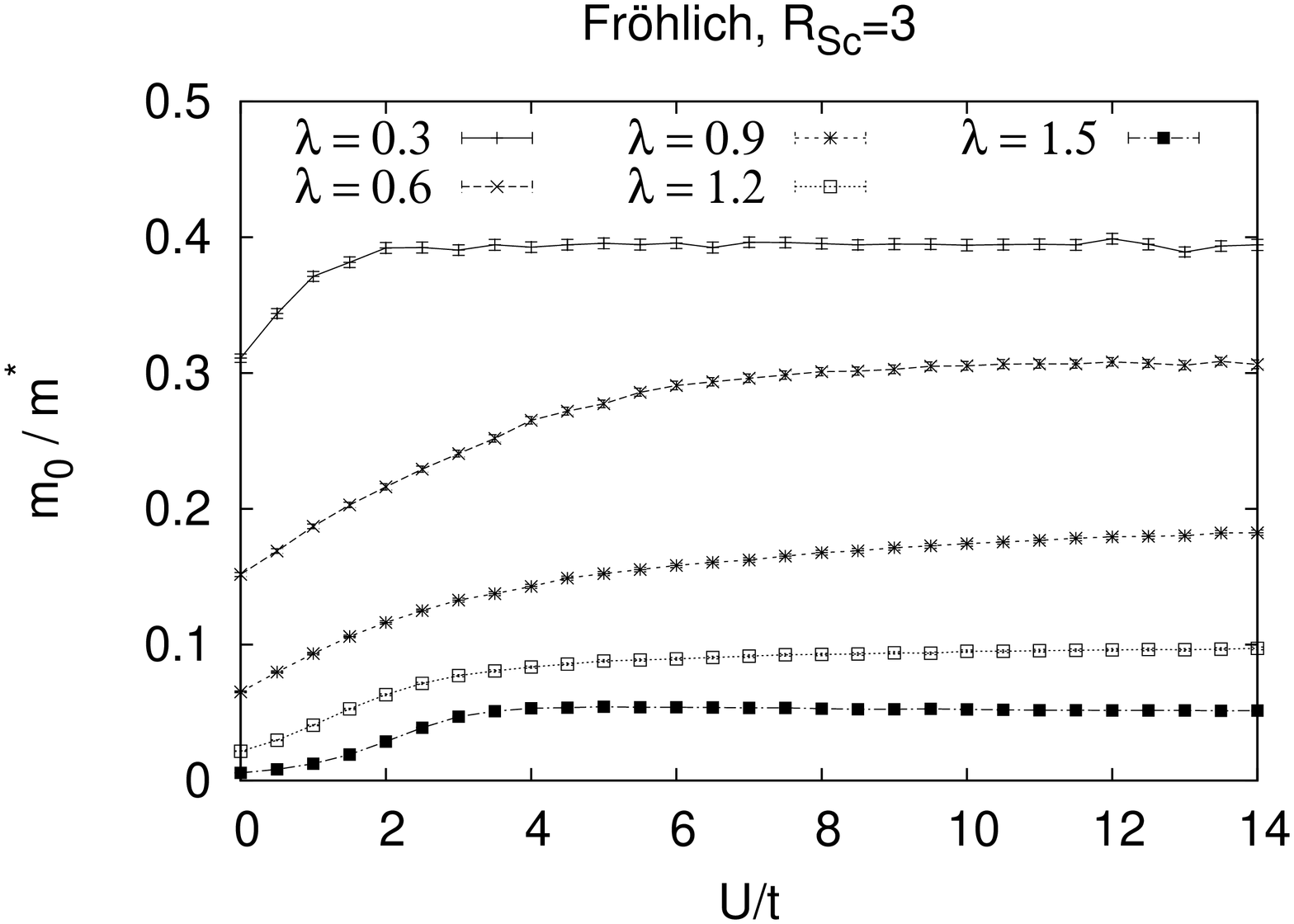}
\includegraphics[height=85mm,angle=270]{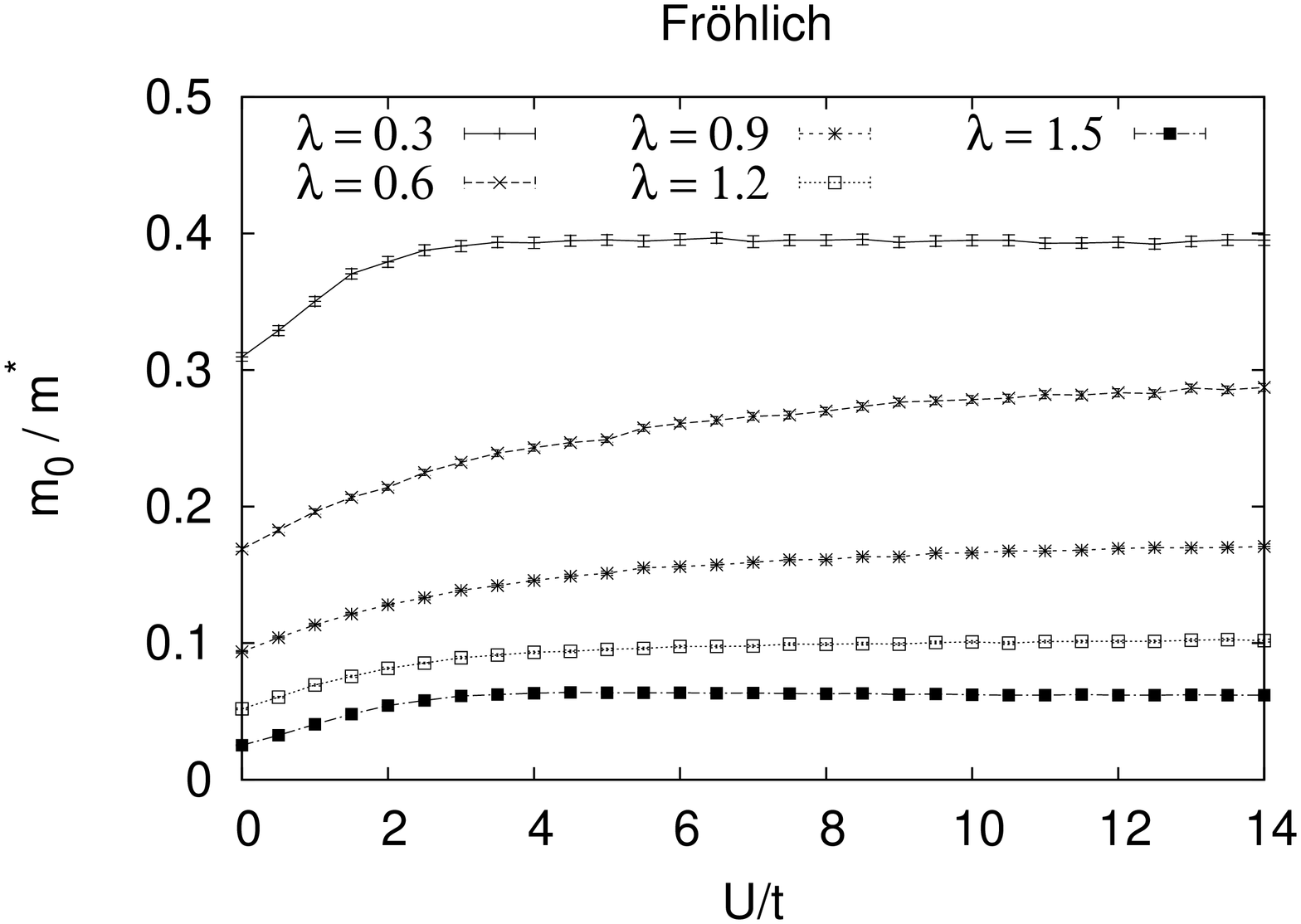}
\includegraphics[height=85mm,angle=270]{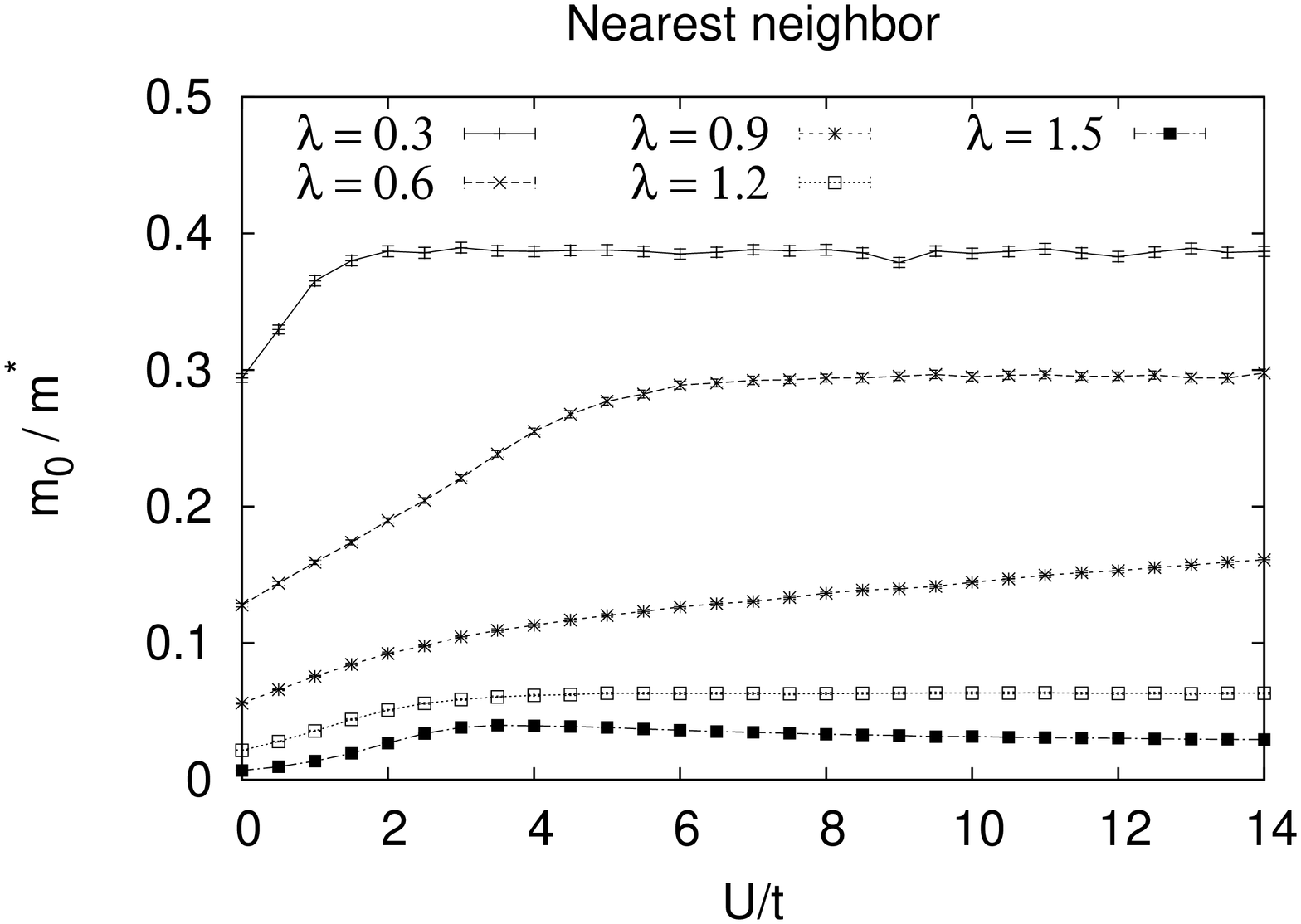}
\caption{Inverse mass of the singlet bipolaron. It is clear that both
the more complicated lattice-Fr\"ohlich bipolaron and the simplified
model with nearest neighbor interaction generate bipolarons with
similar light mass. The inverse mass of the nearest-neighbor model has a small hump around
$U=3$.}
\label{fig:inversemass}
\end{figure*}

\begin{figure*}
\includegraphics[height=85mm,angle=270]{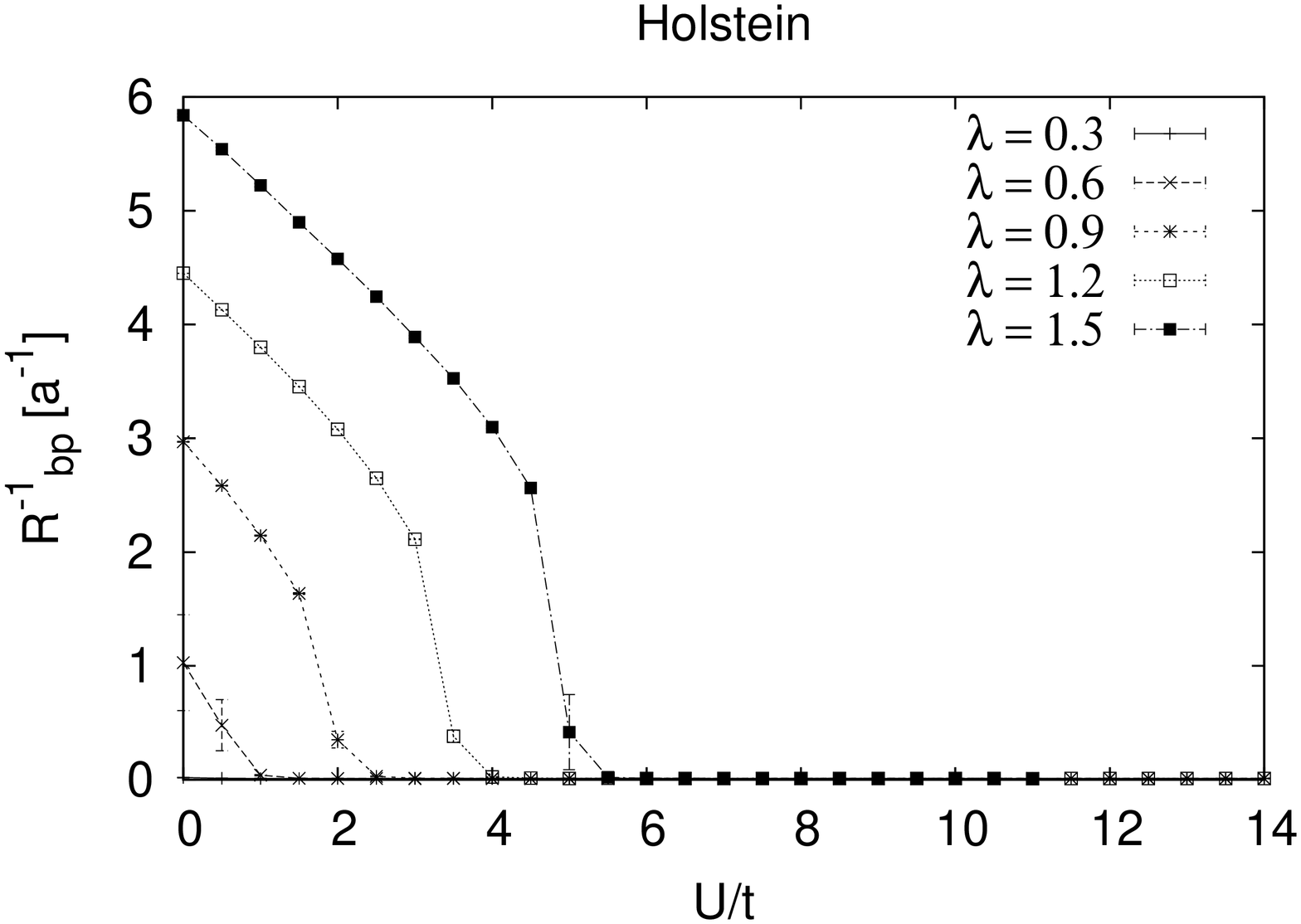}
\includegraphics[height=85mm,angle=270]{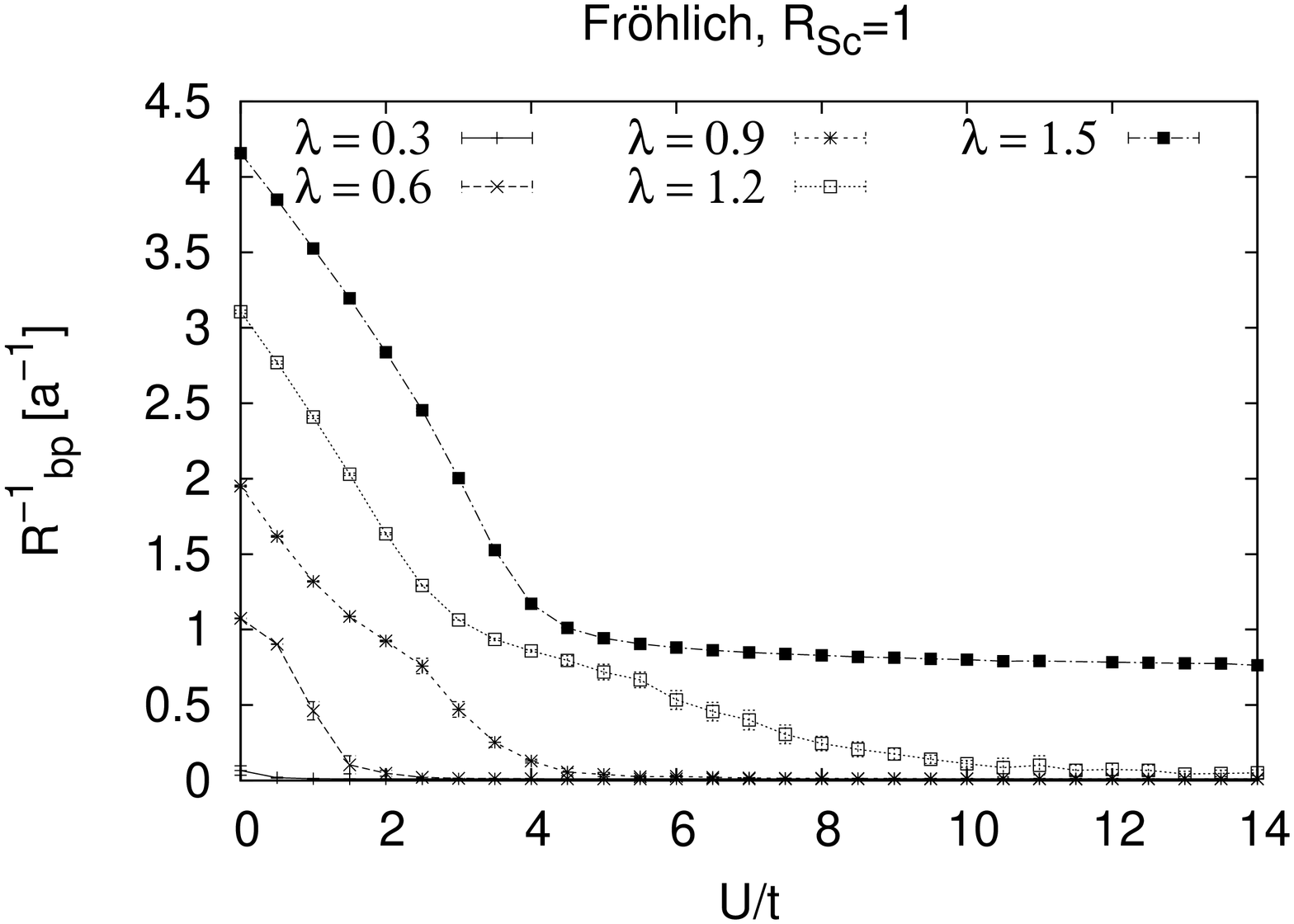}
\includegraphics[height=85mm,angle=270]{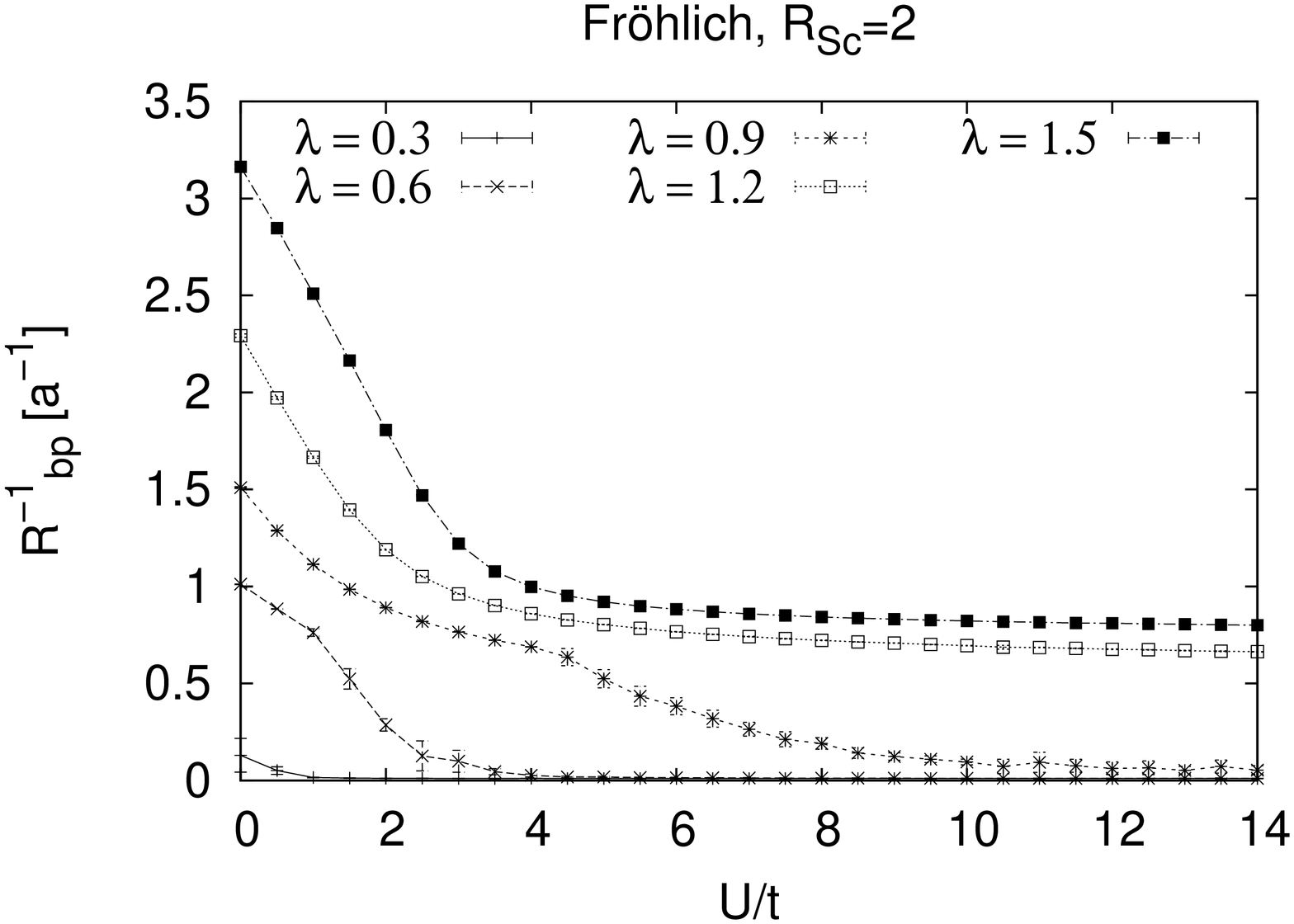}
\includegraphics[height=85mm,angle=270]{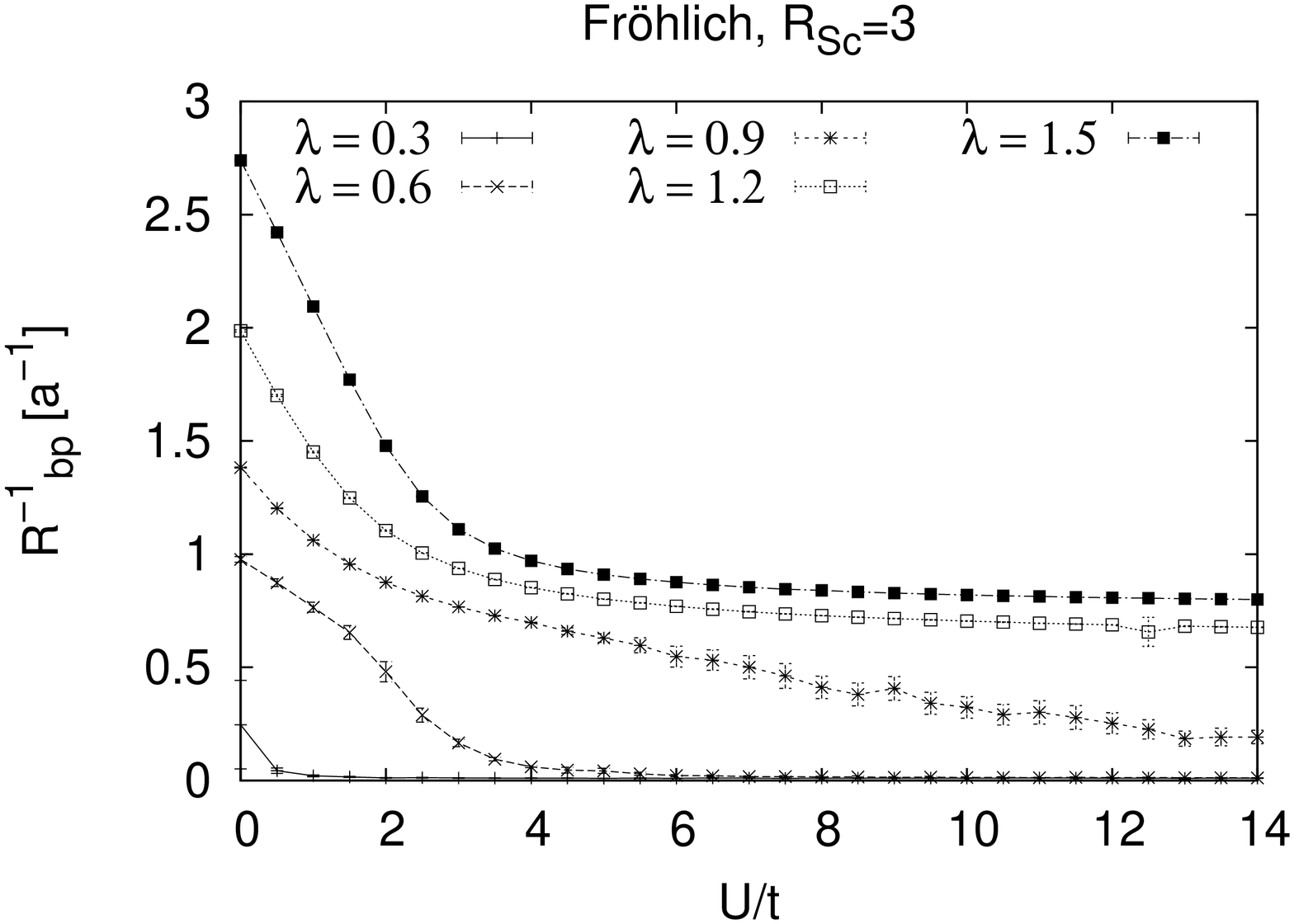}
\includegraphics[height=85mm,angle=270]{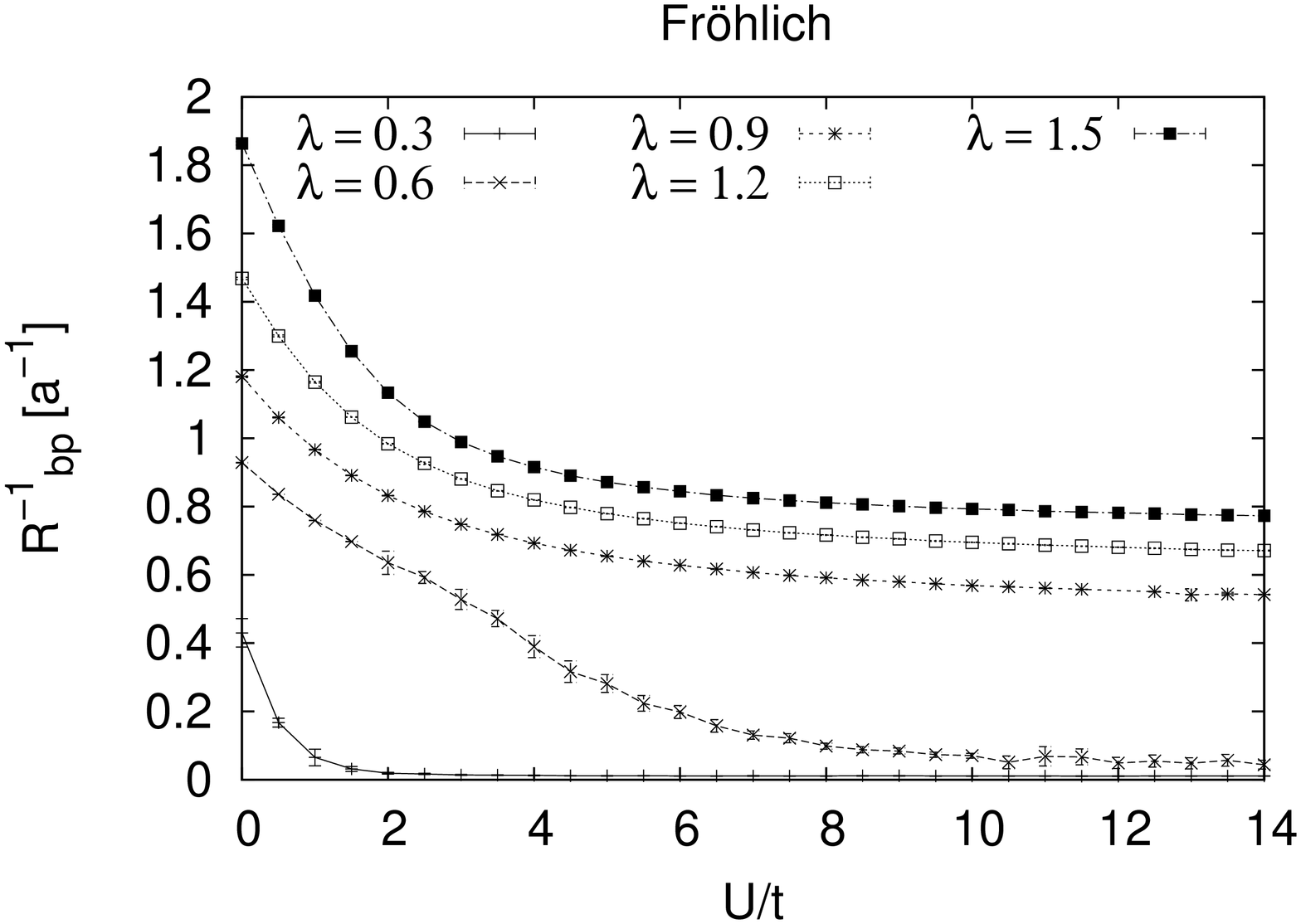}
\includegraphics[height=85mm,angle=270]{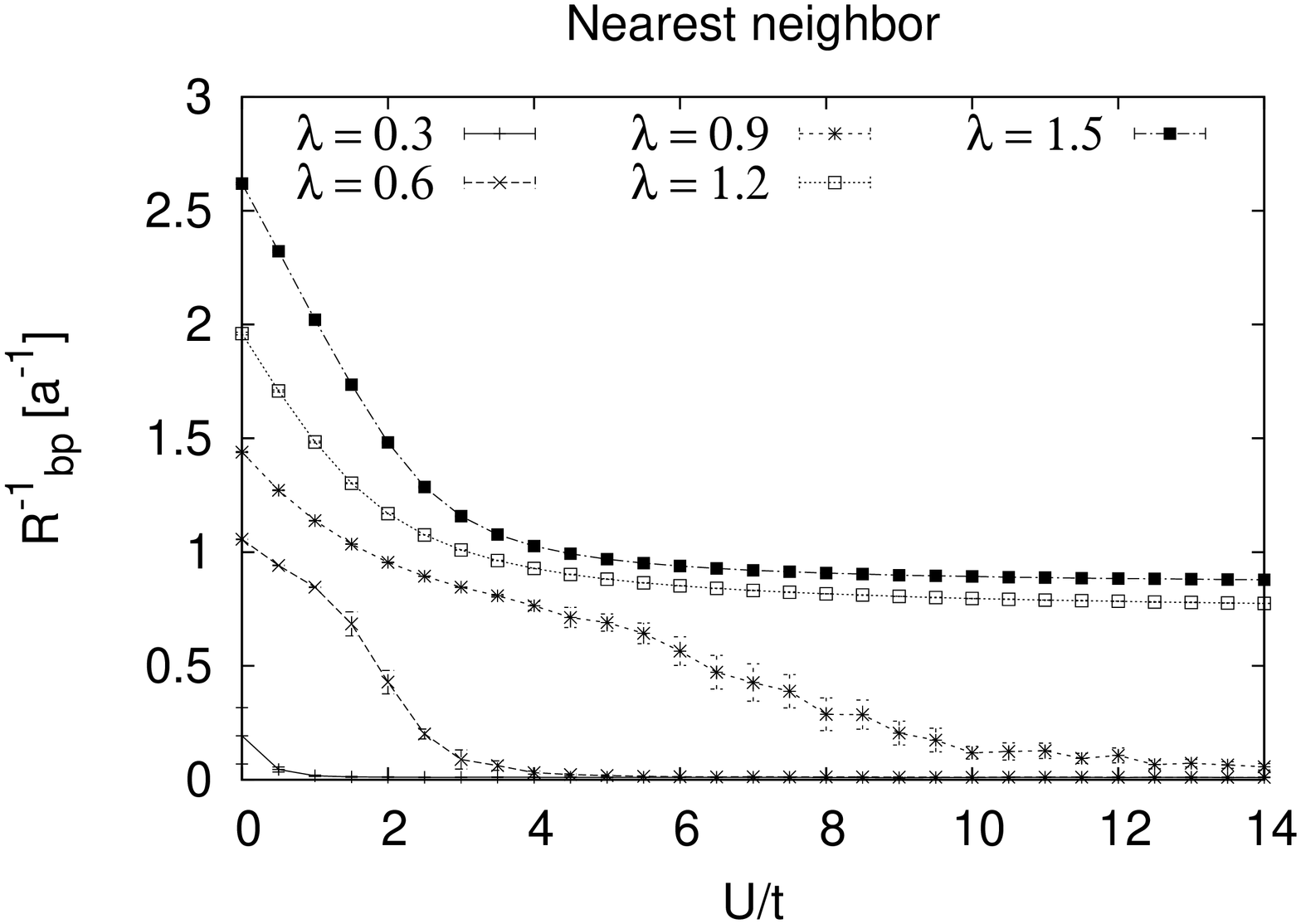}
\caption{Inverse pair size associated with the singlet bipolaron. The
Holstein bipolaron always breaks up at large $U$, indicated by the
vanishing inverse radius. In contrast, all the models with long-range
interaction have bound S1 states at large $U$ when there is sufficient
electron-phonon coupling.}
\label{fig:inversesize}
\end{figure*}

\begin{figure*}
\includegraphics[height=85mm,angle=270]{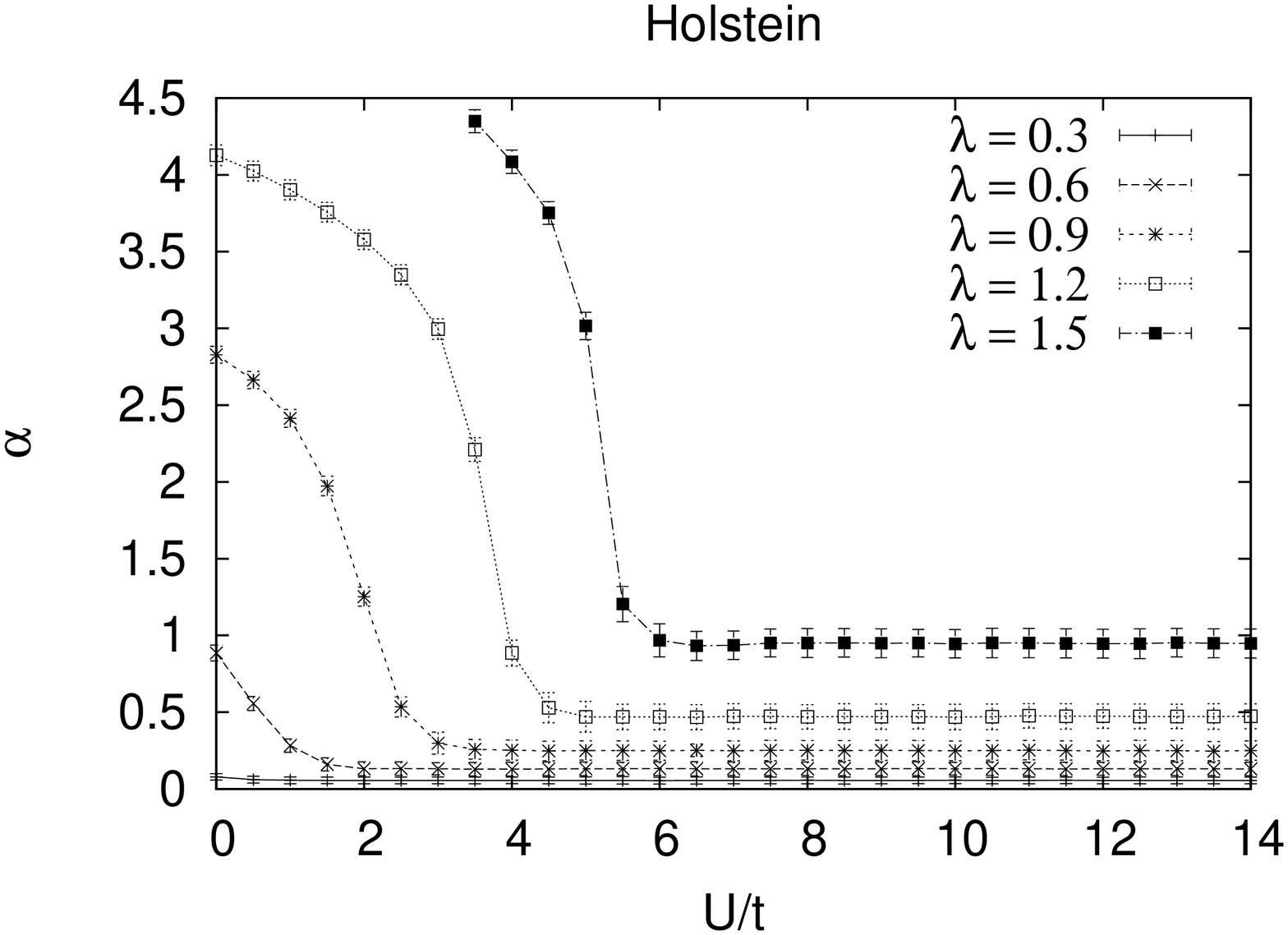}
\includegraphics[height=85mm,angle=270]{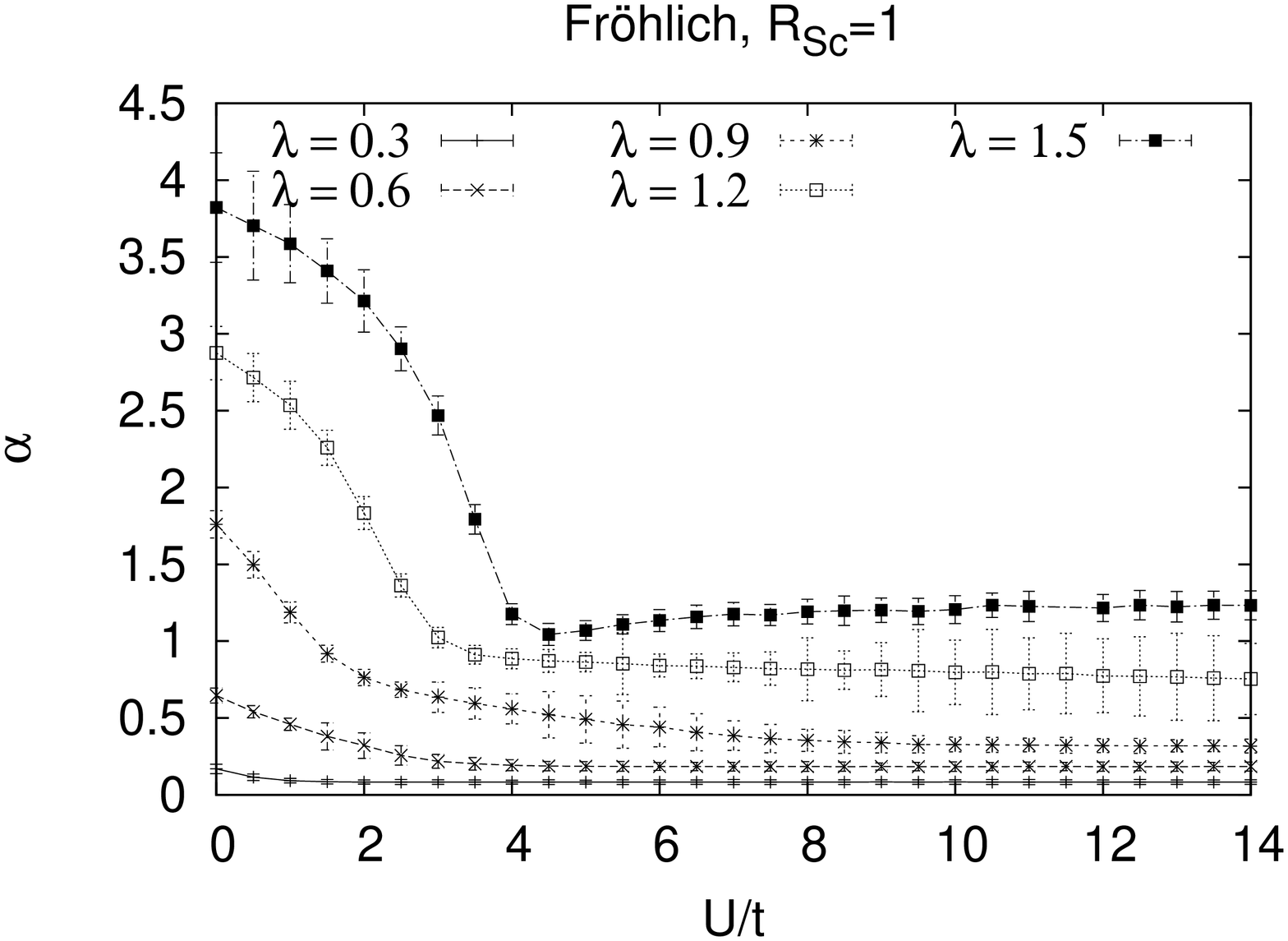}
\includegraphics[height=85mm,angle=270]{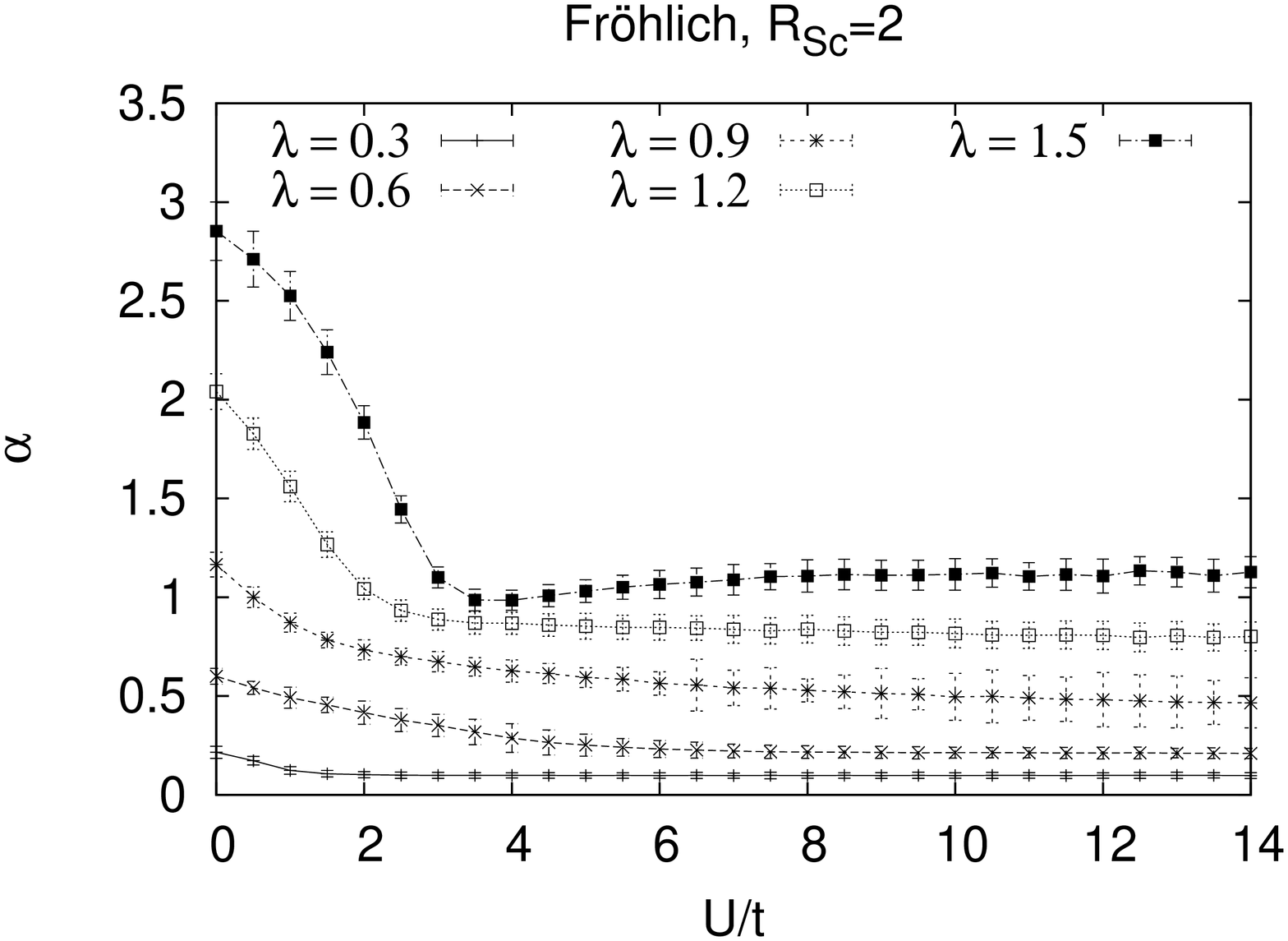}
\includegraphics[height=85mm,angle=270]{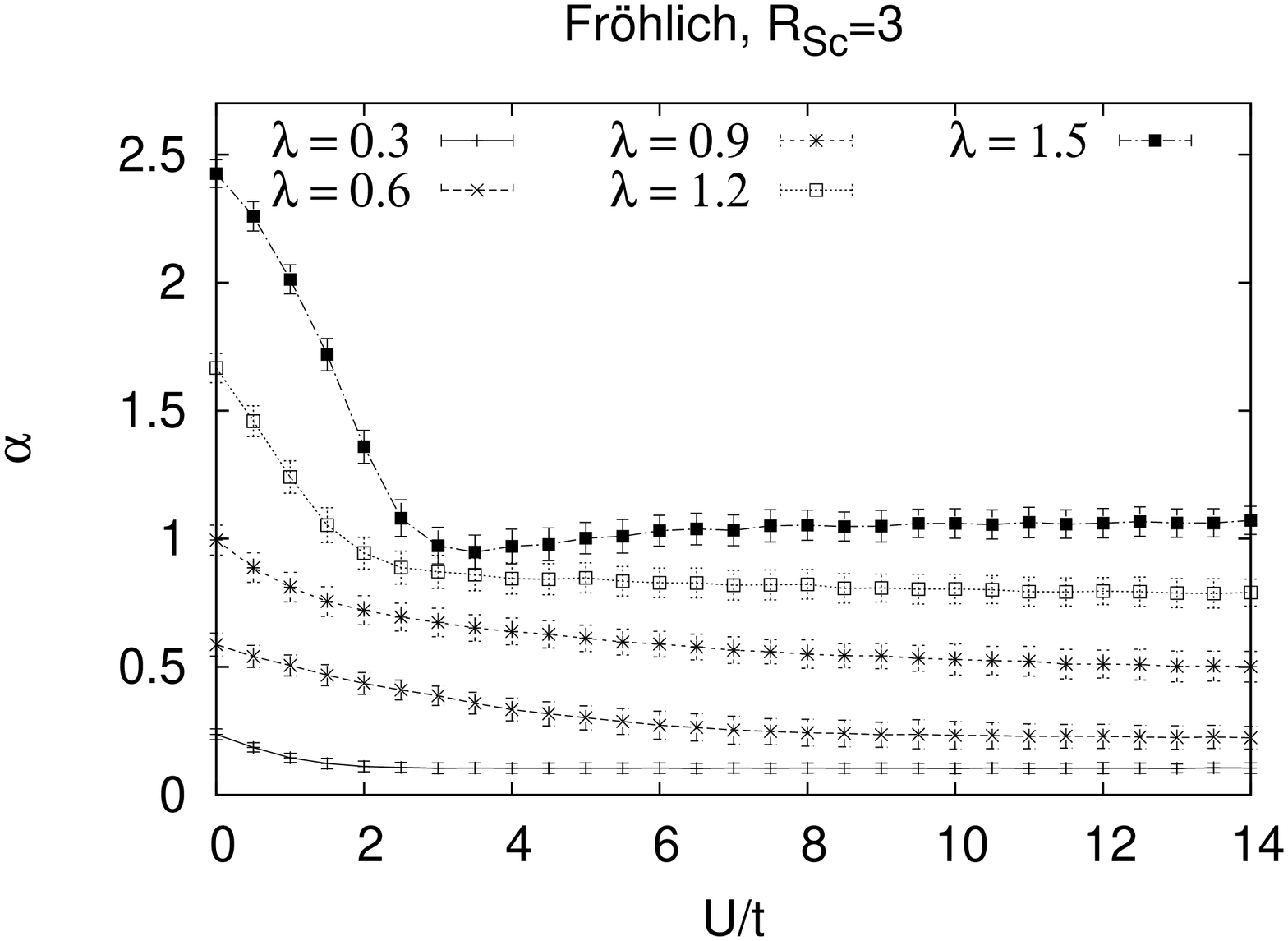}
\includegraphics[height=85mm,angle=270]{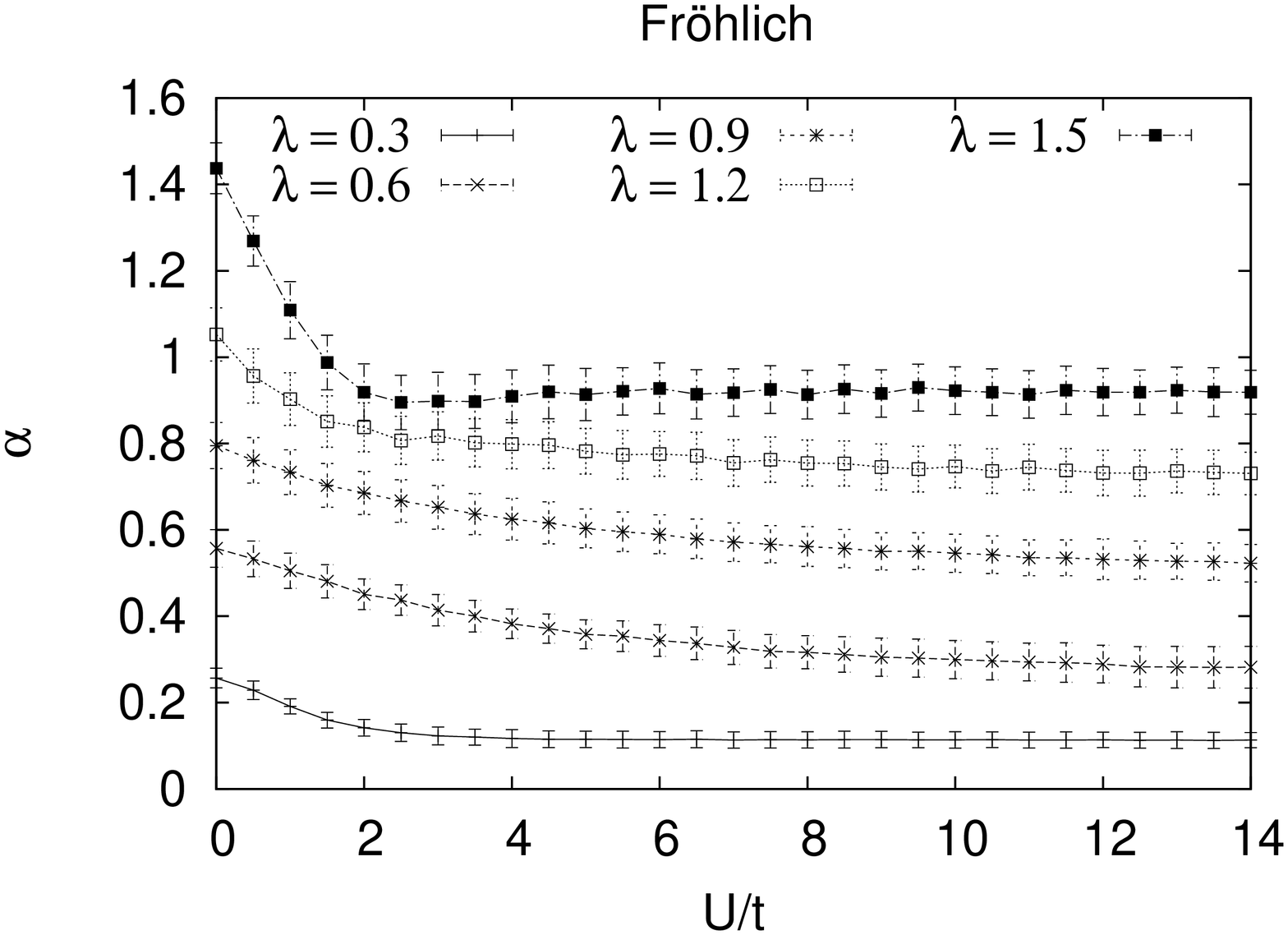}
\includegraphics[height=85mm,angle=270]{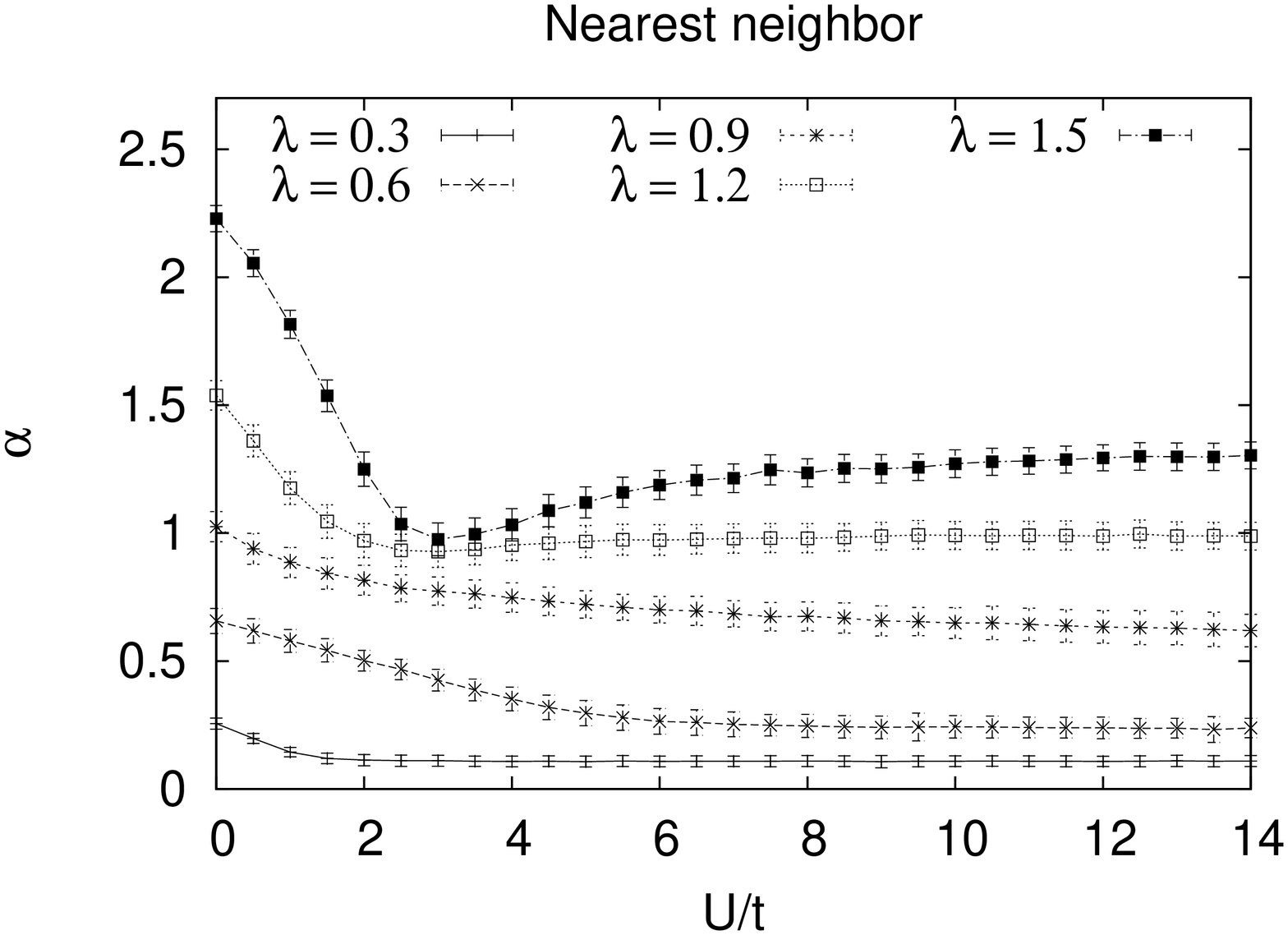}
\caption{Mass isotope exponent of the singlet bipolaron. The dip in
the exponent seen in the models with long range interaction
corresponds to the crossover between S0 to S1 bipolarons. In general,
the mass isotope exponent is much larger for S0 pairs (low $U$) than
S1 pairs (high $U$, $\lambda$ and long range interactions).}
\label{fig:isotopeexponent}
\end{figure*}

\section{Singlet bipolaron}
\label{sec:singlet}

In this section, we examine and compare the properties of the singlet
bipolarons formed in the Hubbard-Holstein model, screened and
unscreened Hubbard-Fr\"ohlich models and in the nearest neighbor
model. We compute the total energy, inverse mass, inverse radius, mass
isotope exponent and number of phonons associated with the bipolaron
cloud.

We begin by examining the total energy associated with the singlet
bipolaron, which can be seen in Fig. \ref{fig:totalenergy}. The energy
is plotted for a range of $U/t$ and $\lambda$ with fixed
$\bar{\omega}=1$. In the Hubbard-Holstein model, there is a rapid
change in the gradient of the plot around the critical coupling for
binding (for large $U$, the graph is flat because the two polarons are
not bound, and so do not interact through the local $U$). The long
range tails in the screened Hubbard-Fr\"ohlich model with $R_{sc}=3$
lead to energies that are similar to the near-neighbor model and the
difference in energies is barely detectable by eye. This is a strong
indicator that the nearest neighbor interaction strength $\Phi(\avec)$
is the most important factor in determining the propoerties of the
bipolaron. $\Phi(\avec)=0.5$ in the near neighbor model and
$\Phi(\avec)=0.4688$ to 4 significant figures in the $R_{Sc}=3$ screened Fr\"ohlich model
($\avec$ is the lattice vector). The unscreened Fr\"ohlich interaction
leads to bipolarons that are more strongly bound. For example, a much
larger $U/t$ is needed to unbind the bipolaron when the Fr\"ohlich
interaction is simulated at $\lambda=0.3$. In the limit of low
screening radius, the energy of the bipolaron approaches the Holstein
limit very slowly, with the very rapid binding close to a critical $U$
associated with the Holstein model not seen in models with long range
interaction. This is a qualitative difference between Holstein models
and those with a small amount of intersite attraction.

In the very strongly bound limit, the effects of electron hopping are
small. An effective atomic Hamiltonian can be found by applying
the Lang-Firsov transformation to the phonon part of the Hamiltonian,
\begin{equation}
\tilde{H}_{\rm at} = -\sum_{ii'}\frac{2t\lambda\Phi_{0}(i,i')}{\Phi_0(0,0)}n_i n_{i'}+\omega\sum_j\left(d^{\dagger}_{j}d_{j}+\frac{1}{2}\right)
\end{equation}
Reintroducing the effects of Coulomb repulsion, the total energy of
the strongly bound onsite bipolaron (formed when $8t\lambda>U$) can be
computed to be,
\begin{equation}
E = U-8t\lambda
\end{equation}
The energy of the strongly bound S1 bipolaron at large $\lambda$ and
$U\gg 2t\lambda$ is,
\begin{equation}
E=-4t\lambda[1+\Phi(\avec)/\Phi_0].
\end{equation}
These limits are also shown in Fig. \ref{fig:totalenergy} (the strong
binding S1 energies are only shown for $\lambda=0.9, 1.2$ and
$1.5$). There is a clear convergence between the numerics and the
strongly bound S0 state at low $U$. Bipolarons with long-range
interactions converge on the limit more slowly than Holstein
bipolarons. The S1 energies at large $U$ begin to agree at large
$\lambda$. This is expected, since the bipolarons formed in models
with long-range interactions are quite mobile, so the effects of
hopping should be considered, even when $\lambda=1.5$. Note that there is
no S1 bipolaron in the very large $U\gg 2t\lambda$ limit of the
Holstein model.

The total number of phonons associated with the singlet bipolaron is
shown in Fig. \ref{fig:numberofphonons}. The similarity between the
pairs in the models with long range interaction and the distinction
with the Holstein bipolaron is clearly visible. The near-neighbor and
lattice-Fr\"ohlich models show only quantitative differences. In
particular, the crossover between weakly and strongly bound cases in
the models with long-range interaction is gentle in comparison with
that seen in the Hubbard-Holstein model; in the case of the Holstein
interaction the bipolaron is rapidly bound on decreasing $U$, causing
an abrupt increase in the number of phonons associated with the
bipolaron.

The total
number of phonons associated with the strongly bound onsite bipolaron
(following the argument in Ref. \onlinecite{hague2007a}) is,
\begin{equation}
N_{\rm ph} = \frac{8\lambda}{\bar{\omega}} \: .
\end{equation}
Note that the number of phonons in the strongly bound onsite bipolaron
does not depend on $U$, so the number of phonons reaches a limiting
value on decreasing $U$. This limit can be seen in
Fig. \ref{fig:numberofphonons} as the arrows on the left-hand
$y$-axis. The Holstein bipolaron rapidly approaches the strongly bound
S0 limit on decreasing $U$, with the S0 limit approached less rapidly
as $R_{sc}$ increases. The reason for
this is that the longer range tails lead to a shallower effective
potential, so S0 bipolarons are harder to bind into purely on-site
pairs. The number of phonons associated with the S1 bipolaron can also
be found,
\begin{equation}
N_{\rm ph} = \frac{4t\lambda}{\omega}[1+\Phi(\avec)/\Phi_0].
\end{equation}
This is represented as arrows on the right hand side of the
graphs. Again it is clear that hopping effects should be included at
intermediate $\lambda$, with the agreement becoming better as
$\lambda$ is increased and the bipolaron becomes more strongly bound.

We show the inverse mass of the singlet bipolaron in figure
\ref{fig:inversemass}. Both the more complicated Hubbard-Fr\"ohlich
bipolaron and the simplified model with only nearest neighbor
interaction generate bipolarons with similar light mass. At large $U$,
the mass of the Holstein bipolaron does not change, because the
bipolaron is not bound. This is also true of the long range models at
small enough $\lambda$. There is a notable feature in the mass at
intermediate $U\sim 3$ and $\lambda=1.5$. The mass associated with the
nearest-neighbor bipolaron has a hump, with the bipolaron becoming
lighter on increasing $U$, before becoming slightly heavier again.

We have not been able to compute analytical expressions for the mass
in the strong coupling limit of the Hubbard-Fr\"ohlich model since
each electron hop excites phonons at an infinite number of sites,
complicating the second order perturbation theory. For the
near-neighbor and Holstein models, the strong binding limits have been
discussed by Bon\v{c}a and Trugman \cite{bonca2000a,bonca2001b}. As
noted by Bon\v{c}a and Trugman, the energies of S0 and S1
configurations in the near-neighbor model with high phonon frequency are
degenerate when $U=W\lambda$ at large $\omega$ \cite{bonca2001b}. This
also occurs in the Hubbard-Fr\"ohlich model when $U-2W\lambda =
2W\lambda\phi(\avec)$. At this point, the motion of the bipolarons is
1st order in $t$ since the electron does not need to tunnel through a
high energy barrier to move, leading to the mass decrease that can be
seen in Fig. \ref{fig:inversemass}. In this sense, the mass decrease
is similar in origin to the superlight small bipolarons seen on
lattices with a triangular component \cite{hague2007a}. The decrease
in mass at intermediate $U$ is less pronounced in the case of the
unscreened Fr\"ohlich interaction, and not visible to the eye on the
plot, but is present, with a peak at around $U/t=4$ for the
$\lambda=1.5$ plot.

The pair size associated with the singlet bipolaron (shown in
Fig. \ref{fig:inversesize}) demonstrates the crossover between S0 and
S1 bipolarons. There is a qualitative difference between models with
short and long range interaction. For larger electron-phonon couplings
and long range interactions, the bipolaron size changes from small
(onsite) $R_{bp}<a$ to intersite $R_{bp}\sim a$ (the lattice constant)
on increasing $U$. The change between these two types of bipolaron
occurs at $U\sim 3$ for $\lambda = 1.5$ in the nearest- neighbor
model. The pair size associated with the unscreened Fr\"ohlich
interaction is not as small as that for the screened Fr\"ohlich
interaction, presumably because the potential well does not change so
rapidly with lattice index. It can be seen that at large $U$, the S1
bipolaron becomes more strongly bound on increasing $\lambda$, with
the inverse radius tending towards unity at $U=14$. Again this
demonstrates that the S1 bipolaron can only be considered to be
strongly bound at large $\lambda$ and $U\gg\lambda$. In contrast, the
Holstein bipolaron is always unbound at large $U$, as the bipolaron
radius becomes infinite ($R_{bp}^{-1}\rightarrow 0$). The modification
of the binding diagram on changing $R_{Sc}$ can be seen in these
figures. For $R_{Sc}=1$, $\lambda=1.5$ is needed to bind the bipolaron
at strong $U$, whereas the Fr\"ohlich bipolaron binds at $\lambda=0.9$.

We finish this section by computing the mass isotope exponent,
$\alpha$, which is plotted in Fig. \ref{fig:isotopeexponent} for a
variety of $\lambda$ and $U$. A dip in the isotope exponent seen in
the nearest neighbor and screened Fr\"ohlich interactions corresponds
to the hump seen in the inverse mass. Again, this dip coincides with
the degeneracy in the S0 and S1 energies.

For completeness, we note that it is also possible to compute
properties of the bipolaron at weak coupling and $U>0$ where neither
singlet nor triplet bipolarons are bound, i.e. the measured properties
correspond to those of two polarons. The weak coupling properties of
the polaron can be determined from second order perturbation theory
and are found to be $E/t=-z-\lambda\Gamma_{E}(\bar{\omega})$,
$m_0/m^{*}=1-\lambda\Gamma_{m^*}(\bar{\omega})$,
$\alpha_{m^*}=\lambda\Gamma_{\alpha_{m^*}}(\bar{\omega})$ and $N_{\rm
ph}=\lambda\Gamma_{N_{\rm ph}}(\bar{\omega})$. For $\bar{\omega}=1$,
the $\Gamma$ coefficients can be found in
Ref. \onlinecite{spencer2005a}, with the exception of the coefficients
for the nearest neighbor model which are $\Gamma_{E}=2.391(5)$,
$\Gamma_{m^*}=0.607(7)$, $\Gamma_{\alpha_{m^*}}=0.294(1)$ and
$\Gamma_{N_{\rm ph}}=1.999(1)$ and for the Hubbard-Fr\"ohlich model
with $R_{sc}=2$, which are $\Gamma_{E}=1.423(7)$,
$\Gamma_{m^*}=0.392(0)$, $\Gamma_{\alpha_{m^*}}=0.1540(0)$ and
$\Gamma_{N_{\rm ph}}=1.116(2)$.

\begin{figure}
\includegraphics[height=80mm,angle=270]{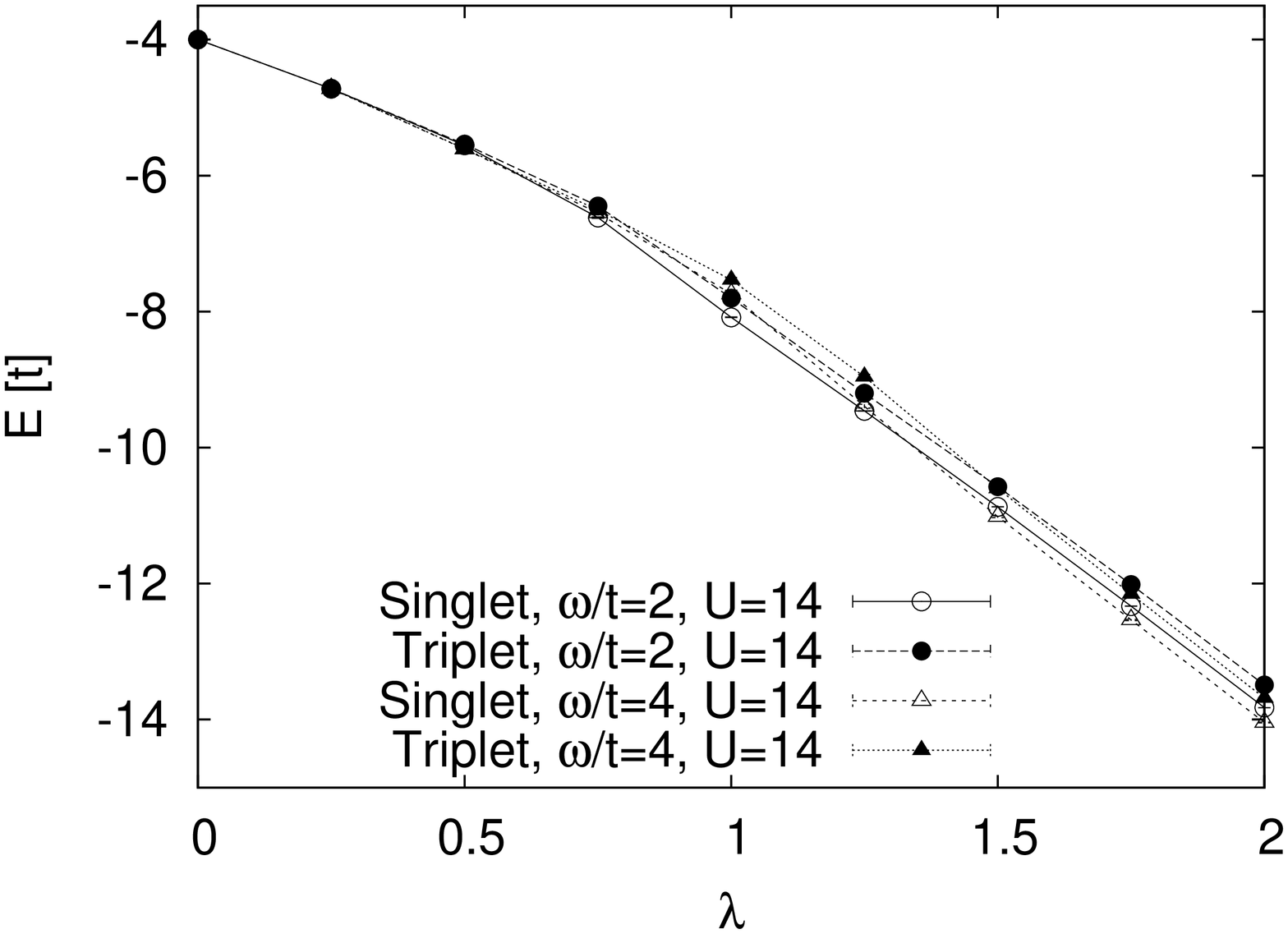}
\includegraphics[height=80mm,angle=270]{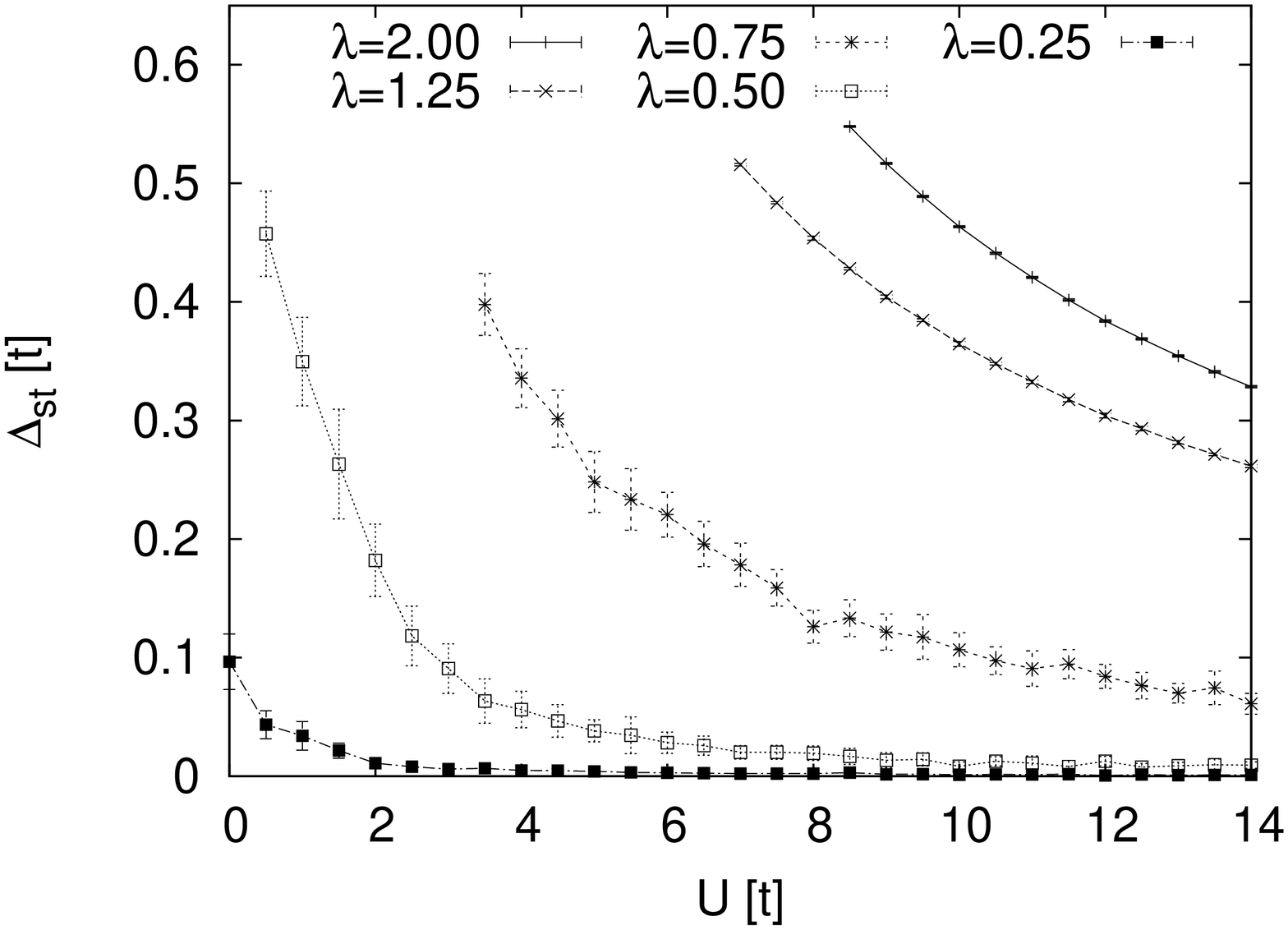}
\caption{Top: Total triplet and singlet energies of Hubbard-Fr\"ohlich
bipolarons. $U=14t$, $\bar{\beta}=7$, $\bar{\omega}=2$ and $U/t=14$,
$\omega/t=4$, $\bar{\beta}=7$. Bottom: Singlet-triplet splitting when
$\bar{\beta}=7$, $\bar{\omega}=2$ and various $U/t$ and $\lambda$.}

\label{fig:tripletenergy}
\end{figure}

\begin{figure}
\includegraphics[height=80mm,angle=270]{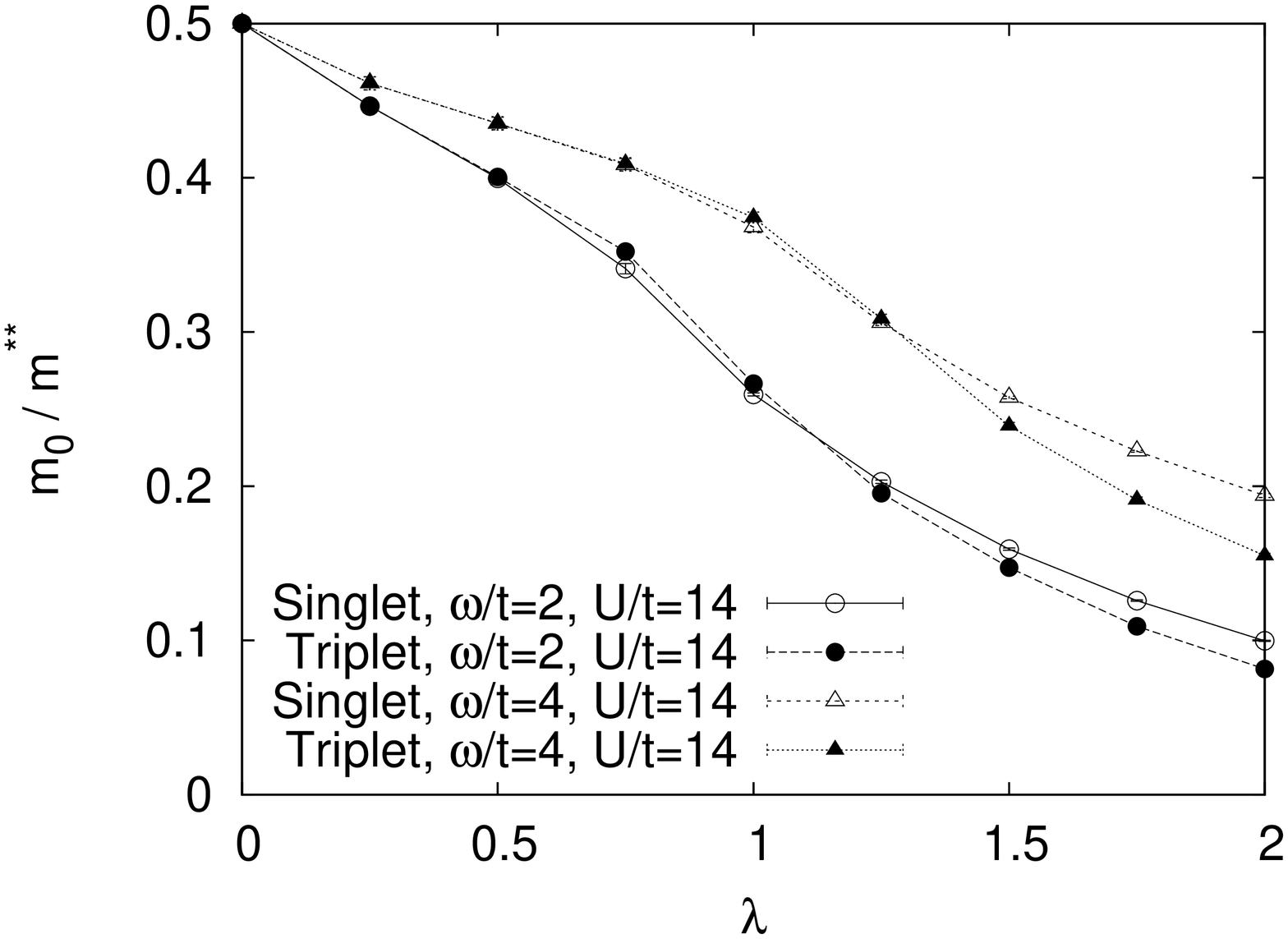}
\includegraphics[height=80mm,angle=270]{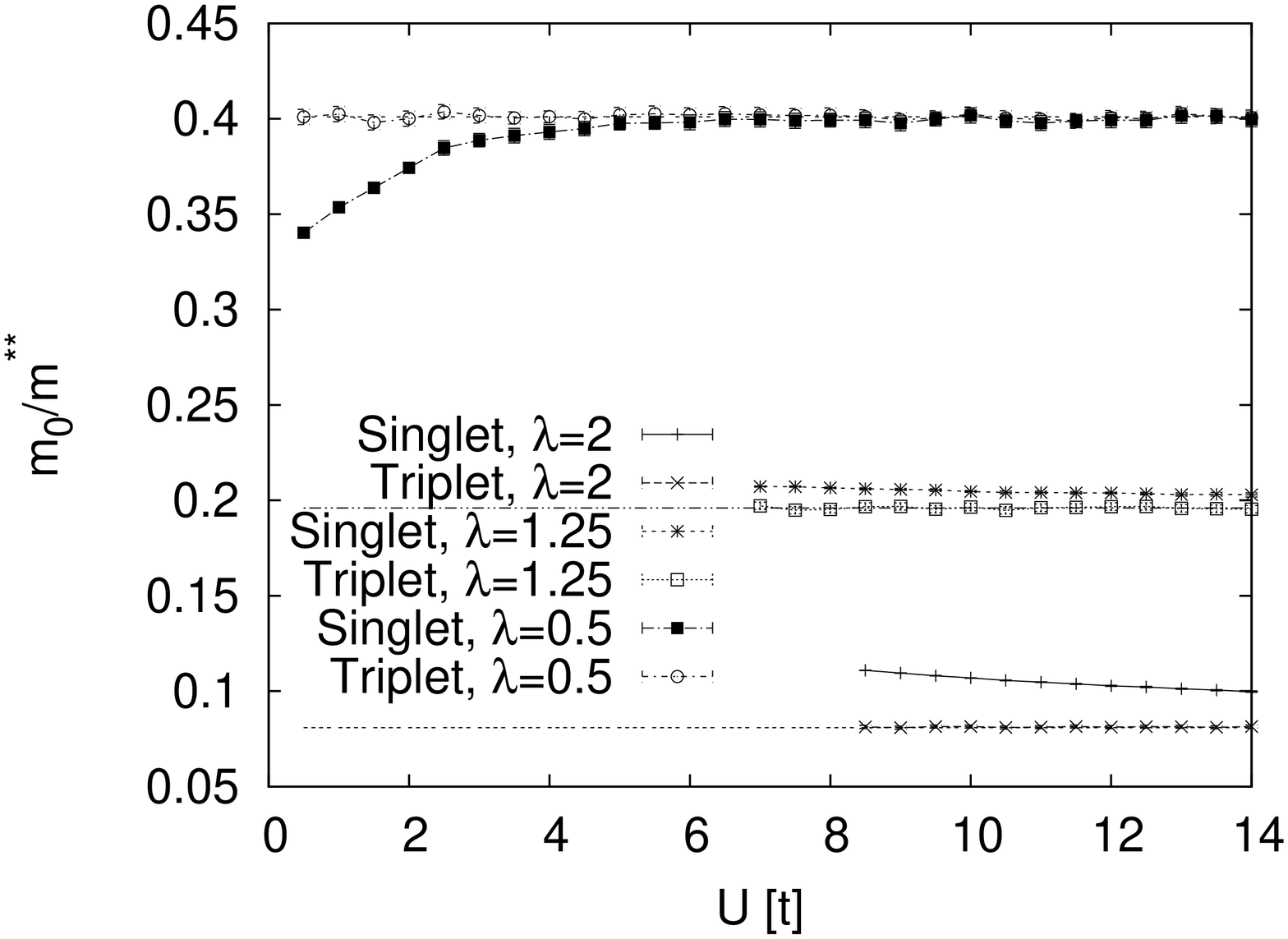}
\caption{Top: Inverse mass of singlet and triplet bipolarons when $\bar{\beta}=7$, $\bar{\omega}=2$, $U=14t$. The Hubbard-Fr\"ohlich model is simulated. N.B. Triplets are heavier than singlets at strong coupling, as in the solution of the UV model. Bottom: Variation of inverse mass with $U$. The triplet mass is constant.}
\label{fig:tripletim}
\end{figure}

\begin{figure}
\includegraphics[height=85mm,angle=270]{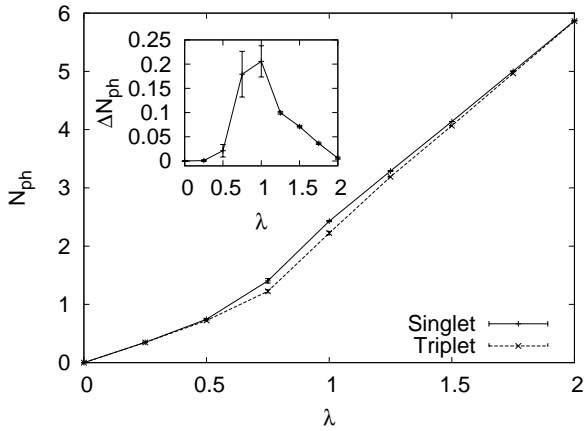}
\caption{Number of phonons associated with singlet and triplet
bipolarons in the Hubbard-Fr\"ohlich model. Inset: the difference
between the number of phonons associated with singlet and triplet
bipolarons. These data were computed for the paramters $\bar{\beta}=7$,
$\bar{\omega}=2$ and $U=14t$. }
\label{fig:tripletnp}
\end{figure}

\begin{figure}
\includegraphics[height=85mm,angle=270]{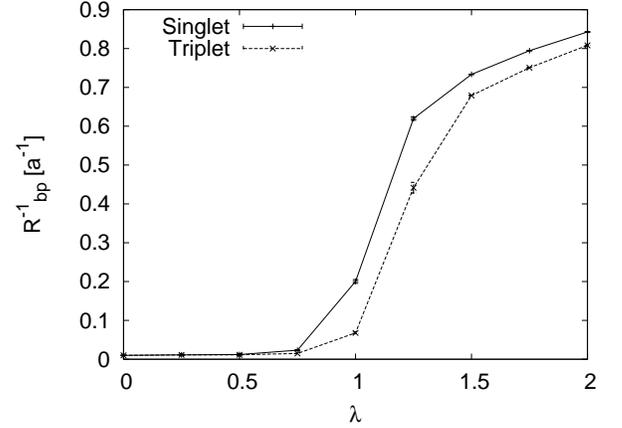}
\caption{Inverse size of singlet and triplet bipolarons in the
Hubbard-Fr\"ohlich model when $\bar{\beta}=7$, $\bar{\omega}=2$,
$U=14t$}
\label{fig:tripletbps}
\end{figure}

\section{Triplet bipolarons in the Hubbard-Fr\"ohlich model}
\label{sec:triplet}

In this section, we compute the properties of triplet bipolarons and
compare with those of singlet bipolarons. Our algorithm can compute
triplet and singlet properties simulaneously since they are related
through the measurement of the exchange sign. Strictly, it is not
necessary to directly calculate the properties of triplet states in 1D
in this way (singlet and triplet bipolarons are degenerate at large
$U$) however this contributes part of the development of our
algorithm, since there is no alternative approach for 2D and
above. The Hubbard-Holstein model in 1D has no bound triplet states,
so the properties associated with triplet symmetries are not examined
for local interactions.

A quantity that is straightforward to measure using our Monte-Carlo
algorithm is the energy splitting between singlet and triplet
bipolaron states. From this and the singlet energy, the total energy
of triplet bipolarons can be determined. The singlet-triplet splitting
and singlet and triplet energies are shown in figure
\ref{fig:tripletenergy}. The top panel shows the total singlet and
triplet energies for various $\lambda$ and $\bar{\omega}=2$ and
$U=14$. Below a certain electron-phonon coupling, the singlets and
triplets are degenerate, as Coulomb repulsion inhibits pairing in the
singlet, and the triplet state (which must be higher in energy than
the singlet) is unbound. This can also be seen for $\omega/t=4$, and
$U/t=14$. The singlet-triplet splitting is relatively small compared
to the total energy of the bipolaron and the total energy only changes
slightly on increase of phonon frequency.

The bottom panel of Fig. \ref{fig:tripletenergy} shows the splitting
between singlet and triplet states for a variety of $U/t$ and
$\lambda$. On decreasing $U$, the singlet-triplet splitting
increases. This is expected as the singlet state becomes increasingly
well bound on decreasing $U$, whereas the binding energy of the
triplet should remain constant. Efficient computation of the splitting
is limited by the temperature, as the higher energy triplet
configurations need to be visited enough to obtain accurate
statistics. For the $\bar{\beta}=7$ shown here, this effectively
limits us to maximum splittings of the order of $t/2$ (otherwise
computation time is excessive). Increase of $\lambda$ also
increases the splitting, which can also be seen in the top panel.

We also examine the relative masses of the triplet and singlet
bipolarons, which can be seen in Fig. \ref{fig:tripletim}. We begin by
computing the inverse effective mass for $\bar{\omega}=2$, $U/t=14$
and $\bar{\beta}=7$ which can be seen in the top panel of the
figure. For large $\lambda$, it is clear that the triplet bipolaron is
heavier than the singlet one. However, for small $\lambda$, the
situation is reversed, with the triplet state slightly lighter than
the singlet one. The slightly lighter mass is persistent to
$\lambda\sim 1.125$, where the masses cross, for large $\lambda$,
triplets are observed to be heavier than singlets.  We also make
computations for $U/t=14$, and $\bar{\omega}=4$. Triplets are heavier
than singlets over a wider range of the parameter space for larger
$\omega$. This is consistent with the large $\omega$ limit, where the
models with long range interaction have similar properties to the $UV$
model (the $UV$ model always has heavier triplets).

The bottom panel of Fig. \ref{fig:tripletim} shows how the masses of
singlets and triplets change with $U$. On decreasing from large $U$,
the singlet mass decreases to become lighter than the triplet as the
S0 and S1 states become degenerate. However at small $U$ this is
reversed, since the singlet becomes tightly bound in an S0 state (see
Fig. 2). This binding into and S0 state, and the associated increase
in mass can also be seen in the $\lambda=0.5$ curves at low $U$.

The number of phonons associated with singlet and triplet bipolarons
is shown in figure \ref{fig:tripletnp}, with the difference between
them shown in the inset. Simulations were run for $\bar{\beta}=7$,
$\bar{\omega}=2$, $U=14t$ and various $\lambda$. The number of phonons
associated with singlet and triplet bipolarons is very similar. For
the parameters shown, there are slightly more phonons associated with
the singlet than the triplet, but the number of phonons is degenerate
when $\lambda$ is very small (as the two particles are not bound). The
number of phonons associated with the cloud is also similar for
singlet and triplet bipolarons at very large $\lambda$. For such a
large $U$ and $\lambda=2$, the singlet bipolaron has some S1
characteristics which explains the similarity between singlet and
triplet. For much larger $\lambda\gg U/t$, the electron-phonon
coupling is expected to overcome the Coulomb repulsion to bind the
singlet bipolaron into an S0 configuration, and the properties of
singlet and triplet will differ.

We complete this section by considering the inverse size of singlet
and triplet bipolarons which can be seen in
Fig. \ref{fig:tripletbps}. As before, $\bar{\beta}=7$,
$\bar{\omega}=2$, and $U=14t$. The triplet is larger for all
$\lambda$. This makes the heavier triplet states slightly
counterintuitive, because in the case of singlets, larger
wavefunctions tend to be associated with lighter bipolarons.

\begin{figure}
\includegraphics[height=85mm, angle=270]{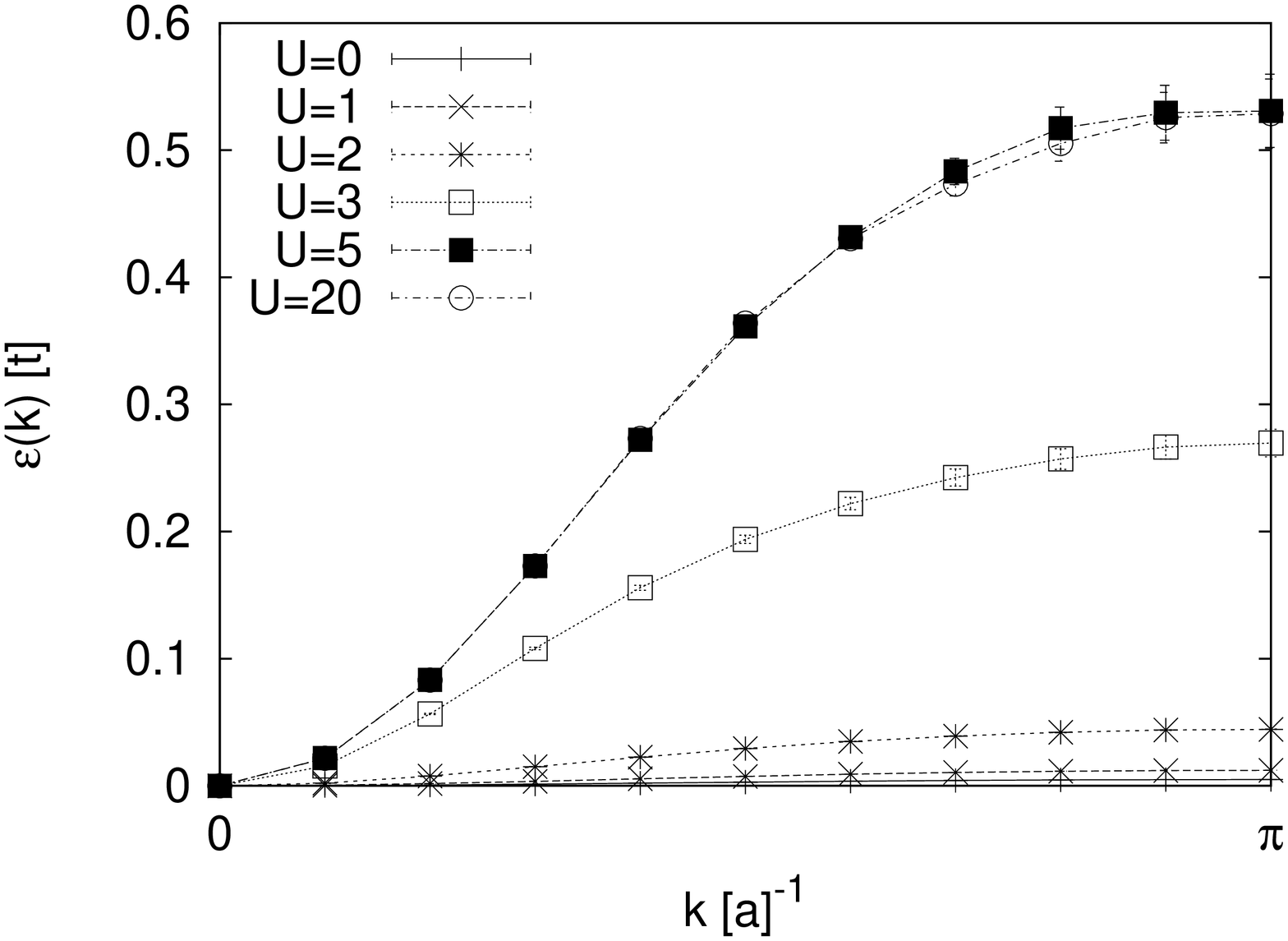}
\includegraphics[height=85mm, angle=270]{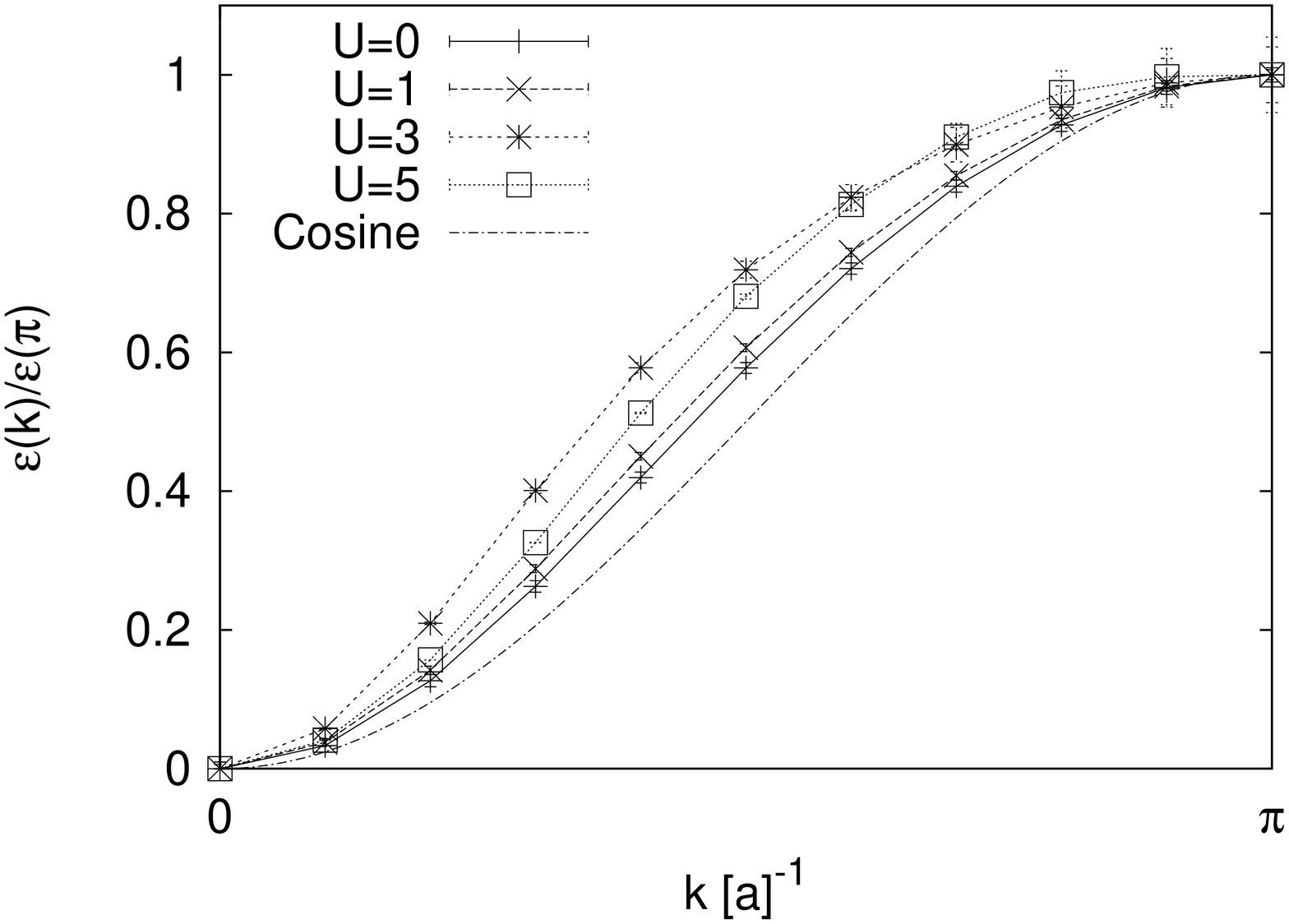}
\caption{Top: Dispersion of the Hubbard-Holstein for $\lambda=1$,
$\bar{\omega}=1$, $\bar{\beta}=14$ and various $U$. Bottom: Comparison
between the shape of the dispersion, $\epsilon(k)/\epsilon(\pi)$ and
the bare cosine band.}
\label{fig:hhdispersionvaru}
\end{figure}

\begin{figure}
\includegraphics[height=85mm, angle=270]{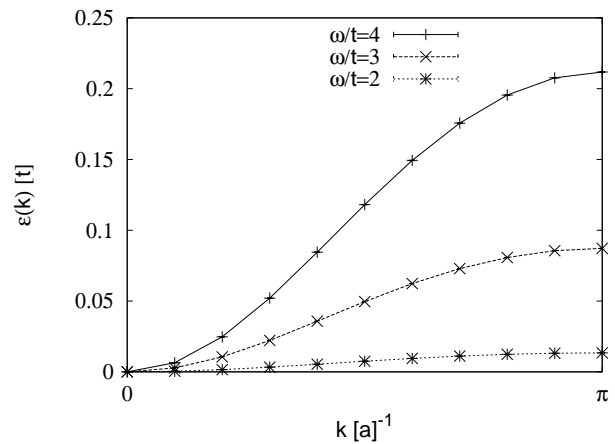}
\caption{Dispersion of the Hubbard-Holstein model for $\lambda=2$,
$U/t=5$, $\bar{\beta}=7$ and various $\omega$. Errorbars are too small to be visible.}
\label{fig:hhdispersionvaromega}
\end{figure}

\begin{figure}
\includegraphics[height=85mm, angle=270]{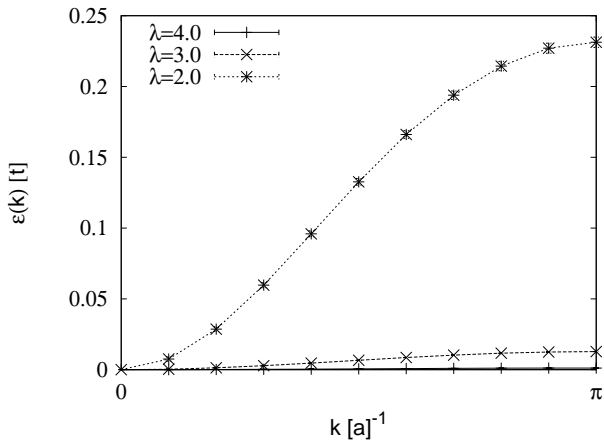}
\caption{Dispersion of the Hubbard-Holstein model for $\omega/t=4$,
$U/t=5$, $\beta/t=3.5$ and various $\lambda$.}
\label{fig:hhdispersionvarlambda}
\end{figure}

\begin{figure*}
\includegraphics[height=85mm,angle=270]{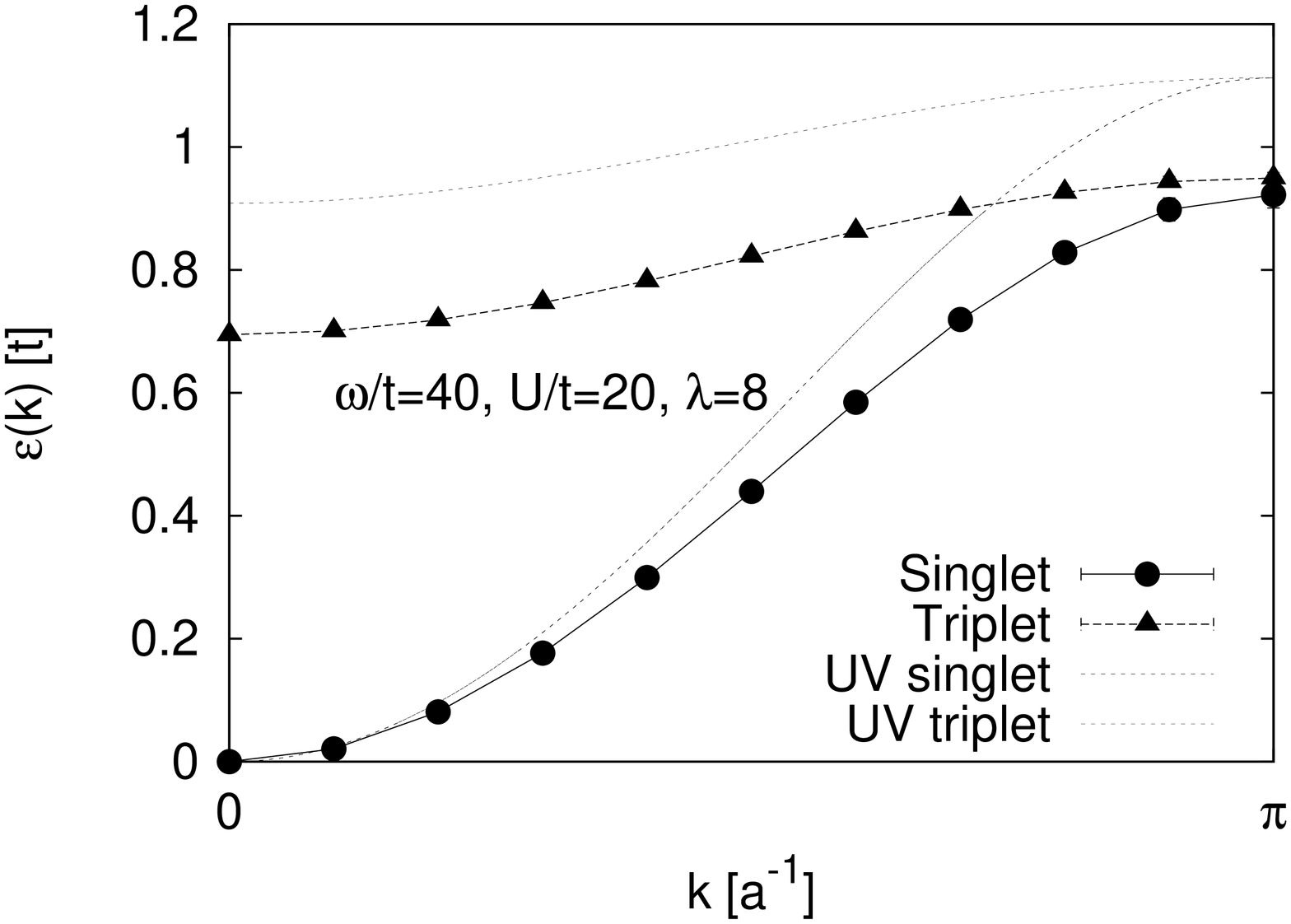}
\includegraphics[height=85mm,angle=270]{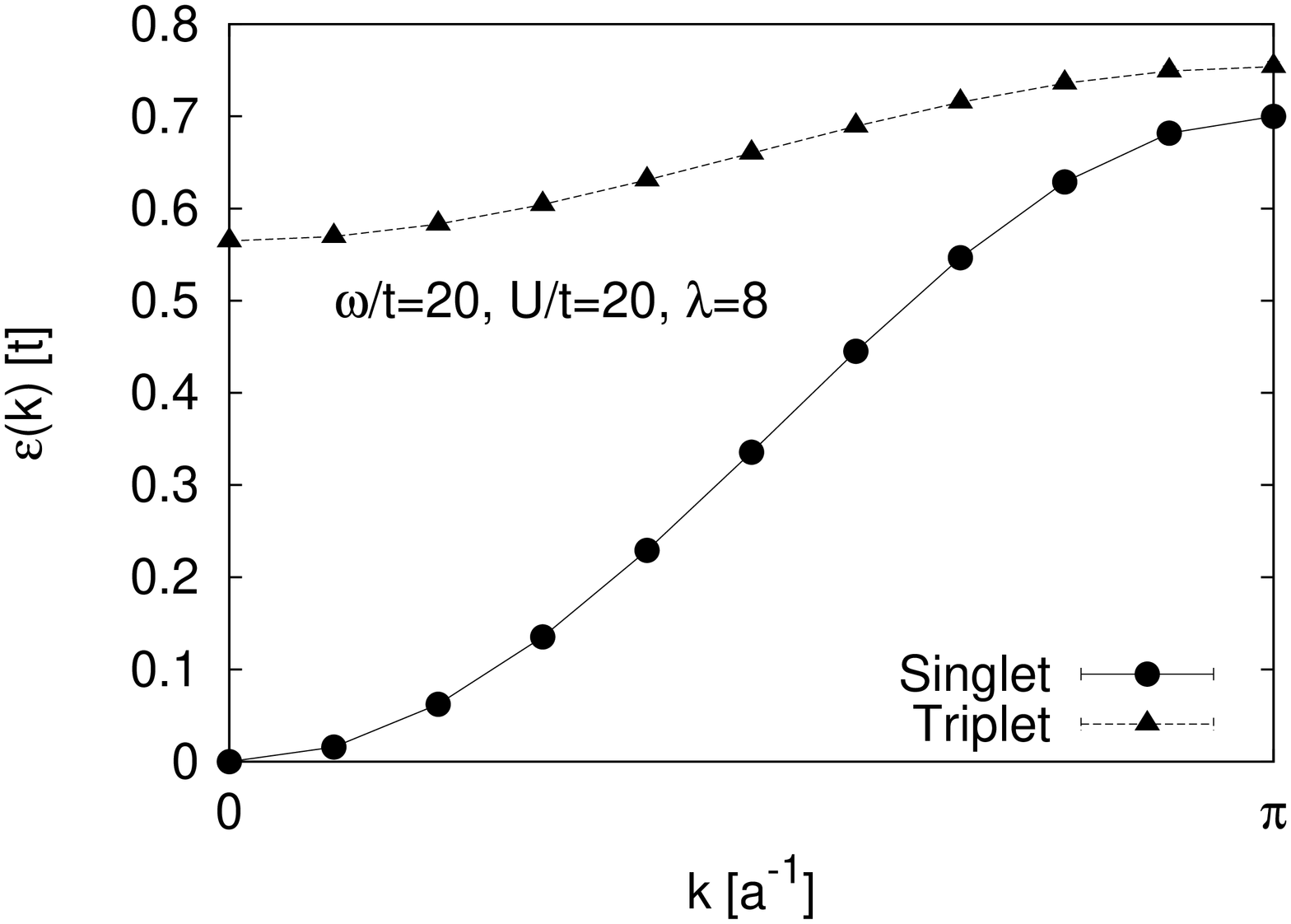}
\includegraphics[height=85mm,angle=270]{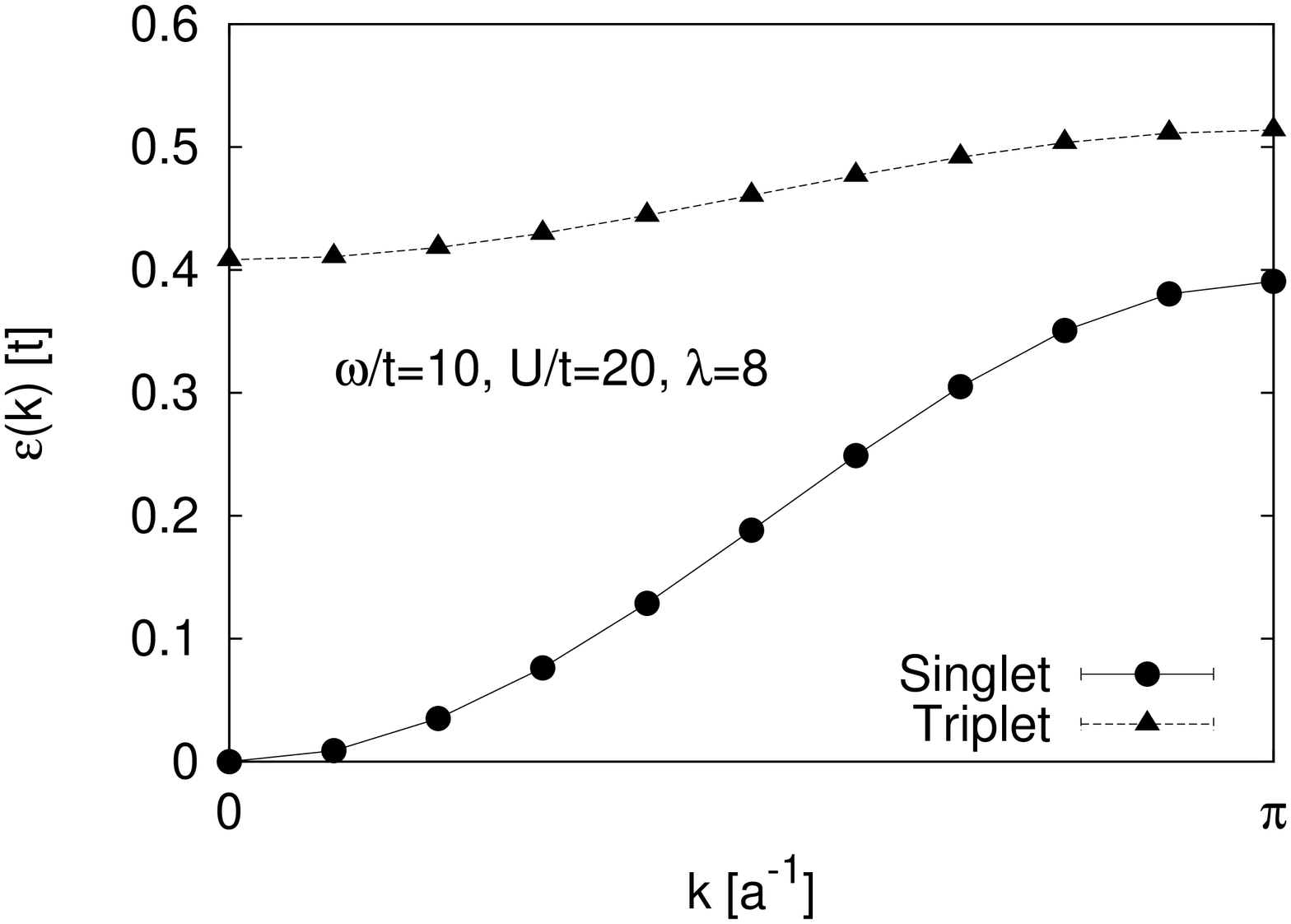}
\includegraphics[height=85mm,angle=270]{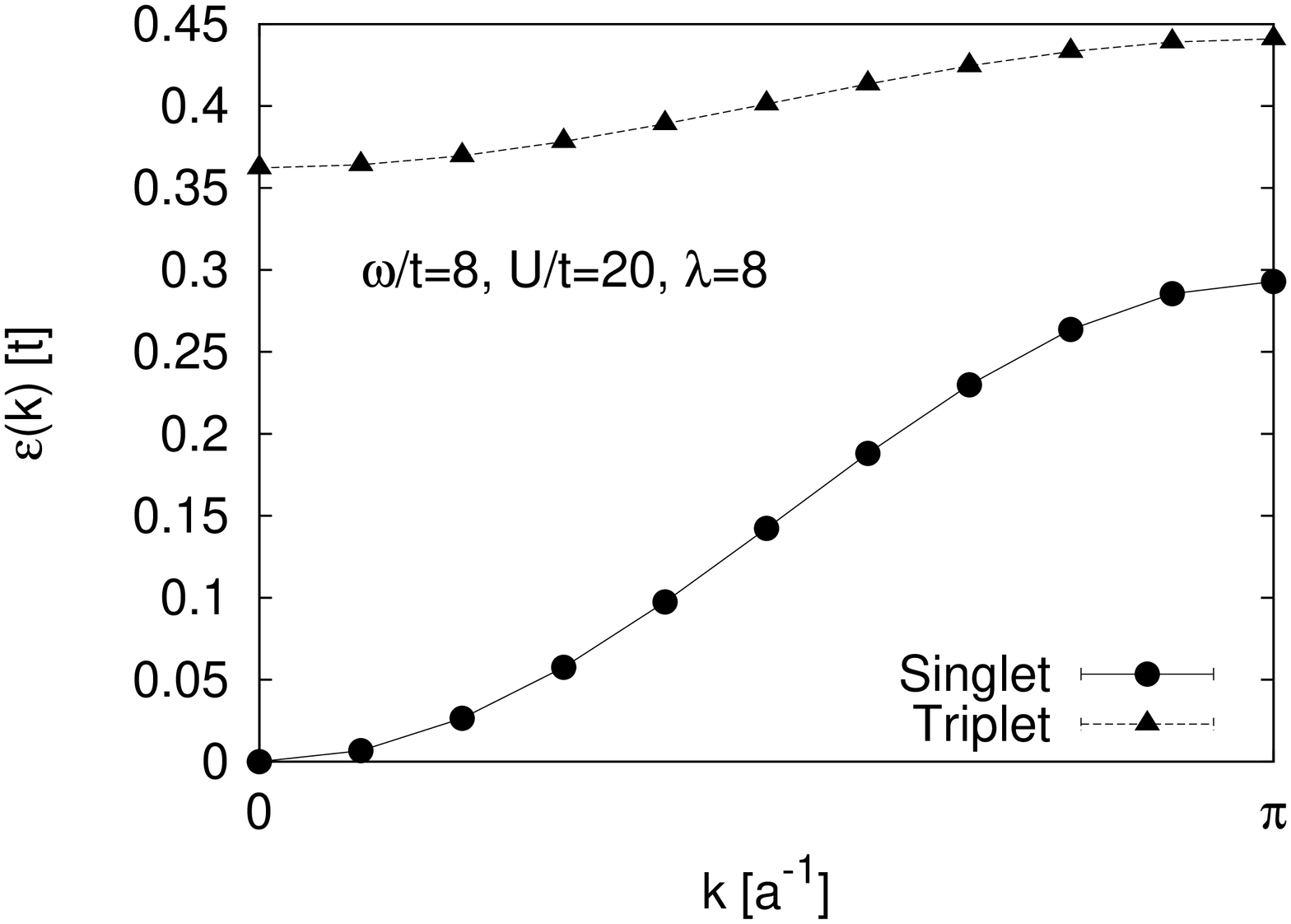}
\includegraphics[height=85mm,angle=270]{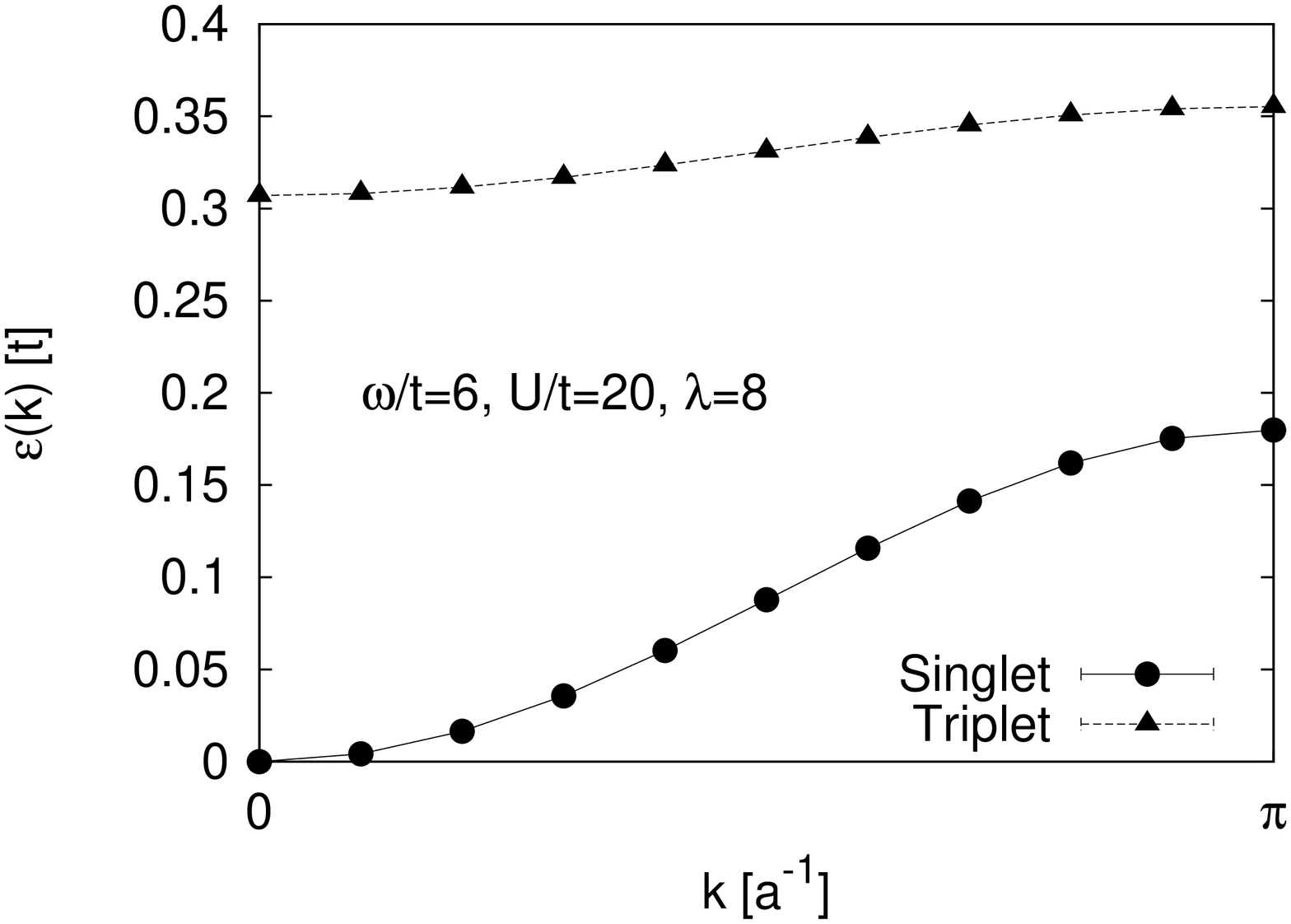}
\includegraphics[height=85mm,angle=270]{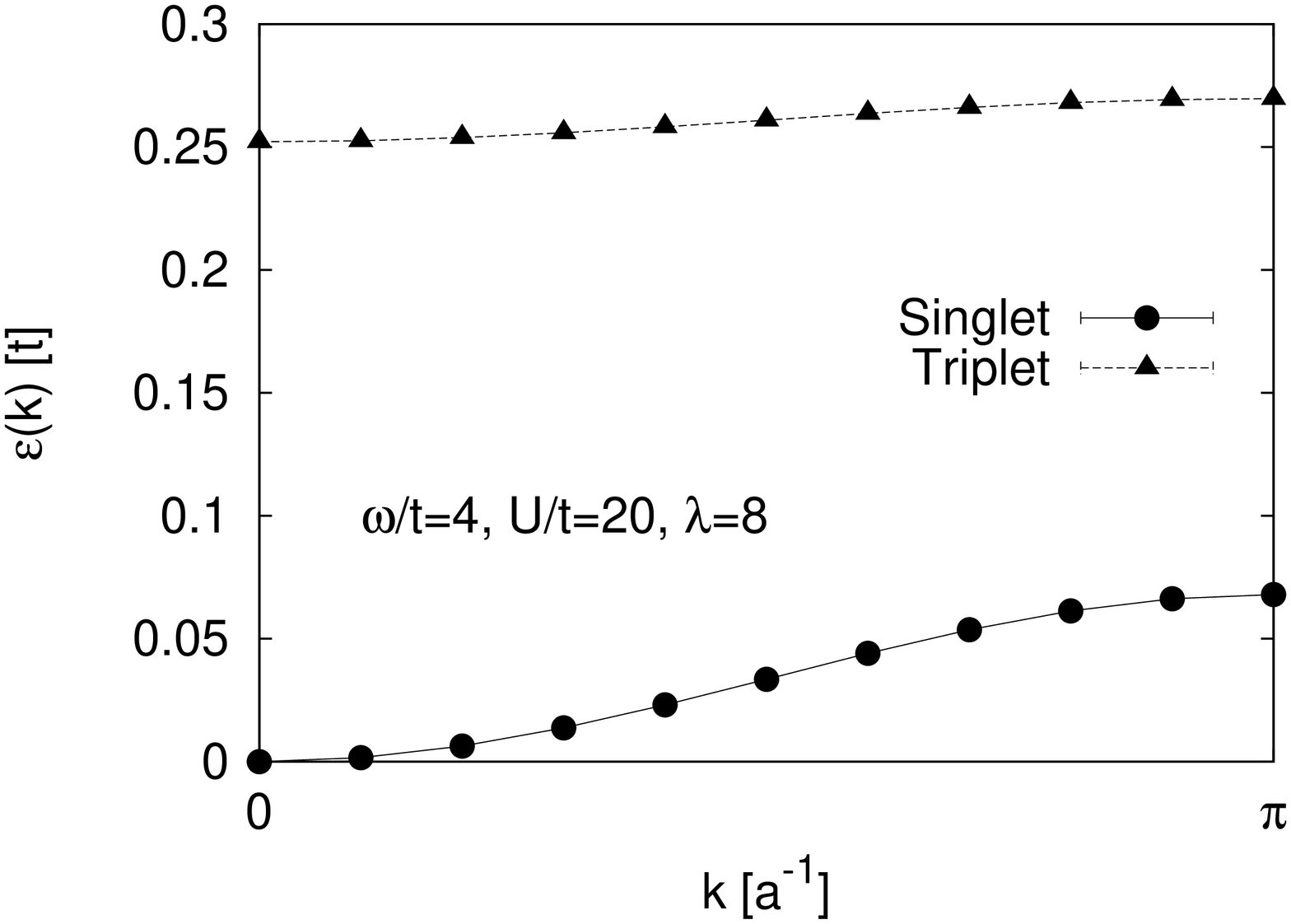}
\caption{Singlet and triplet dispersions. Data are shown for various
$\omega/t$, $U/t=20$, $\lambda = 8$, $\beta/t = 3.5$ and a Fr\"ohlich
interaction. Error bars are smaller than the points. The dispersions
are plotted relative to the zone center singlet energy. Note that the
triplet spectrum is much flatter. The spectra are near degenerate at
the zone edge for large phonon frequencies, but are separated in
energy for lower phonon frequencies indicating that the separation is
a dynamical effect. The dispersions of the correponding $UV$ model are
also shown on the $\omega/t=40$ plot.}
\label{fig:dispersionvaromega}
\end{figure*}

\begin{figure*}
\includegraphics[height=85mm,angle=270]{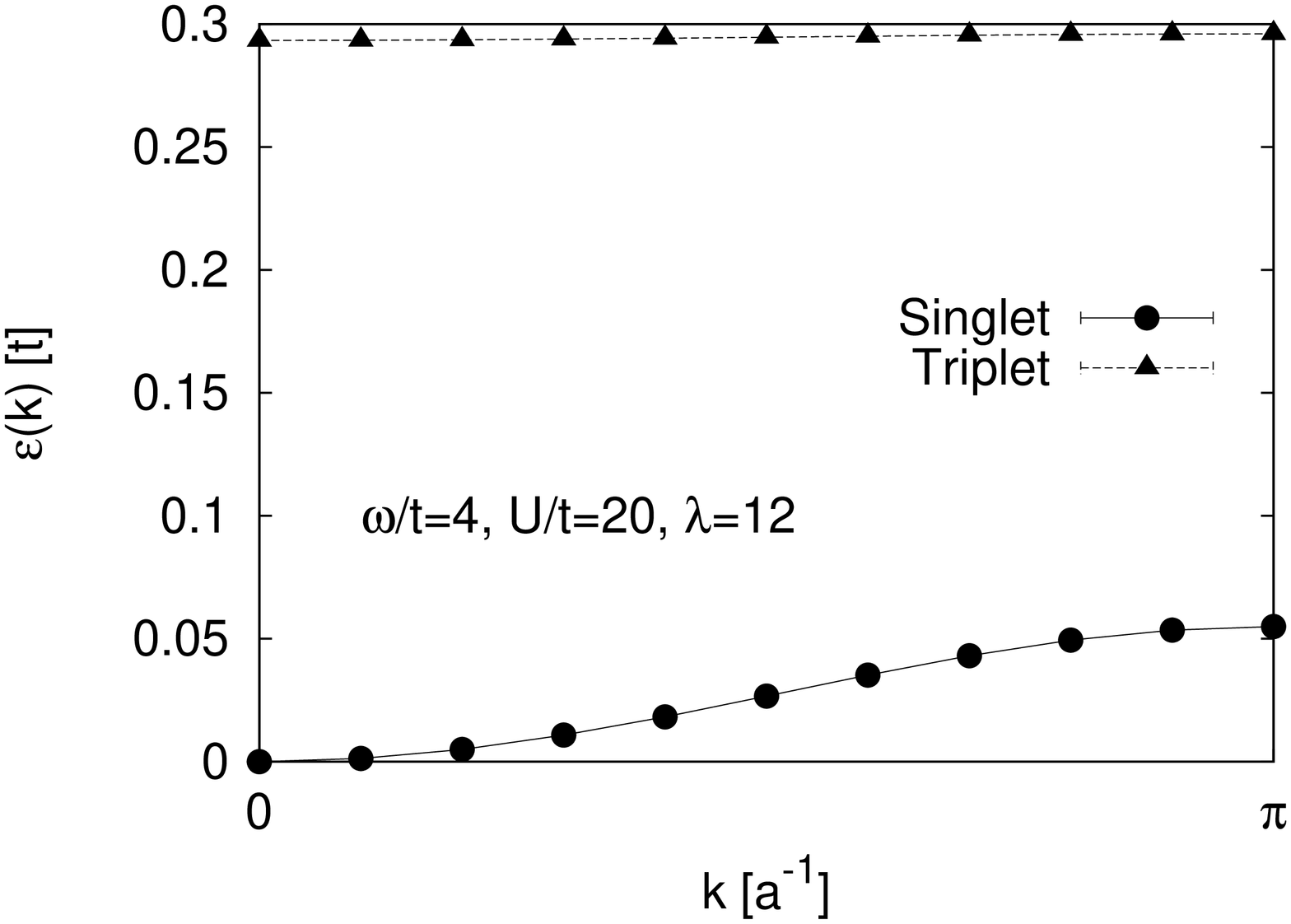}
\includegraphics[height=85mm,angle=270]{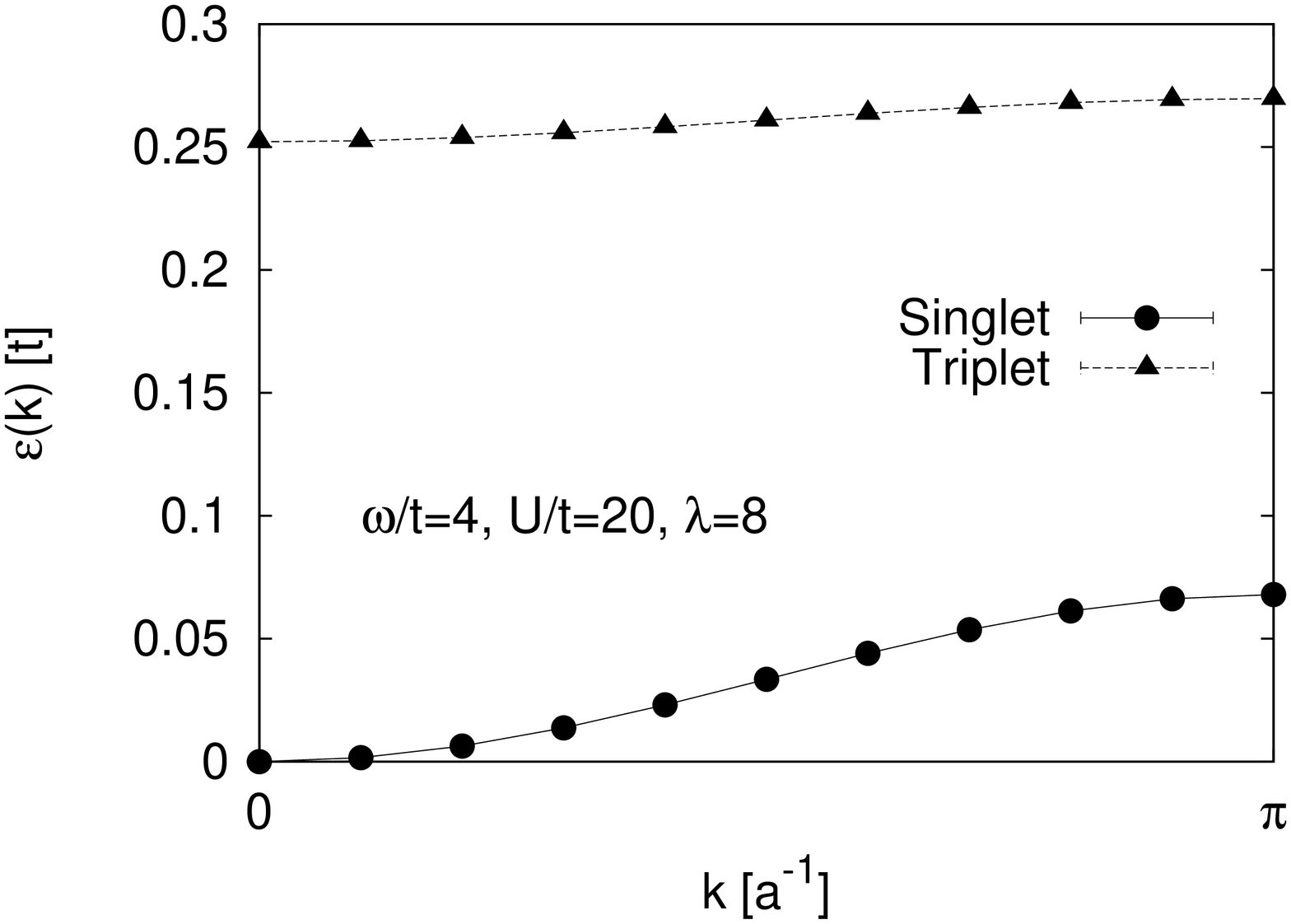}
\includegraphics[height=85mm,angle=270]{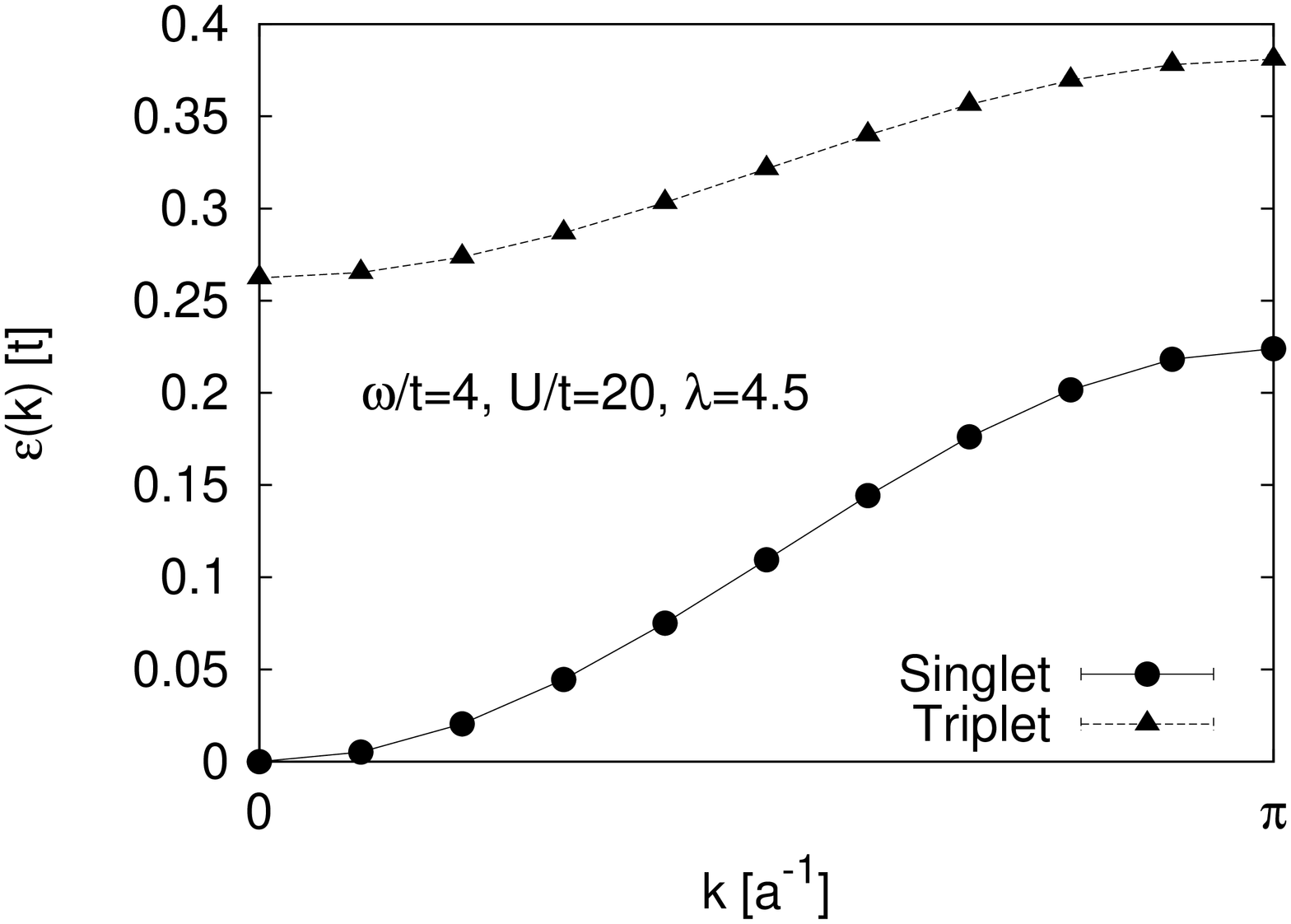}
\includegraphics[height=85mm,angle=270]{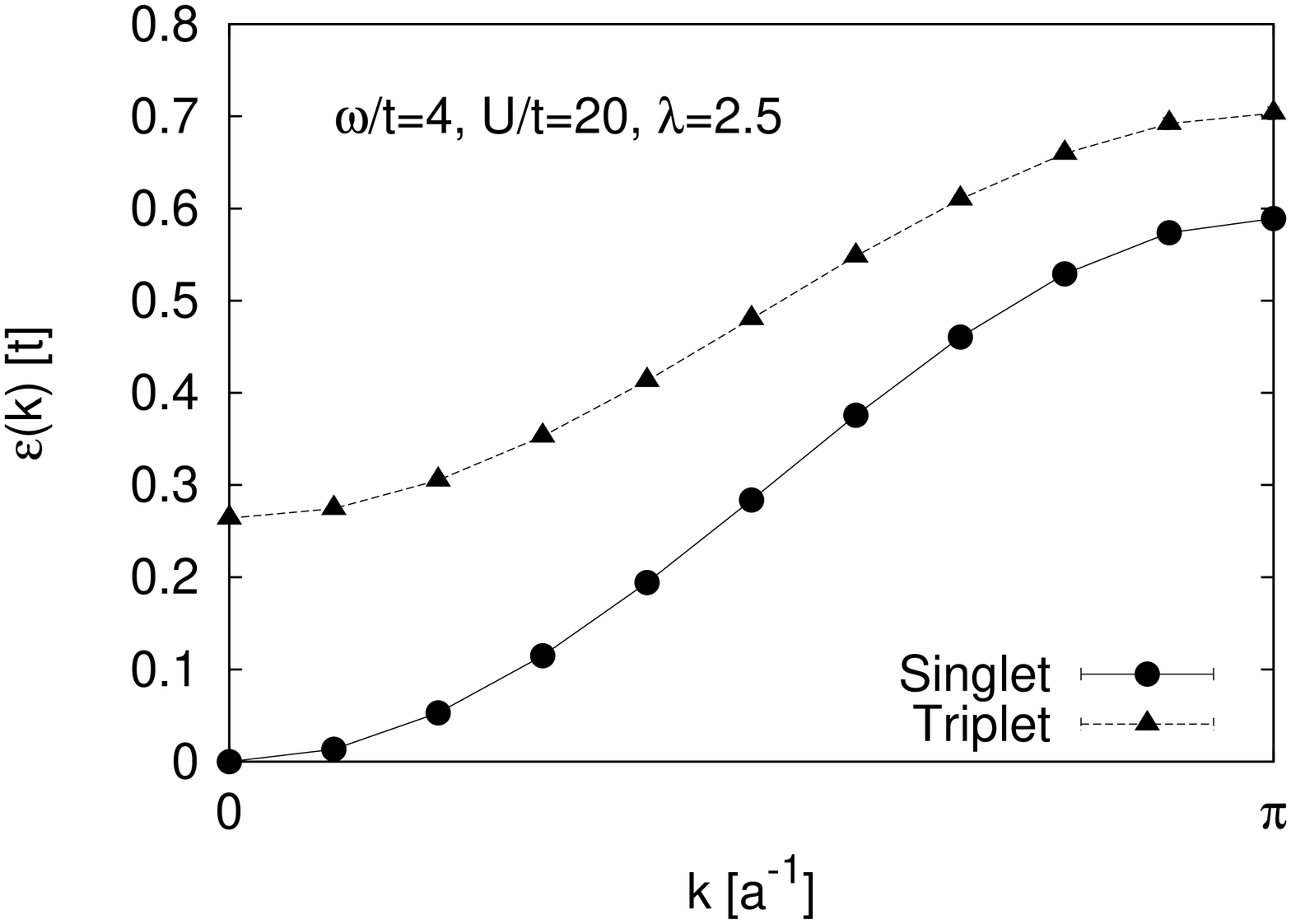}
\includegraphics[height=85mm,angle=270]{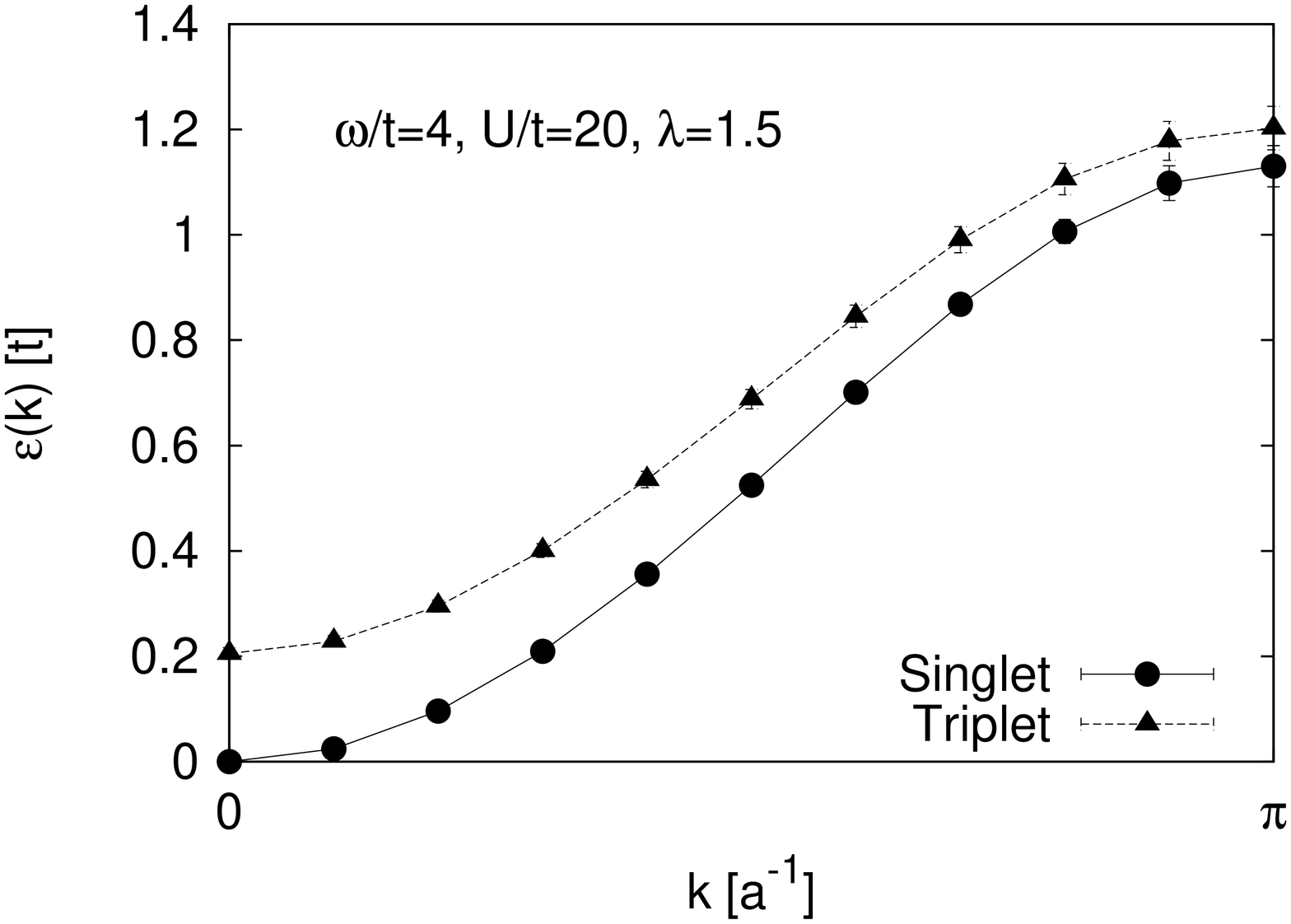}
\includegraphics[height=85mm,angle=270]{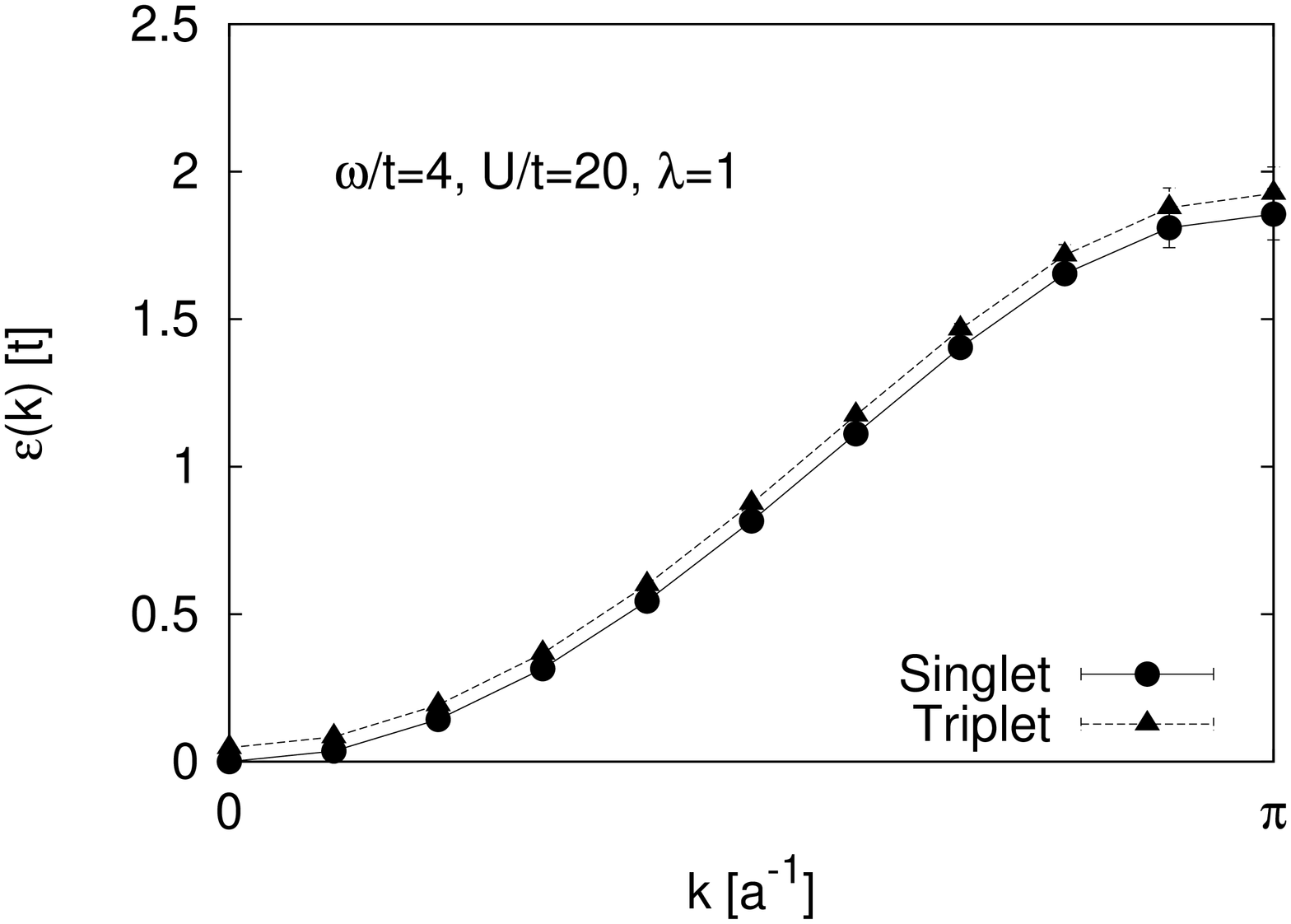}
\caption{Singlet and triplet dispersions. Data are plotted for the
parameters, $\omega/t = 4$, $U/t=20$, various $\lambda$, $\beta/t =
3.5$ and a Fr\"ohlich interaction. Where no errorbars are visible,
they are smaller than the points. The dispersions are plotted relative
to the zone center singlet energy. Note that the triplet dispersion is
much flatter than the singlet one.}
\label{fig:dispersionvarlambda}
\end{figure*}

\begin{figure}
\includegraphics[height=85mm,angle=270]{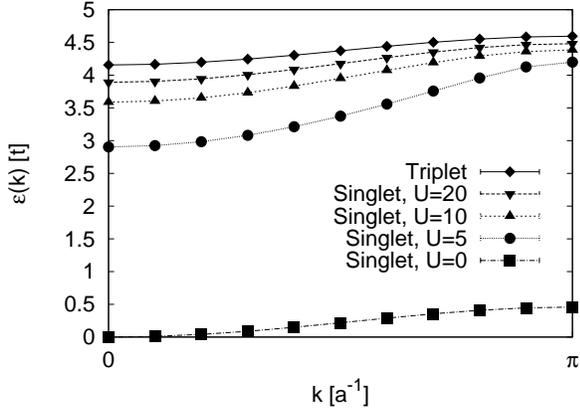}
\caption{Variation of singlet spectrum with varying $U$. Data are
plotted for parameters $\omega/t = 4$, $\lambda=2.5$, $\beta=3.5$ and
a Fr\"ohlich interaction. Errorbars are smaller than the points.}
\label{fig:dispersionvaru}
\end{figure}

\begin{figure}
\includegraphics[height=85mm, angle=270]{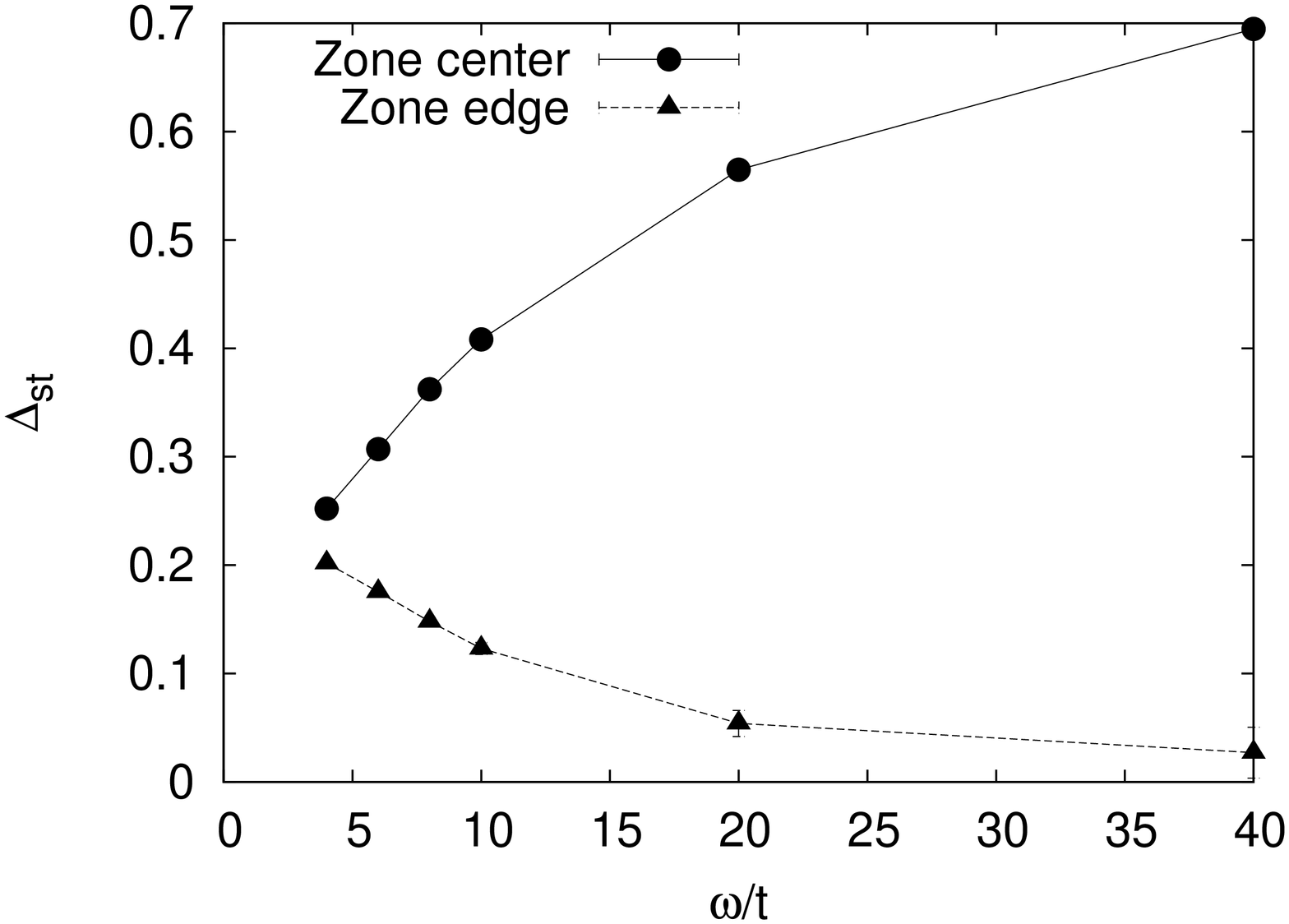}
\caption{Dependence of the zone center and zone edge singlet-triplet
splitting with $\omega/t$. $\lambda=8$, $U/t=20$, $\beta = 3.5$.}
\label{fig:kdepsplitting}
\end{figure}

\section{Bipolaron dispersion}
\label{sec:dispersion}

One of the powers of our Monte-Carlo algorithm is the possibility to
compute the dispersions of the bipolaron efficiently. In this section,
we discuss spectra computed with our algorithm. The estimator for our
spectra is given in equation \ref{eqn:spectrum}. We compute triplet
and singlet spectra at the same time.

\subsection{Hubbard--Holstein model}

In the Hubbard-Holstein model, triplets are never bound, so we do not
compute the dispersion associated with triplet symmetry, concentrating
instead on singlets. We show the effects of varying $U/t$ on the
dispersion of the Hubbard-Holstein bipolaron for $\lambda=1$,
$\bar{\omega}=1$ in Fig. \ref{fig:hhdispersionvaru}. On decreasing
$U$, the electron-phonon interaction begins to dominate, resulting in
strongly bound on-site pairs. This is characterised by a significant
decrease in the band-width, which is consistent with a large increase
in the effective mass. For $U/t\gtrsim 5$, the dispersion does not
change significantly on increasing $U$ (the bipolaron is not bound at
larger $U$). This is consistent with the estimate from the large
frequency limit of the model, where bipolarons unbind if
$U>4t\lambda$.

The bottom panel of Fig. \ref{fig:hhdispersionvaru} shows the shape of
the dispersion, $\epsilon(k)/\epsilon(\pi)$ as compared to the cosine
dispersion associated with the non-interacting limit. The dispersion
is skewed away from the cosine and is flatter at larger momenta. The
dispersion also becomes shallower at larger $U$ where the bipolaron
unbinds into two polarons, demonstrating that the band flattening in
bipolarons is significantly larger than in polarons.

The effects of varying $\omega$ are shown in
Fig. \ref{fig:hhdispersionvaromega}. As the phonon frequency is
increased, the bipolaron band gets wider, which is consistent with the
decrease in effective mass. The change in band width is quite
significant as $\omega/t$ changes from 2 to 4, and is around a factor
of 10. We also compute the dispersion of the Hubbard-Holstein model
for $\omega/t=4$, $U/t=20$ and various $\lambda$, which can be seen in
Fig. \ref{fig:hhdispersionvarlambda}. On increased $\lambda$, the band
width decreases, consistent with the significant increase in effective
mass. The bipolaron binds into an S0 state on increasing $\lambda$,
with the bandwidth changing by more than three orders of magnitude
from $\lambda=1$ (not shown in figure) to $\lambda=4$. We note that
similar calculations were presented in
Ref. \onlinecite{hohenadler2005a}.

\subsection{Hubbard--Fr\"ohlich model}

The form of bipolaron dispersions in the Hubbard-Fr\"ohlich model has
not previously been computed. At large coupling and large phonon
frequency, the Hubbard-Fr\"ohlich model maps onto a UV-model, which is
a convenient starting point for analytics. The singlet dispersion of
the UV model can be found from the equation \cite{kornilovitch2004a},
\begin{eqnarray}
\left(\sqrt{\epsilon_s(k)^2-16t^2\cos^2(k/2)}+\bar{U}\right)\times &&\\
\left(\sqrt{\epsilon_s(k)^2-16t^2\cos^2(k/2)}-\epsilon_s(k)\right)-2\bar{V}\left(\bar{U}-\epsilon_s(k)\right)&=0&\nonumber
\end{eqnarray}
the triplet dispersion is given by
\begin{equation}
\epsilon_t(k)=-\bar{V}-\frac{4t^2}{\bar{V}}\cos^2(k/2).
\end{equation}
here, we take $\bar{U}=U-2W\lambda$ and
$\bar{V}=2W\lambda\phi(\avec)$, which is an acceptable approximation
for a very tightly bound state of the Hubbard-Fr\"ohlich model at very
large $\omega$ (since at very large $\lambda$ the magnitude of the
pair wavefunction tends to zero too rapidly to sample next-nearest
neighbor interactions). 

We plot dispersions for various $\omega/t$ with $U/t=20$, $\lambda =
8$, $\beta/t = 3.5$ and a Fr\"ohlich interaction in figure
\ref{fig:dispersionvaromega}. All dispersions are plotted relative to
the zone center singlet energy. We also plot the UV model dispersions
on the panel showing the $\omega/t=40$ curves. We have already determined
that triplet bipolarons are significantly heavier than singlets in the
large $U$ large $\lambda$ regime, and this is consistent with the much
flatter triplet dispersion. In analytics of the $UV$ model
\cite{kornilovitch2004a}, singlet and triplet dispersions are found to
be degenerate at $k=\pi/a$. In agreement with results from the
$UV$-model, the spectra are nearly degenerate at the zone edge for
large phonon frequencies, but become increasingly separated in energy
for lower phonon frequencies. This is clear evidence that the
degeneracy at the $\pi$ point in the $UV$-model is lifted by dynamical
effects.

We also investigate the effect of varying $\lambda$ on the singlet and
triplet dispersions in Fig. \ref{fig:dispersionvarlambda}. Data are
plotted for the parameters $\omega/t = 4$, $\beta/t = 3.5$, $U/t=20$
and for various $\lambda$. Where no errorbars are visible, they are
smaller than the points. The dispersions are plotted relative to the
zone center singlet energy. The triplet spectrum is much flatter than
the singlet one, which is consistent with the larger triplet masses at
higher $\omega$ and $\lambda$. In particular, the triplet dispersion
is almost flat for $\lambda = 12$ in comparison with the singlet
dispersion. For $\lambda = 1$, where singlets and triplets are only
just bound, the dispersions are near degnenerate. In contrast to the
Hubbard-Holstein bipolaron, the bandwidth decreases by only one order
of magnitude on increasing $\lambda$ from 1 to 4.5, which is a feature
of the long-range interaction.

We are also interested in how the dispersion varies as the Coulomb
repulsion is increased. Variation of the singlet spectrum with varying $U$
is shown in Fig. \ref{fig:dispersionvaru}. Data are plotted for
parameters $\omega/t = 4$, $\lambda=2.5$, $\beta=3.5$ and a Fr\"ohlich
interaction. The triplet dispersion does not vary with $U$, and as $U$
is increased to very large values singlet and triplet dispersions
converge.

In figure \ref{fig:kdepsplitting}, we compute the zone center and zone
edge singlet-triplet splitting as $\omega/t$ is varied. $\lambda=8$,
$U/t=20$, $\beta = 3.5$. As phonon frequency increases, the zone edge
singlet and triplet states become closer in energy with singlet and
triplet expected to become degenerate at very large $\omega$ as is
found in the $UV$ model. Contrary to this, the zone center
singlet-triplet splitting increases with increased $\omega$. This is mainly
due to decrease in singlet energy. The triplet also
becomes lower in energy on increasing $\omega$, but the decrease is
smaller than for the singlet bipolaron.

\section{Summary}
\label{sec:summary}

In this article, we presented the results of simulations of bipolaron
properties for a series of electron-phonon models in 1D. The models
vary in terms of the interaction range, from the simple on-site
Holstein interaction, through the model suggested by Bon\v{c}a and Trugman
with interaction only between particles on nearest neighbor sites to
the full lattice Hubbard--Fr\"ohlich model. The continuous time
quantum Monte Carlo algorithm was used to simulate the models, and
full details of our algorithm have been presented. A wide range of bipolaron properties have been calculated, with a detailed overview of the parameter space and comparison made between models with differing ranges of electron-phonon interaction. In general, only a small long range interaction is necessary to significantly modify model properties from the strictly local Holstein case, with light masses and strong pairing present for even relatively strong screening.

The strength of the nearest neighbor effective interaction is found to
be the most important factor that influences bipolaron
properties. This increases the possibility of engineering a bipolaron
condensate in higher dimensions, since with shorter tails, clustering
corresponding to phase separation is less likely. Given that the
simplified near neighbor model is much better suited to analytics than
the lattice Fr\"ohlich approach, it is suggested that it should be
adopted as a more general model to investigate electron-phonon
problems in higher dimensions. Models involving ions oscillating above
the plane or chain of electrons have been suggested to derive this and
similar forms. This type of configuration is not possible in
3D. However, in $D>2$ a long range component to electron-phonon models
is expected, since a perfectly momentum independent interaction is
unlikely. Simulation of such models will form part of future
publications.

\begin{acknowledgments}
We acknowledge useful discussions with John Samson and Sasha
Alexandrov.
\end{acknowledgments}

\bibliography{coulombfrohlich_1d}

\appendix*
\section{}

The formal derivation of the continuous time limit for updates (V) to
(VIII) by considering detailed balance is very similar to that for
updates (I) to (IV), which has been described in detail elsewhere
\cite{hague2007b} (although here we make a slight modification for
weighted insertions, where kinks are more likely to have similar
values of $\tau$ according to the weighting $p(\tau,\tau')$). Thus, we
simply summarize the rules needed for implementation of the algorithm
here. We use only updates I, II, V and VIII as a minimal update set
that allows the algorithm to efficiently explore all path
configurations. A detailed derivation is then given for the multiple
kink update relating exchanged and direct configurations in subsection
\ref{update:exchange}, since there are some subtle considerations.

\subsection{Update (I): Kink addition on both paths}

Consider two path configurations $C$ and $D$, where $D$ has two
additional kinks, $\lvec$ on path A and $\lvec$ on path B. The update
proceeds by selecting kink shifts, paths and kinks to add and remove
according to the scheme,

\renewcommand{\labelenumi}{(\roman{enumi})}
\begin{enumerate}
\item Choose kink types shifts and paths according to the probability rules R1-R3.
\item Propose pair removal with probability $1/2$. Otherwise attempt
pair addition.
\item If $N_{A\lvec}=0$ or $N_{B\lvec}=0$ in the initial
configuration and pair removal was proposed, then the update is rejected.
\item For pair addition, select times with equal probability density according to rules R\ref{rule4} and R\ref{rule4a}.
\item For pair removal, select an $\lvec$ kink from path A and a $\lvec$ kink from path B according to rules R\ref{rule5} and R\ref{rule5a}.
\end{enumerate}

\begin{equation}
P({\rm addition}) =\mathrm{min}\left\{ 1,  \frac{t^2\beta e^{A(D)-A(C)}}{N_{\lvec A}(D)\sum_{i=1}^{N_{\lvec B}(D)}p(\tau,\tau_i)}\right\}
\end{equation}

\subsection{Update (II): Addition of kink and antikink pair on one path}

Consider two path configurations $C$ and $D$ that differ only in that
$D$ has additional kinks $\lvec$ on path A and $-\lvec$ on path A.

Using the following rules:

\renewcommand{\labelenumi}{(\roman{enumi})}
\begin{enumerate}
\item Choose kink types shifts and the path according to the probability rules R1-R3
\item Propose pair removal with probability $1/2$. Otherwise attempt
pair addition.
\item If $N_{A\lvec}=0$ or $N_{A-\lvec}=0$ in the initial
configuration and removal was proposed, then reject the update.
\item For pair addition, select times with probability density according to rules R\ref{rule4} and R\ref{rule4a}.
\item For pair removal, select an $\lvec$ kink from path A and a $-\lvec$ kink from path A according to rules R\ref{rule5} and R\ref{rule5a}.
\end{enumerate}

\begin{equation}
P({\rm addition}) =\mathrm{min}\left\{ 1,  \frac{t^2\beta e^{A(D)-A(C)}}{N_{\lvec A}(D)\sum_{i=1}^{N_{-\lvec A}(D)}p(\tau,\tau_i)}\right\}
\end{equation}

\subsection{Update (V): Addition of kink and antikink on different paths}

Consider two path configurations $C$ and $D$ that differ only in that $D$ has an additional hop $\lvec$ on path A and $-\lvec$ on path B.

Using the following rules:

\renewcommand{\labelenumi}{(\roman{enumi})}
\begin{enumerate}
\item Choose kink types shifts and paths according to the probability rules R1-R3
\item Propose pair removal with probability $1/2$. Otherwise attempt
pair addition.
\item If removal is chosen and $N_{A\lvec}=0$ or $N_{B-\lvec}=0$ in
the initial configuration then removal is aborted as a rejection.
\item For pair addition, select times with equal probability density according to rules R\ref{rule4} and R\ref{rule4a}.
\item For pair removal, select an $\lvec$ kink from path A and a $-\lvec$ kink from path B according to rule R\ref{rule5}.
\end{enumerate}

\begin{equation}
P(C\rightarrow D)=\mathrm{min}\left\{ 1,  \frac{t^2\beta e^{A(D)-A(C)}}{N_{\lvec A}(D)\sum_{i=1}^{N_{-\lvec B}(D)}p(\tau,\tau_i)}\right\}
\end{equation}

\subsection{Update (VIII): Addition/removal of kink pair to a single path}

Consider two configurations $C$ and $D$, where configuration $D$ has
two more kinks on path A than configuration $C$ and the same number of
kinks on path B. There is a slight complication here because there are
two ways of inserting kinks to reach the end configuration, so we
always use equal time probabilities on this update (i.e. no weighted
kink insertions).

\begin{enumerate}
\item Choose kink types shifts and paths according to the probability rules R\ref{rule1}-R\ref{rule3}.
\item Propose pair removal with probability $1/2$. Otherwise attempt
pair addition.
\item If $N_{A\lvec}< 2$ and removal has been chosen, the update is aborted.
\item For pair addition, select times with equal probability density
according to rule R\ref{rule4}. We do not correlate kink times on this
update due to ambiguities with kink insertion ordering.
\item For pair removal, select two different $\lvec$ kinks from path A.
\end{enumerate}

If the Metropolis scheme is used,
\begin{equation}
P(C\rightarrow D)=\mathrm{min}\left\{ 1, \frac{t^2\beta^2}{(N_{\lvec}(D)+1)N_{\lvec}(D)}e^{A(D)-A(C)} \right\}
\end{equation}

\subsection{Exchange by kink insertion and removal}
\label{update:exchange}

In the path-integral formalism, an exchange of particles corresponds
to swapping the ends of the paths at $\tau=\beta$. An example of an
exchanged configuration is shown in Fig. \ref{fig:examplepath}. To
make the exchange, the $\tau=\beta$ end of path B is shifted by
$-\Delta$ lattice sites, and that of path A through $+\Delta$ sites. From
here on, we refer to the $\tau=\beta$ end of a path as the `top' of
the path. This section explains how to take the continuous time limit
of this more complicated class of update.

\begin{widetext}

In the conventional way, update probabilities are determined from the
detailed balance equation,
\begin{equation}
W(C) \cdot Q({\rm forward}) \cdot P(C \rightarrow D) = W(D) \cdot Q({\rm inverse}) \cdot P(D\rightarrow C),
\end{equation}
where $W(I)$ represents the statistical weight of configuration $I$
and Q is the probability of selecting a specific update from all the
available updates (for example a specific hop direction is chosen and
a value for the number of antikinks to remove). Configuration $D$ is
the exchanged configuration, and $C$ represents the non-exchanged
configuration.

The total difference in hopping quanta between initial and final
configurations is $\Delta-2m$ since $n=\Delta-m$ kinks are added
and $m$ antikinks are removed. Therefore, the ratio of statistical weights in
the partition function is,
\begin{equation}
\frac{W(D)}{W(C)}=(t\Delta\tau)^{\Delta-2m}e^{A_A(D)-A_A(C)}
\end{equation}
where the subscript $A$ indicates that only path $A$ is considered for
now.

One way of making the $+\Delta$ shift at the top of path A is to
insert $\Delta-m$ kinks and to remove $m$ antikinks. The maximum
number of antikinks that can be removed is $m=N_{A-\lvec}$ if $\Delta
> N_{A-\lvec}$ (since it is not possible to remove more antikinks than
exist) or $m=\Delta$ otherwise. The minimum number of antikinks that
can be removed is $m=0$. Therefore, there are $N_P$ possible choices
of $m$, where $N_{P}=\Delta+1$ if $\Delta\leq N_{A-\lvec}$, otherwise
$N_{P}=N_{A-\lvec}+1$. It is convenient to write the number of
possible updates as the single expression,
\begin{equation}
N_{P}=\mathrm{min}(N_{A-\lvec},\Delta)+1.
\end{equation}
The value of $m$ is chosen with equal weighting $1/N_{P}$.

Now that a value of $m$ has been chosen, it is necessary to determine
the probability that specific kinks are inserted into or removed from
a specific configuration. The probability of removing $m$ of the
available $N_{A-\lvec}$ antikinks is simply
$1/_{N_{A-\lvec}}C_{m}$. The probability of choosing to introduce
$n=\Delta-m$ kinks into configuration $C$ with equal weighting $1/\beta$
is $(\Delta-m)!(\Delta\tau/\beta)^{\Delta-m}$ (the probability is
increased because there are $(\Delta-m)!$ possible ways to order the
kink insertion that are not distinguishable in the final
configuration).

All of these considerations contribute to the probability of updating
from configurations $C$ to $D$ via the product,
\begin{equation}
Q(\mathrm{forward}) = \frac{(\Delta-m)!}{\mathrm{min}(N_{A-\lvec},\Delta)+1}\left(\frac{\Delta\tau}{\beta}\right)^{\Delta-m}\frac{1}{_{N_{A-\lvec}}C_m}
\end{equation}

The inverse process changes the path from configuration $D$ to
$C$. Similar to the forward process, up to $\Delta$ {\rm kinks} can be
removed if they exist, otherwise, only
$N_{A\lvec}+\Delta-m=N_{A\lvec}+n$ kinks may be removed. Therefore
there are $N_P=\mathrm{min}(N_{A\lvec}+n,\Delta)+1$ possible ways to
choose how many kinks to remove. In the reciprocal process, $m$
antikinks are inserted, and $\Delta-m$ kinks are removed from
$N_{A\lvec}+\Delta-m$ kinks, so the total probability for choosing a
specific inverse process is,
\begin{equation}
Q(\mathrm{inverse})=\frac{m!}{\mathrm{min}(N_{A\lvec}+n,\Delta)+1}\left(\frac{\Delta\tau}{\beta}\right)^{m}\frac{1}{_{N_{A\lvec}+\Delta-m}C_{\Delta-m}}
\end{equation}
Substituting these probabilities into the detailed balance equation,
it is possible to write the ratio update probabilities for a single
path as,
\begin{equation}
\frac{P_A(C\rightarrow D)}{P_A(D\rightarrow C)} = \frac{m!}{(\Delta-m)!}\frac{\mathrm{min}(N_{A-\lvec},\Delta)+1}{\mathrm{min}(N_{A\lvec}+n,\Delta)+1}\left(\frac{\Delta\tau}{\beta}\right)^{2m-\Delta}(t\Delta\tau)^{\Delta-2m}\frac{_{N_{A-\lvec}}C_m}{_{N_{A\lvec}+\Delta-m}C_{\Delta-m}}
\end{equation}
\end{widetext}
There is a cancellation of the discrete time interval $\Delta\tau$, so
the continuous time limit can be taken immediately,
\begin{equation}
\frac{P_A(C\rightarrow D)}{P_A(D\rightarrow C)} = (t\beta)^{2n-\Delta}\frac{\mathrm{min}(N_{A-\lvec},\Delta)+1}{\mathrm{min}(N_{A\lvec}+n,\Delta)+1}\frac{_{N_{A-\lvec}}P_{\Delta-n}}{_{N_{A\lvec}+n}P_{n}}
\end{equation}

A similar result is obtained for path B, with kinks replaced by
antikinks and vice-versa. The probabilities for the two paths are
independent, and so the total probability is simply a product of the
two. Once the ratio of update probabilities has been calculated, the
Metropolis update scheme can be applied to obtain a suitable
Monte-Carlo procedure. We note that the factorials in this expression
are a potential cause of overflow, so we work with logarithms in our
code.

\subsection{Common segment exchange}

There is another possible exchange type, which involves swapping the
allocation of kinks between paths for $\tau>\tau_{CS}$ for any common
segment at $\tau_{CS}$, i.e. kinks at times $\tau>\tau_{CS}$
associated with path A are reassigned to path B and those assigned to
path B are reassigned to path A (common segment means that the two
paths occupy the same lattice site at the same imaginary time). Common
segment exchange is not applicable for all path configurations, but
has potential to speed up computation. In our algorithm, we attempt a
common segment exchange with probability $p_{CS}$ every time an
exchange is attempted. Once the attempt is started, we test for the
existence of a common segment. If there are no shared segments, then
the update is rejected and no further exchange is attempted on that
time step. Otherwise, the common segment exchange is accepted with
probability $P(C\rightarrow D) = {\rm min}\left\{1,\exp(A(D) -
A(C))\right\}$.

\subsection{Path shift}

To encourage rapid exploration of the path configurations, it is also
convenient to introduce another update that is attempted with a small
independent probability relative to the other updates. In this update,
one of the paths is shifted by a random number of lattice
spacings. These updates are useful when the bipolaron is only just
bound. The update proceeds as follows:

Attempt the update with an independent probability on each step. Test
the configuration of the paths. If the paths are in an exchanged
configuration, then the update is rejected. Otherwise, select one of
the paths with equal probability. Choose a displacement for the path
at random, and choose whether to move the path left or right with
equal probability (important for the detailed balance so the move is
reversible). The move is accepted with probability $P(C\rightarrow D)
= {\rm min}\left\{1,\exp(A(D) - A(C))\right\}$.

%

\end{document}